\documentclass{jfm}

\usepackage{graphicx}
\usepackage{epstopdf,epsfig}
\usepackage{natbib}
\usepackage{hyperref}
\usepackage{bm}
\usepackage{subcaption}
\usepackage{threeparttable}
\usepackage{comment}

\hypersetup{
    colorlinks = true,
    urlcolor   = blue,
    citecolor  = black,
}

\newcommand{\RomanNumeralCaps}[1]
\linenumbers

\title{Enhancement of wave transmissions in multiple radiative and convective zones}

\author{Tao Cai\aff{1}
  \corresp{\email{tcai@must.edu.mo}},
  Cong Yu\aff{2,1}
  \corresp{\email{yucong@mail.sysu.edu.cn}}
 \and Xing Wei\aff{3}}

\affiliation{\aff{1}State Key Laboratory of Lunar and Planetary Sciences, Macau University of Science and Technology, Macau, People's Republic of China
\aff{2}School of Physics and Astronomy, Sun Yat-sen University, Zhuhai, 519082, People's Republic of China
\aff{3}Department of Astronomy, Beijing Normal University, Beijing, People's Republic of China}

\begin{document}
\maketitle

\begin{abstract}
In this paper, we study wave transmission in a rotating fluid with multiple alternating convectively stable and unstable layers. We have discussed wave transmissions in two different circumstances: cases where the wave is propagative in each layer and cases where wave tunneling occurs. We find that efficient wave transmission can be achieved by `resonant propagation' or `resonant tunneling', even when stable layers are strongly stratified, and we call this phenomenon `enhanced wave transmission'. Enhanced wave transmission only occurs when the total number of layers is odd (embedding stable layers are alternatingly embedded within clamping convective layers, or vise versa). For wave propagation, the occurrence of enhanced wave transmission requires that clamping layers have similar properties, the thickness of each clamping layer is close to a multiple of the half wavelength of the corresponding propagative wave, and the total thickness of embedded layers is close to a multiple of the half wavelength of the corresponding propagating wave (resonant propagation). For wave tunneling, we have considered two cases: tunneling of gravity waves and tunneling of inertial waves. In both cases, efficient tunneling requires that clamping layers have similar properties, the thickness of each embedded layer is much smaller than the corresponding e-folding decay distance, and the thickness of each clamping layer is close to a multiple-and-a-half of half wavelength (resonant tunneling).
\end{abstract}



\section{Introduction}
\label{sec:intro}
Inertial and gravito-inertial waves are important phenomena in rotating stars and planets. Wave propagation can transport momentum and energy, therefore it may have significant impact on stellar or planetary structures and evolutions. For example, internal-gravity waves (IGWs) play an important role in transporting angular momentum when they propagate in the radiative zones of stars \citep{belkacem15a,belkacem15b,pincon17,aerts19}. Study reveals that IGWs can reduce differential rotation in low mass stars in a short timescale \citep{rogers13}. It can also explain the misalignment of exoplanets around hot stars \citep{rogers12}. Apart from IGWs, inertial waves can also be generated in rotating planets \citep{ogilvie04,wu05a,goodman09}. It has been found that the resonantly excited inertial wave has important impact on the tidal dissipation in planets \citep{wu05b}.

It is quite common for a star or planet to have a multi-layer structure. For example, superadiabatic region embedded in radiative layers may appear in neutron star's atmosphere because of the ionization of $^{56}Fe$ \citep{miralles97}. Waves generated by convective motions can transport energy to the chromosphere and corona, which may drive the stellar wind. In A-type stars, it is possible for them to have complex internal structures with multiple convection zones. Interaction between these convective zones has important implications in material mixing and energy transport \citep{silvers07}. It is also known that layered semiconvection zones can be formed in stars in the process of double-diffusive convection \citep{mirouh12,wood13,garaud18}. For main sequence stars slightly more massive than the Sun, it has been found that the efficiency of mixing in layered semiconvection zones sensitively depends on the layer height \citep{moore16}. Layered convection zones also exist in planets. Multi-layer structure has been detected in a region several hundred meters below the surface of the Arctic ocean \citep{rainville2008mixing}. Observation shows that a shallow convective region is embedded within stable layers in the atmosphere of Venus \citep{tellmann2009structure}. Seismology on Saturn's ring reveals the layered stable stratification in the deep interior of Saturn \citep{fuller14a}. If stable stratification exists in the deep interior of Saturn, then g-mode can be excited in this region. Layered convection probably also exists in Jupiter. New model with layered convection on Jupiter and Saturn indicates that the heavy elements in our Jovian planets are more enriched than previously thought \citep{leconte12}. One interesting question is whether g-mode waves can transmit to the surface, so that they can be possibly captured by observations.

Wave can transmit in a double barrier system through a tunneling process. \citet{sutherland2004internal} studied transmissions of internal wave tunneling for both $N^2$-barrier (low $N^2$ layer embedded in high $N^2$ layers and horizontal mean density varies continuously) and mixed-$N^2$ (low $N^2$ layer embedded in high $N^2$ layers but horizontal mean density varies discontinuously) profiles, where $N^2$ is the square of buoyancy frequency. They found that wave transmission can be efficient by resonant transfers. \citet{sutherland2016internal} investigated the transmission of internal waves in a multi-layer structure separated by discontinuous density jumps. He deduced an analytical solution for wave transmission when the steps are evenly spaced, and predicted that waves with longer horizontal wavelength and larger frequencies are more likely to transmit in the density staircase profile. \citet{sutherland1996internal} considered wave propagation in a profiles of piecewise linear stratified layers with weaker stratification at the top. He discovered that large-amplitude IGWs incident from the bottom can partially transmit energy into the top layer by the generation of lower frequency wave packet.
Resonant tunneling of electron transmission in double barriers is familiar in quantum physics, and has been widely used in designing semiconductor devices, such as tunnel diode, NPN (negative-positive-negative) and PNP (positive-negative-positive) triodes \citep{singh2010semiconductor}. In comparison with tunneling of electron transmission, it is expected that resonant tunneling also occurs for wave transmission in multi-layer structures.

Wave transmission in a three-layer structure with rotational effects has been considered by \citet{gerkema2008internal}. They compared wave transmissions with and without traditional approximations (the horizontal component of rotation is neglected when traditional approximation is adopted on a {\it f}-plane). For a three-layer structure with a convective layer embedded in strongly stratified layers, waves cannot survive in both convective and stratified layers under the traditional approximation, while it is possible if non-traditional effects are taken into account. They also showed that near-inertial waves are always transmitted efficiently for stratified layers of any stratification.  \citet{belyaev15} investigated the free modes of a multi-layer structure wave propagation with rotation at the poles and equator. They found that g-modes with vertical wavelengths smaller than the layer thickness are evanescent. \citet{andre17} studied the effects of rotation on free modes and wave transmission in a multi-layer structure at a general latitude. They showed that transmission can be efficient when the incident wave is resonant with waves in adjacent layers with half-wavelengths equal to the layer depth. They also discovered that perfect wave transmission can be obtained at the critical latitude. \citet{pontin20} studied the wave propagation in semiconvective regions of non-rotating giant planets in the full sphere. They found that wave transmissions are efficient for very large wavelength waves.

In previous works \citep{wei20,wei2020erratum,cai20}, we have discussed the efficiency of inertial and gravito-inertial wave transmissions in a two-layer structure on {\it f}-planes. For a step stratification near the interface, we have found that the transmission generally is not efficient if the stable layer is strongly stratified. In this paper, we investigate the wave transmission in a multi-layer structure on {\it f}-planes. Specifically, we will consider wave transmissions in two different mechanisms: wave propagation and wave tunneling. We find that wave transmission in a multi-layer structure can be significantly different from that in a two-layer structure.

\section{The model and result}
For the Boussinesq flow in a rotating {\it f}-plane, the hydrodynamic equations can be synthesized into a partial differential equation on vertical velocity $w$ \citep{gerkema05}:
\begin{eqnarray}
\nabla^2 w_{tt}+(\bm{f}\bm{\cdot} \bm{\nabla})^2 w +N^2(z) \nabla_{h}^2 w=0~, \label{eq1}
\end{eqnarray}
where $\bm{f}=(0,\tilde{f},f)$ is the vector of coriolis parameters, $N^2$ is the square of the buoyancy frequency, $\nabla^2$ is Laplacian operator and $\nabla_{h}^2$ is its horizontal component, and the subscript $t$ represents taking derivative with respect to time. By letting $w=W(\chi,z)\exp(-i\sigma t)$, \citet{gerkema05} found (\ref{eq1}) can be transformed into the following equation:
\begin{eqnarray}
AW_{\chi\chi}+2BW_{\chi z}+CW_{zz}=0, \label{eq2}
\end{eqnarray}
where $A=\tilde{f}_{s}^2-\sigma^2+N^2$, $B=f\tilde{f}_{s}$, $C=f^2-\sigma^2$, $\tilde{f}_{s}=\tilde{f}\sin\alpha$, and $\chi$ is a variable satisfying $x=\chi\cos\alpha$ and $y=\chi\sin\alpha$. Here $x,y,z$ are the west-east, south-north, and vertical directions, respectively. We define $A_{0}=\tilde{f}_{s}^2-\sigma^2$, so that $A=A_{0}+N^2$. Taking advantage of plane waves and assuming $W(\chi,z)=\psi(z)\exp(i\delta z+ik\chi)$, \citet{gerkema05} have further simplified (\ref{eq2}) into an equation of wave amplitude $\psi$:
\begin{eqnarray}
\psi_{zz}+k^2 \frac{B^2-AC}{C^2}\psi=0~.\label{eq3}
\end{eqnarray}
where $\delta=-kB/C$. A wave solution then requires $B^2-AC>0$, and the squared wavenumber is $r^2=k^2(B^2-AC)/C^2$. If $B^2-AC>0$, then $r$ is a real number and the flow propagates along the vertical direction as a wave. On the other hand, if $B^2-AC<0$, then $r$ is a pure imaginary number, and wave amplitude increases or decreases exponentially.

In previous works \citep{wei20,cai20}, we have investigated wave transmission in a two-layer setting {\it f}-plane (a convective layer with $N^2=0$ and a convectively stable layer with $N^2>0$). Note that the actual $N^2$ should be smaller than zero. In real stars, however, convection is generally efficient on transporting energy, leading to a nearly adiabatic thermal structure. Thus $N^2$ in convective layers are only slightly smaller than zero. For this reason, we choose $N^2=0$ for convective layers in our model. In this paper, we extend previous works to study the wave transmission among multi-layer setting {\it f}-plane. At this beginning stage, we use an ideal model by assuming $N^2$ is a constant in each layer. In all convective layers, $N^2$ are equal and set to be zeros. In stable layers, $N^2$ can be different but remains a constant in each layer, and its minimum and maximum values are $N_{min}^2$ and $N_{max}^2$, respectively. We also assume that the propagation of inertial waves is not affected by convection. The validity of this assumption requires that the nonlinear and viscous effects are small. Detailed discussion on relations and differences between convection and inertial waves could be found in \citet{zhang2017theory}. Here we attempt to use a toy model to gain some insights on wave transmissions in a multi-layer structure.

\citet{cai20} have made a detailed discussion on the frequency main of wave solutions.
If a wave could survive in a convective layer, then the following condition must be satisfied:
\begin{eqnarray}
&&\Delta_{c}=B^2-A_{0}C>0~.\label{eq4}
\end{eqnarray}
Similarly, condition for wave propagation in a stable layer is
\begin{eqnarray}
&&\Delta_{s}=B^2-A_{0}C-N_{s}^2C>0~.\label{eq5}
\end{eqnarray}
Let us define
\begin{eqnarray}
&&\sigma_{1,4}^2=\left[(f^2+\tilde{f}_{s}^2)\mp \sqrt{(f^2+\tilde{f}_{s}^2)^2}\right]/2~,\\
&&\sigma_{2,5}^2=\left[(f^2+\tilde{f}_{s}^2+N_{min}^2)\mp \sqrt{(f^2+\tilde{f}_{s}^2+N_{min}^2)^2-4N_{min}^2 f^2}\right]/2~,\\
&&\sigma_{3,6}^2=\left[(f^2+\tilde{f}_{s}^2+N_{max}^2)\mp \sqrt{(f^2+\tilde{f}_{s}^2+N_{max}^2)^2-4N_{max}^2 f^2}\right]/2~.
\end{eqnarray}
where $\sigma_{1}=0$ and $\sigma_{4}=f^2+\tilde{f}_{s}^2$ are roots of $\Delta_{c}=0$, and $\sigma_{2,5}$ and $\sigma_{3,6}$ are roots of $\Delta_{s}=0$ when $N_{s}^2=N_{min}^2$ and $N_{s}^2=N_{max}^2$, respectively. It is not difficult to verify that the relation $\sigma_{1}\leq\sigma_{2}\leq\sigma_{3}\leq\sigma_{4}\leq\sigma_{5}\leq\sigma_{6}$ holds.

\begin{figure}
\centering
\includegraphics[width=\linewidth]{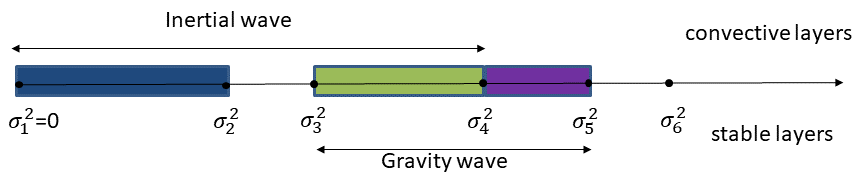}
\caption{Sketch plot of wave propagations in multiple convectively stable and unstable layers. Wave frequency ranges are shown above and below the middle arrow for convective and stable layers, respectively. Wave propagation occurs in the green region, tunneling of inertial wave occurs in the blue region, and tunneling of gravity wave occurs in the purple region. \label{fig:f1}}
\end{figure}

Fig.\ref{fig:f1}(a) shows the frequency ranges for different waves in convectively unstable and stable layers, respectively. In the green region ($\sigma_{3}^2<\sigma^2<\sigma_{4}^2$), waves can survive and propagate in both convective and stable layers, and we term this phenomenon `wave propagation'. In the blue region $\sigma_{1}^2<\sigma^2<\sigma_{2}^2$, inertial waves can survive in convective layers but gravity waves cannot survive in stable layers. Inertial waves can transmit through a tunneling process, and we term this phenomena `tunneling of inertial wave'. In the purple region $\sigma_{4}^2<\sigma^2<\sigma_{5}^2$, gravity waves can survive in stable layers but inertial waves cannot survive in convective layers. Similarly, gravity waves can transmit through a tunneling process, and we term this phenomena `tunneling of gravity wave'.

In \citet{cai20}, we have also deduced that $c_{p0}c_{g}$ has the same sign as $C$, where $c_{p0}$ is the modified vertical component of wave phase velocity and $c_{g}$ is the vertical component of the wave group velocity. The vertical component of phase velocity should be computed by $\sigma/(\delta\pm r)$, but here the tilted effect is excluded in the modified one $c_{p0}=\sigma/(\pm r)$. Since the wave direction of energy propagation is determined by $c_{g}$, a proper choice of wave direction depends on whether the wave is sub-inertial ($C>0$) or super-inertial ($C<0$).

\subsection{Wave propagation}
\begin{figure}
\centering

\begin{subfigure}{0.45\textwidth}
\includegraphics[width=\linewidth]{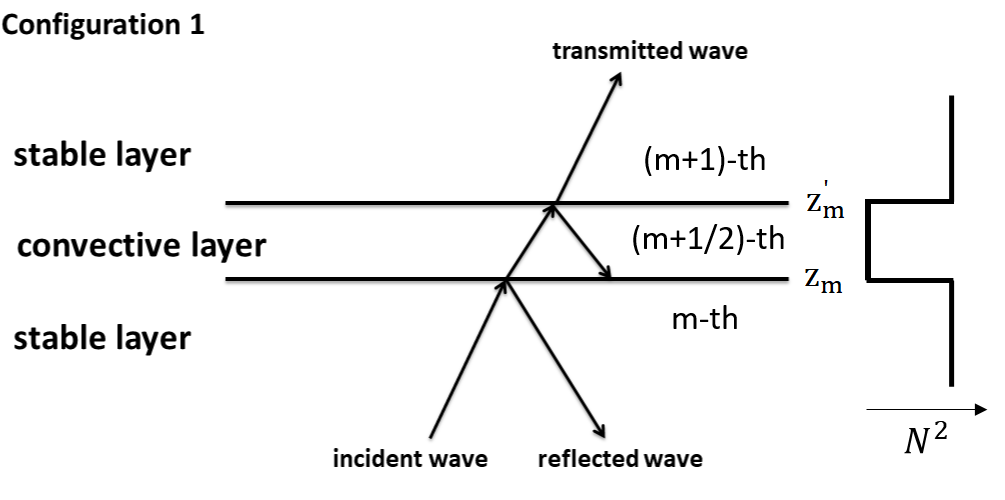}
\caption{}
\end{subfigure}
\begin{subfigure}{0.45\textwidth}
\includegraphics[width=\linewidth]{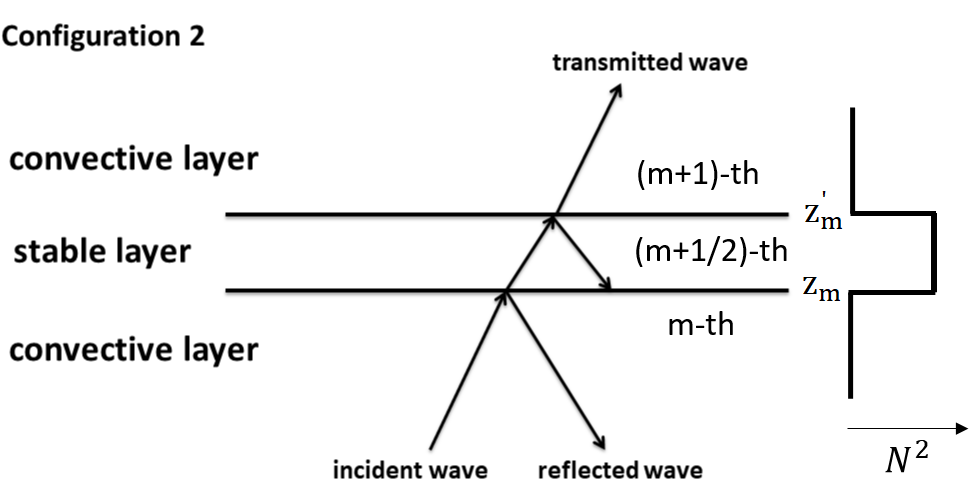}
\caption{}
\end{subfigure}

\caption{Sketch plots of wave propagations in multiple convectively stable and unstable layers. (a)The convective layer is embedded between two stable layers. (b)The stable layer is embedded between two convective layers. Structure of the square of buoyancy frequency $N^2$ is shown by the side. \label{fig:f2}}
\end{figure}

In this section, we discuss wave propagation in both convective and stable layers, which requires the frequency is in the range $\sigma_{3}^2<\sigma^2<\sigma_{4}^2$.
We consider different configurations with different combinations of layer structures and wave directions. The incident wave can propagate from the convective layer or stable layer, and the wave propagating direction can be upward or downward. Since up/down symmetry holds in Boussinesq flow, it is sufficient to discuss the cases with waves incident from the bottom. The cases with waves incident from the top can be inferred from the up/down symmetry. Fig.~\ref{fig:f2} shows sketch plots of the two configurations in a three-layer setting {\it f}-plane.  The structures with more layers are similar. At each interface, two boundary conditions have to be satisfied: the vertical velocity is continuous; and the first derivative of the vertical velocity is continuous \citep{wei20,cai20}.

We start the discussion on configuration 1 (fig.~\ref{fig:f2}(a)), and assume that the number of interfaces is even. We label the convective layer with the half grid number ($m+1/2$). For the $(m+1/2)$-th convective layer , we label its lower and upper neighbouring stable layers as $m$-th and $(m+1)$-th stable layers, respectively; and we set the locations of the lower and upper interfaces at $z_{m}$ and $z'_{m}$, respectively. The thickness of the convective layer is $\Delta z'_{m}=z'_{m}-z_{m}$, and the thickness of the stable layer is $\Delta z_{m}=z_{m}-z'_{m-1}$. For the $m$-th stable layer, we define its wavenumber square as $s_{m}^2=k^2(B^2-A_{0}C-N_{m}^2 C)/C^2$. For the $(m+1/2)$-th convective layer, we define its wavenumber square as $q_{m}^2=k^2(B^2-A_{0}C)/C^2$.
In this paper, we only consider a simple case, in which the square of buoyancy frequency is a constant within each convective or stable layer. Under this assumption, we find wave solutions in the $m$-th stable layer is
\begin{eqnarray}
\psi_{m}(z)=a_{m}e^{-is_{m}z}+b_{m}e^{is_{m}z}, \quad z\in (z'_{m-1},z_{m})~,\label{eq9}
\end{eqnarray}
and in the $(m+1/2)$-th convective layer is
\begin{eqnarray}
\psi_{m+1/2}(z)=c_{m}e^{-iq_{m}z}+d_{m}e^{-iq_{m}z}, \quad z\in (z_{m},z'_{m})~,\label{eq10}
\end{eqnarray}
respectively. As mentioned earlier, $c_{p0}c_{g}$ has the same sign as C, from which we conclude $Sgn(c_{g})=Sgn(C)Sgn(c_{p0})$. The sign of the modified vertical phase velocity $c_{p0}$ is determined by the sign of $q_{m}$ or $s_{m}$. If $Sgn(C)Sgn(c_{p0})>0$, then the vertical group velocity $c_{g}$ is positive (wave direction is outgoing); on the other hand, if $Sgn(C)Sgn(c_{p0})<0$, then $c_{g}$ is negative (wave direction is incoming). We choose $q_{m}>0$ for $C>0$, and $q_{m}<0$ for $C<0$. For either case, the waves with wavenumbers $q_{m}$ and $s_{m}$ are outgoing waves, and the waves with wavenumbers $-q_{m}$ and $-s_{m}$ are incoming waves. Note that all $q_{m}$ and $s_{m}$ always have the same sign. Matching the boundary conditions at the interfaces $z_{m}$ and $z'_{m}$, we have
\begin{eqnarray}
&&a_{m}e^{-is_{m}z_{m}}+b_{m}e^{is_{m}z_{m}}=c_{m}e^{-iq_{m}z_{m}}+d_{m}e^{iq_{m}z_{m}}~,\label{eq11}\\
&&(-a_{m}e^{-is_{m}z_{m}}+b_{m}e^{is_{m}z_{m}})s_{m}=(-c_{m}e^{-iq_{m}z_{m}}+d_{m}e^{iq_{m}z_{m}})q_{m}~,\\
&&c_{m}e^{-iq_{m}z'_{m}}+d_{m}e^{iq_{m}z'_{m}}=a_{m+1}e^{-is_{m+1}z'_{m}}+b_{m+1}e^{is_{m+1}z'_{m}}~,\\
&&(-c_{m}e^{-iq_{m}z'_{m}}+d_{m}e^{iq_{m}z'_{m}})q_{m}=(-a_{m+1}e^{-is_{m+1}z'_{m}}+b_{m+1}e^{is_{m+1}z'_{m}})s_{m+1}~,\label{eq14}
\end{eqnarray}
Wave propagations in multi-layer structures have been investigated in \citet{belyaev15}, \citet{andre17} and \citet{pontin20}. A useful approach on modeling wave propagations in a multi-layer structure is to build on relations of wave amplitudes by transfer matrices.
From the above boundary conditions, we can derive the following transfer relations (see Appendix \ref{appendixa}):
\begin{eqnarray}
\left[
\begin{array}{c}
a_{m}\\
b_{m}
\end{array}
\right]=\bm{T}_{m,m+1}
\left[
\begin{array}{c}
a_{m+1}\\
b_{m+1}
\end{array}
\right]
=\left[
\begin{array}{cc}
  \widehat{T}_{11} & \widehat{T}_{12} \\
  \widehat{T}_{21} & \widehat{T}_{22}
\end{array}
\right]
\left[
\begin{array}{c}
a_{m+1}\\
b_{m+1}
\end{array}
\right]~,
\label{eq15}
\end{eqnarray}
where
\begin{eqnarray}
&&\widehat{T}_{11}=\frac{1}{4}e^{i(s_{m}z_{m}-s_{m+1}z'_{m})}[(1+\frac{q_{m}}{s_{m}}+\frac{s_{m+1}}{s_{m}}+\frac{s_{m+1}}{q_{m}})e^{iq_{m}\Delta z'_{m}} \label{eq16}\\\nonumber
&& \quad +(1-\frac{q_{m}}{s_{m}}+\frac{s_{m+1}}{s_{m}}-\frac{s_{m+1}}{q_{m}})e^{-iq_{m}\Delta z'_{m}}]~,\\
&&\widehat{T}_{12}=\frac{1}{4}e^{i(s_{m}z_{m}+s_{m+1}z'_{m})}[(1+\frac{q_{m}}{s_{m}}-\frac{s_{m+1}}{s_{m}}-\frac{s_{m+1}}{q_{m}})e^{iq_{m}\Delta z'_{m}}\\\nonumber
&& \quad +(1-\frac{q_{m}}{s_{m}}-\frac{s_{m+1}}{s_{m}}+\frac{s_{m+1}}{q_{m}})e^{-iq_{m}\Delta z'_{m}}]~,
\end{eqnarray}
with $\widehat{T}_{21}=\widehat{T}_{12}^{*}$ and $\widehat{T}_{22}=\widehat{T}_{11}^{*}$. Here the asterisk symbol represents the conjugate of a complex number.

\subsubsection{The three-layer case}
If we consider the special three-layer case (outgoing transmitted wave requires $a_{2}=0$), we can quickly obtain
\begin{eqnarray}
\left[
\begin{array}{c}
a_{1}\\
b_{1}
\end{array}
\right]=\left[
\begin{array}{cc}
T_{11} & T_{12}\\
T_{21} & T_{22}
\end{array}
\right]
\left[
\begin{array}{c}
0\\
b_{2}
\end{array}
\right]~,
\end{eqnarray}
where
\begin{eqnarray}
&&\left[
\begin{array}{cc}
  T_{11} & T_{12} \\
  T_{21} & T_{22}
\end{array}
\right]=\bm{T}_{1,2}~,
\end{eqnarray}
Thus we have
\begin{eqnarray}
&&a_{1}=T_{12}b_{2}~,\\
&&a_{2}=T_{22}b_{2}~.
\end{eqnarray}
In \citet{cai20}, we found that the averaged energy flux of a wave is
\begin{eqnarray}
\langle F \rangle=\frac{C\Im(\psi_{z}/\psi)}{2k^2\sigma}|\psi|^2~,
\end{eqnarray}
where $\Im$ denotes the imaginary part of a complex number.
Let us define the overall transmission ratio as
\begin{eqnarray}
\eta=\frac{|\langle F \rangle_{t}|}{|\langle F \rangle_{i}|}
\end{eqnarray}
where $\langle F \rangle_{i}$ is the averaged energy flux of the incident wave at the lowermost interface, and $\langle F \rangle_{t}$ is the averaged energy flux of the transmitted wave at the uppermost interface. For this three-layer case, we have
\begin{eqnarray}
\eta=|\frac{s_{2}}{s_{1}}|\frac{|b_{2}|}{|a_{2}|}=|\frac{s_{2}}{s_{1}}|\frac{1}{|T_{22}|^2}~.
\end{eqnarray}
From (\ref{eq16}) and the relation between $T_{22}$ and $T_{11}$, we obtain
\begin{eqnarray}
|T_{22}|^2&&=\frac{1}{4}[(1+\frac{s_{2}}{s_{1}})^2\cos^2(q_{1}\Delta z'_{1})+(\frac{q_{1}}{s_{1}}+\frac{s_{2}}{q_{1}})^2\sin^2(q_{1}\Delta z'_{1})]~\nonumber\\
&&=\frac{1}{4}(1+\frac{s_{2}}{s_{1}})^2\cos^2 (q_{1}\Delta z'_{1})+\frac{1}{4}(\frac{q_{1}}{s_{1}}+\frac{s_{2}}{q_{1}})^2\sin^2(q_{1}\Delta z'_{1})~.
\end{eqnarray}
Therefore, the overall transmission ratio is
\begin{eqnarray}
\eta=\left[\frac{1}{4}\left(\sqrt{\frac{s_{1}}{s_{2}}}+\sqrt{\frac{s_{2}}{s_{1}}}\right)^2\cos^2(q_{1}\Delta z'_{1})+\frac{1}{4}\left(\sqrt{\frac{q_{1}^2}{s_{1}s_{2}}}+\sqrt{\frac{s_{1}s_{2}}{q_{1}^2}}\right)^2\sin^2(q_{1}\Delta z'_{1})\right]^{-1}~.\label{eq28}
\end{eqnarray}
Note that the terms inside square roots are always positive, no matter what direction of the propagating wave is. It is easy to prove $(\sqrt{s_{1}/s_{2}}+\sqrt{s_{2}/s_{1}})^2/4\geq 1$, and so as the term before $\sin^2(q_{1}\Delta z_{1}')$. Thus we can show $\eta \leq [\cos^2(q_{1}\Delta z_{1}')+\sin^2 (q_{1}\Delta z_{1}')]^{-1}=1$, which means that $\eta$ is always smaller than or equal to 1. \citet{gerkema2008internal} has obtained similar formula in their study on internal-wave transmission in weakly stratified layers. Their equation (38) is a special case of our formula (\ref{eq28}) with $s_{1}=s_{2}$. Some interesting conclusions can readily be drawn from (\ref{eq28}). In the previous investigation of a two-layer structure \citep{wei20,cai20}, it has been found that wave transmission is hindered (because vertical wavelengths vary significantly across the interface) when the stable layer is strongly stratified ($N^2/(2\Omega)^2 \gg 1 $). However, it is not always the case in the three-layer structure. For example, when $|q_{1}|\Delta z'_{1} \rightarrow \ell\pi$ ($\ell$ is a positive integer number) and $s_{1}/s_{2} \rightarrow 1$, we see that $\eta \rightarrow 1$. For this case, there is no reflection and all of the incident wave is transmitted. This result is independent of $N^2/(2\Omega)^2$, and it holds for both weakly ($N^2/(2\Omega)^2 \ll 1 $) and strongly ($N^2/(2\Omega)^2 \gg 1 $) stratified rotating fluids. To better understand the behavior of $\eta$, we separate $\eta$ into two parts: the first part is the solution at $\sin^2(q_{1}\Delta z'_{1})=0$ and the second part is the solution at $\cos^2(q_{1}\Delta z'_{1})=0$. The transmission ratio $\eta$ of the general case is a weighted harmonic mean of $\eta_{1}$ and $\eta_{2}$.

For the first part, the condition $\sin^2(q_{1}\Delta z'_{1})=0$ is equivalent to $\Delta z'_{1}=\ell \pi/|q_{1}|$. In other words, it requires that the thickness of the middle convective layer is a multiple of the half wavelength of the propagating wave.
In such case, the overall transmission ratio is
\begin{eqnarray}
\eta_{1}=\left[\frac{1}{4}\left(\sqrt{\frac{s_{1}}{s_{2}}}+\sqrt{\frac{s_{2}}{s_{1}}}\right)^2\right]^{-1}~.
\end{eqnarray}
From this equation, we see that $\eta_{1}$ only depends on the wavesnumber ratio $s_{2}/s_{1}$ of the stable layers. $\eta_{1}$ decreases with $s_{2}/s_{1}$ when $s_{2}/s_{1}>1$, and increases with $s_{2}/s_{1}$ when $s_{2}/s_{1}<1$. The maximum value $\eta_{1}=1$ is achieved at $s_{2}/s_{1}=1$. Thus the transmission is efficient when $|N_{1}^2-N_{2}^2|/(2\Omega)^2$ is small, or when the wave is at a critical colatitude $\theta_{c}=\cos^{-1}\pm \sigma/(2\Omega)$ (because $C\rightarrow 0$ when $\theta\rightarrow \theta_{c}$), or when both stable layers are weakly stratified (because $N_{1,2}^2C \ll B^2-A_{0}C$). The latter two points have also been observed in the two-layer structure \citep{cai20}, while the first point is new in the three-layer structure. Enhancement of (near-inertial) wave transmission near the critical colatitude was also reported in \citet{gerkema2008internal} and \citet{andre17}. Efficient wave transmission at critical colatitude and weakly stratified flow can be explained by a common reason: the inertial and gravity waves separated by an interface have almost the same vertical wavelengths, and thus these waves are `resonant' at the interface. Similar reason can be used to explain the enhanced transmission when the degree of stratifications in both clamping layers (in configuration 1, the {\it embedding} convective layer is embedded within two neighboring {\it clamping} stable layers) are similar: the incident wave is `resonant' with waves in adjacent layers with wavelengths equal to free modes of the multi-layer structure \citep{andre17}.

\begin{figure}
\centering

\begin{subfigure}{0.3\textwidth}
\includegraphics[width=\linewidth]{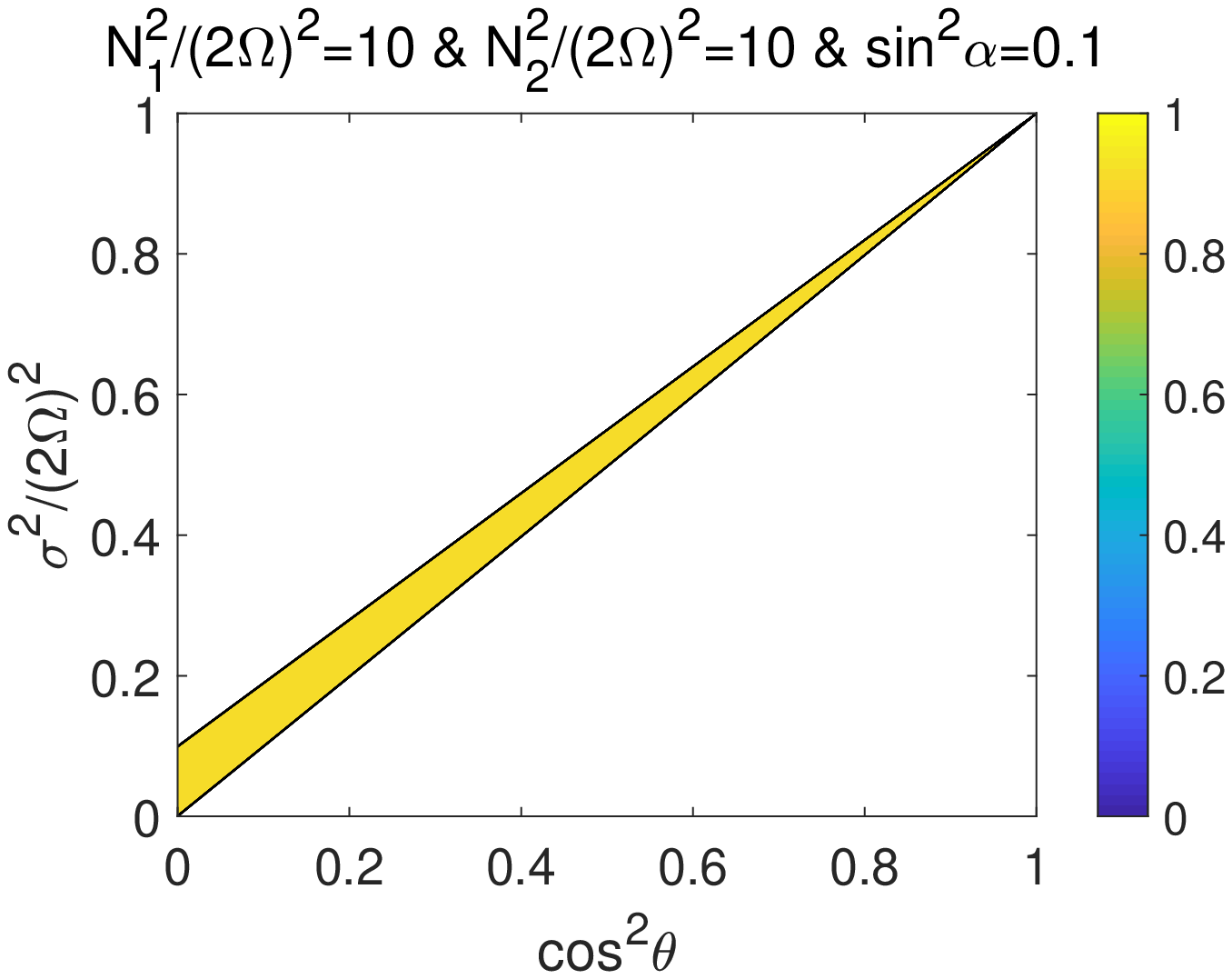}
\caption{}
\end{subfigure}
\begin{subfigure}{0.3\textwidth}
\includegraphics[width=\linewidth]{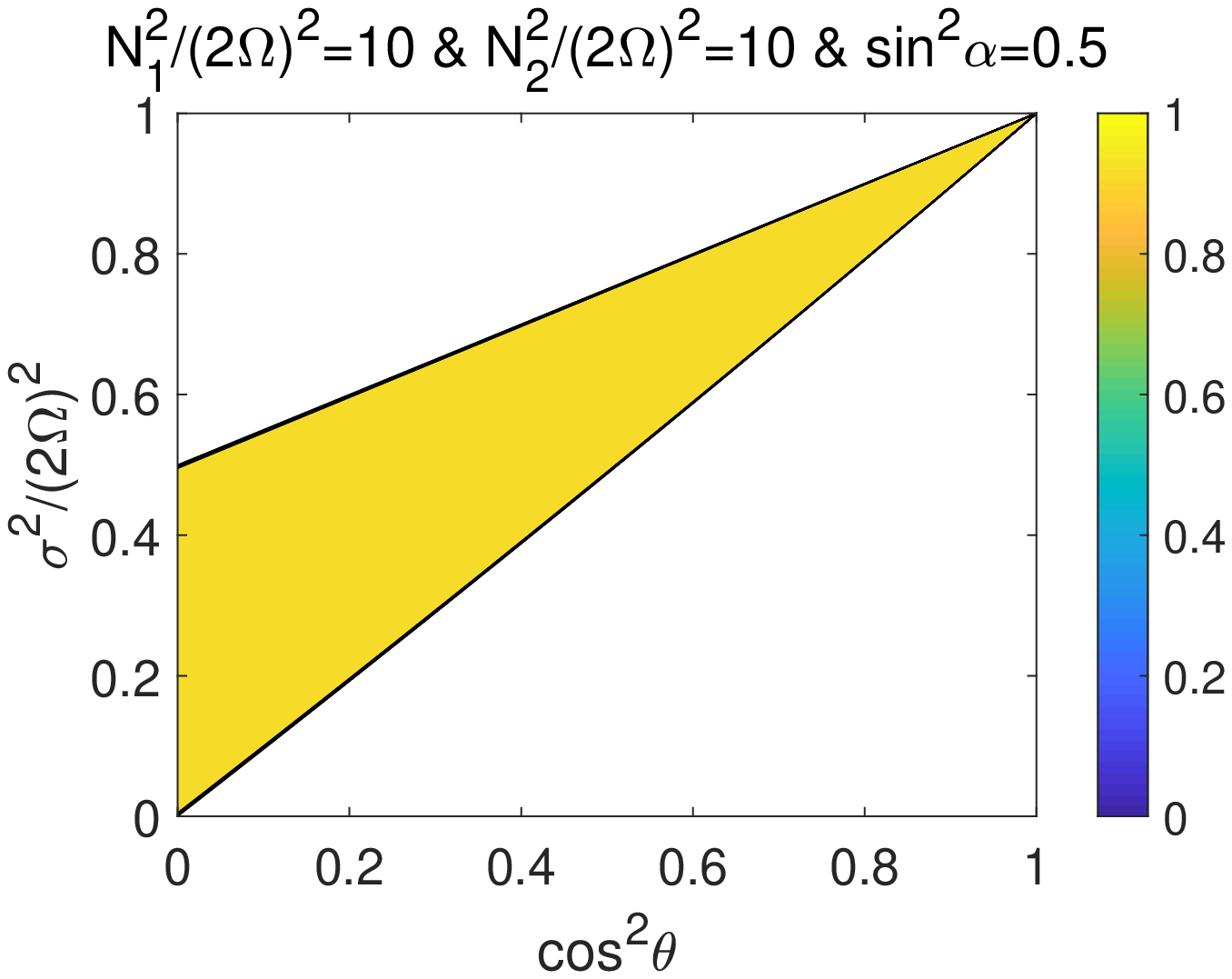}
\caption{}
\end{subfigure}
\begin{subfigure}{0.3\textwidth}
\includegraphics[width=\linewidth]{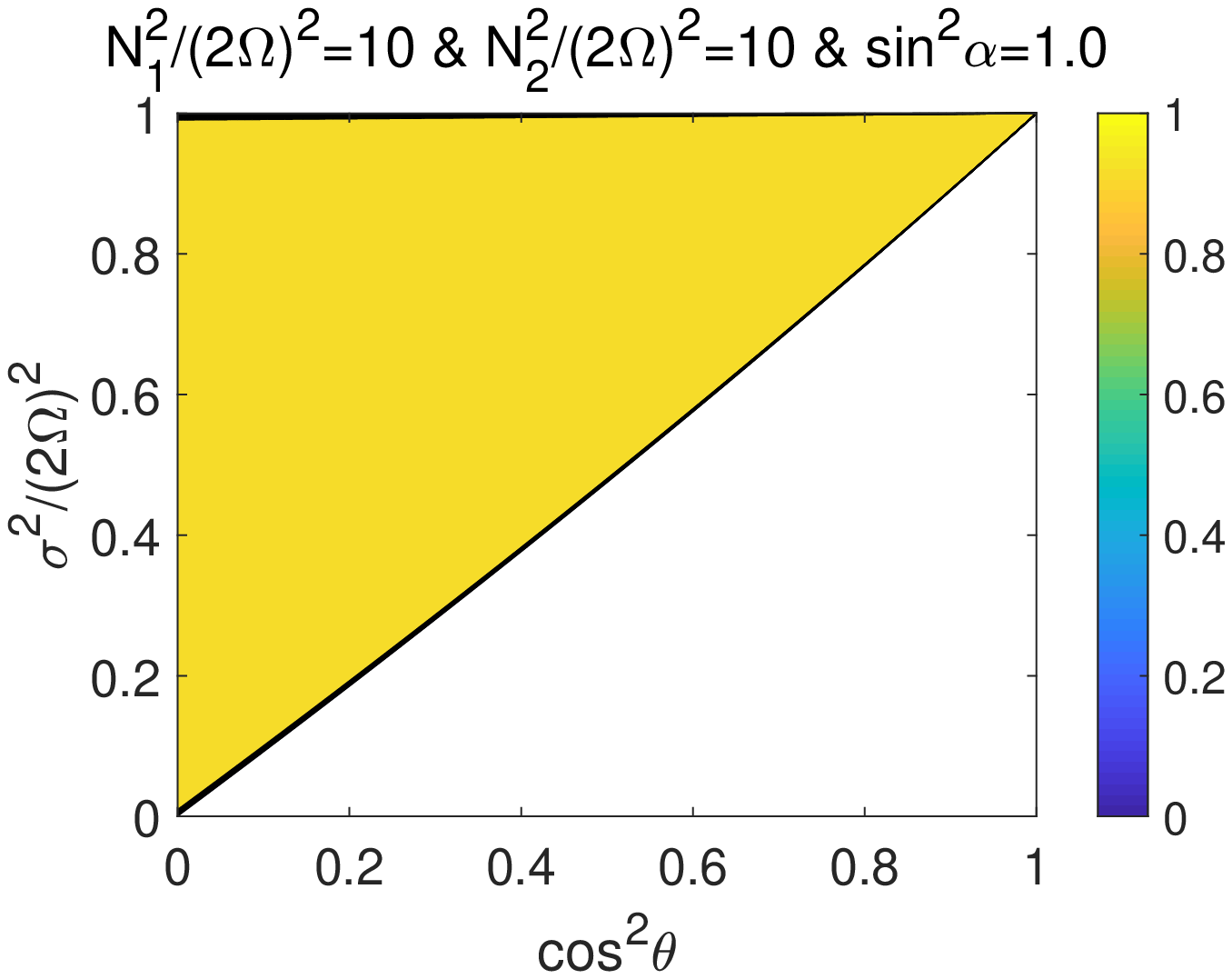}
\caption{}
\end{subfigure}

\medskip

\begin{subfigure}{0.3\textwidth}
\includegraphics[width=\linewidth]{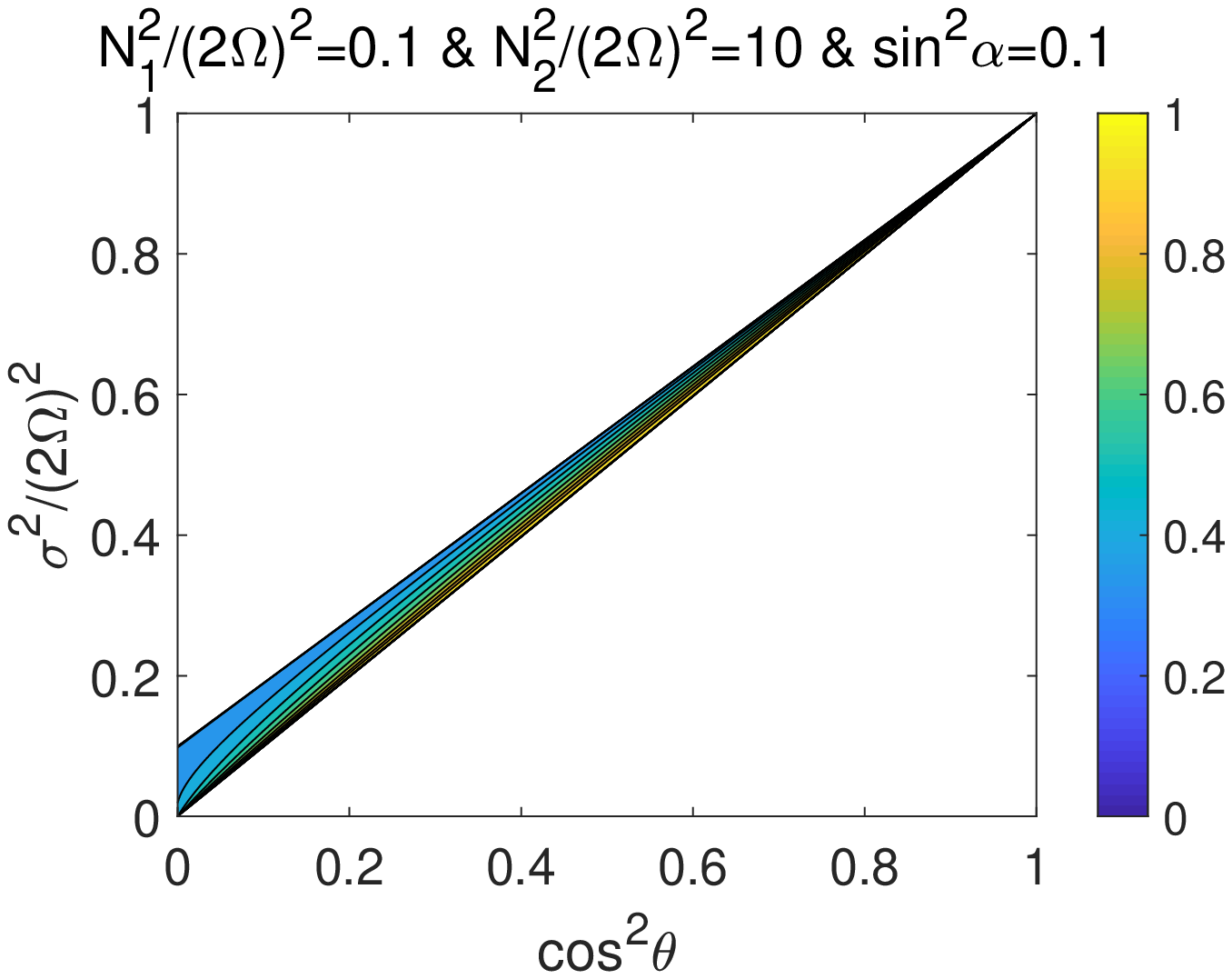}
\caption{}
\end{subfigure}
\begin{subfigure}{0.3\textwidth}
\includegraphics[width=\linewidth]{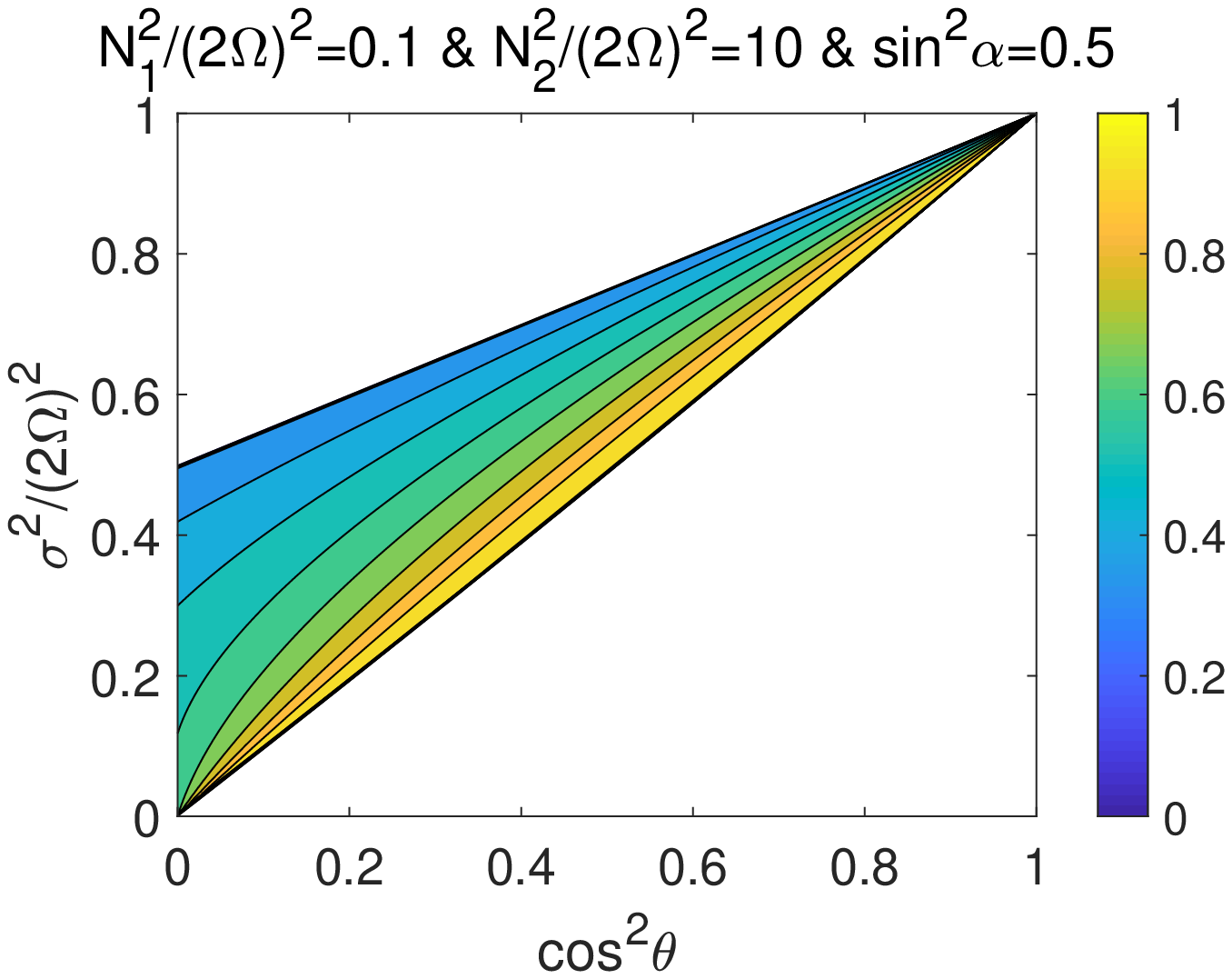}
\caption{}
\end{subfigure}
\begin{subfigure}{0.3\textwidth}
\includegraphics[width=\linewidth]{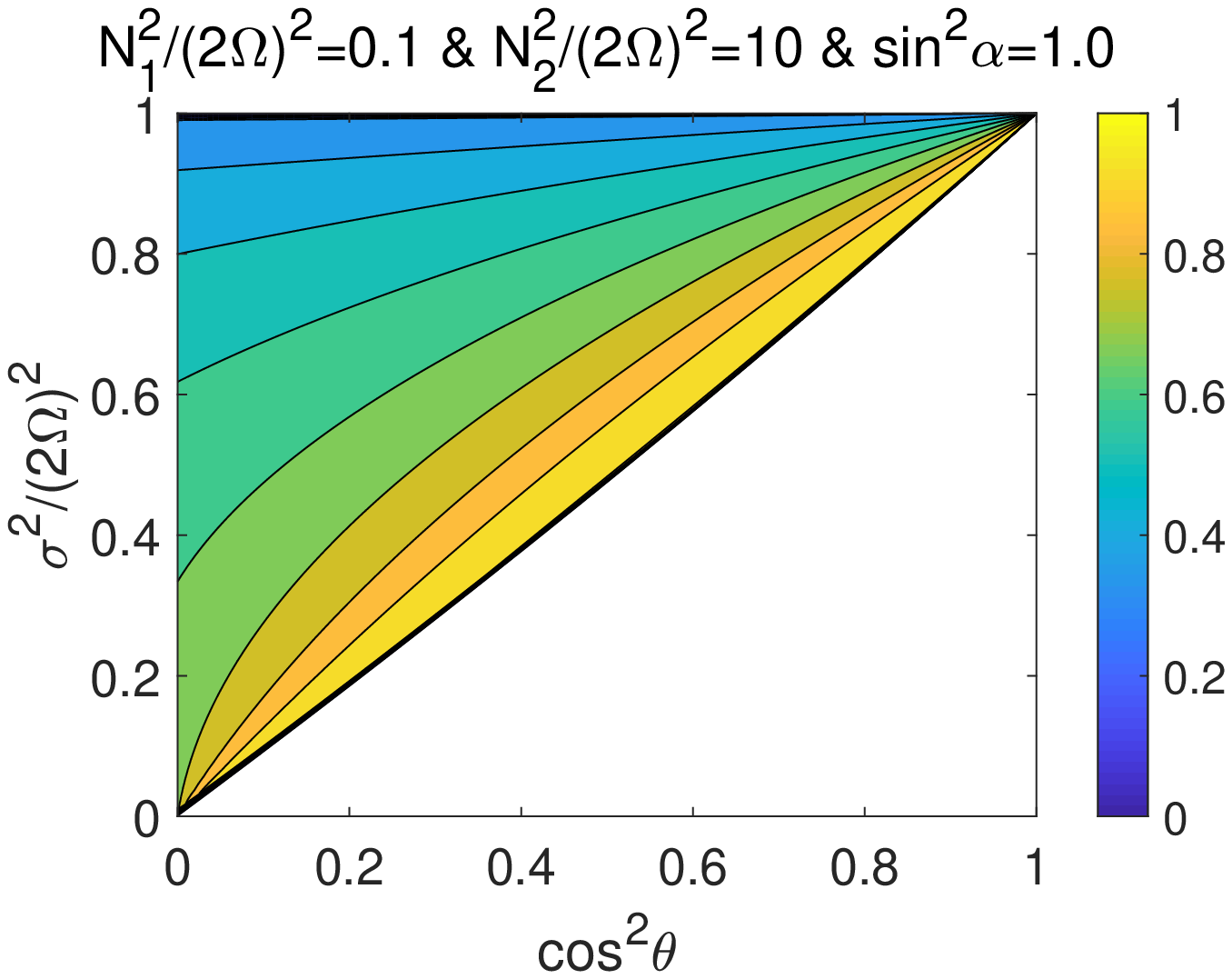}
\caption{}
\end{subfigure}

\medskip

\begin{subfigure}{0.3\textwidth}
\includegraphics[width=\linewidth]{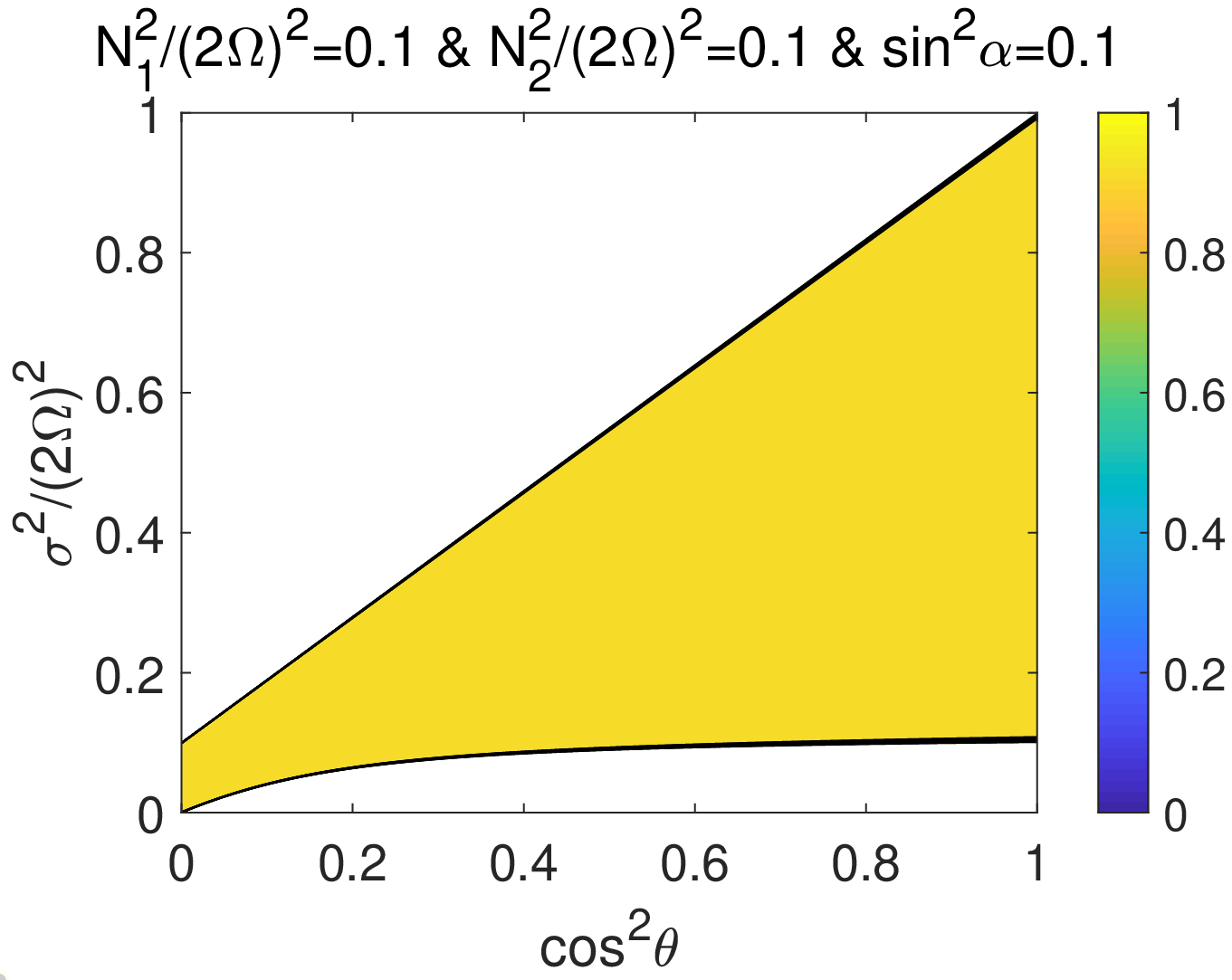}
\caption{}
\end{subfigure}
\begin{subfigure}{0.3\textwidth}
\includegraphics[width=\linewidth]{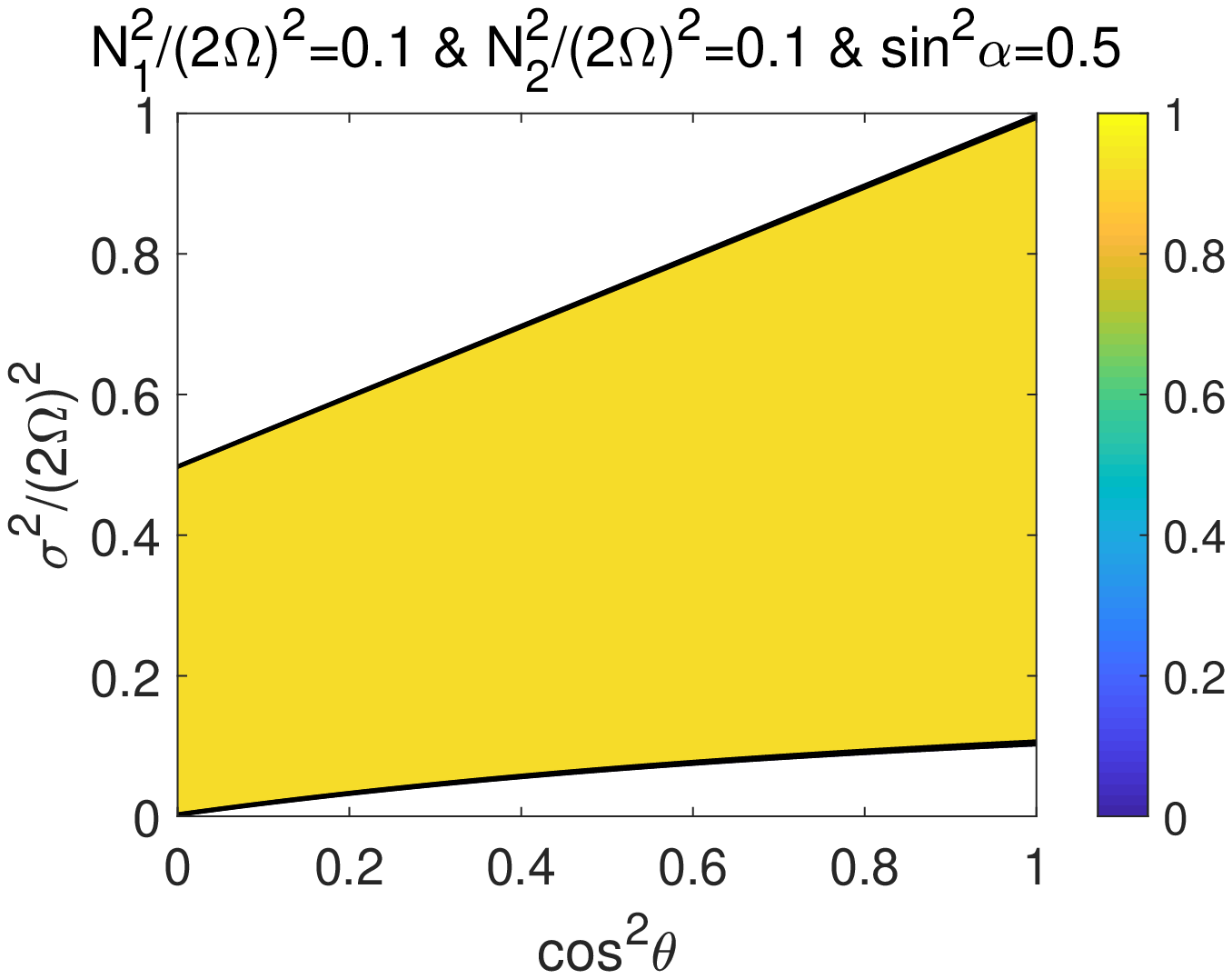}
\caption{}
\end{subfigure}
\begin{subfigure}{0.3\textwidth}
\includegraphics[width=\linewidth]{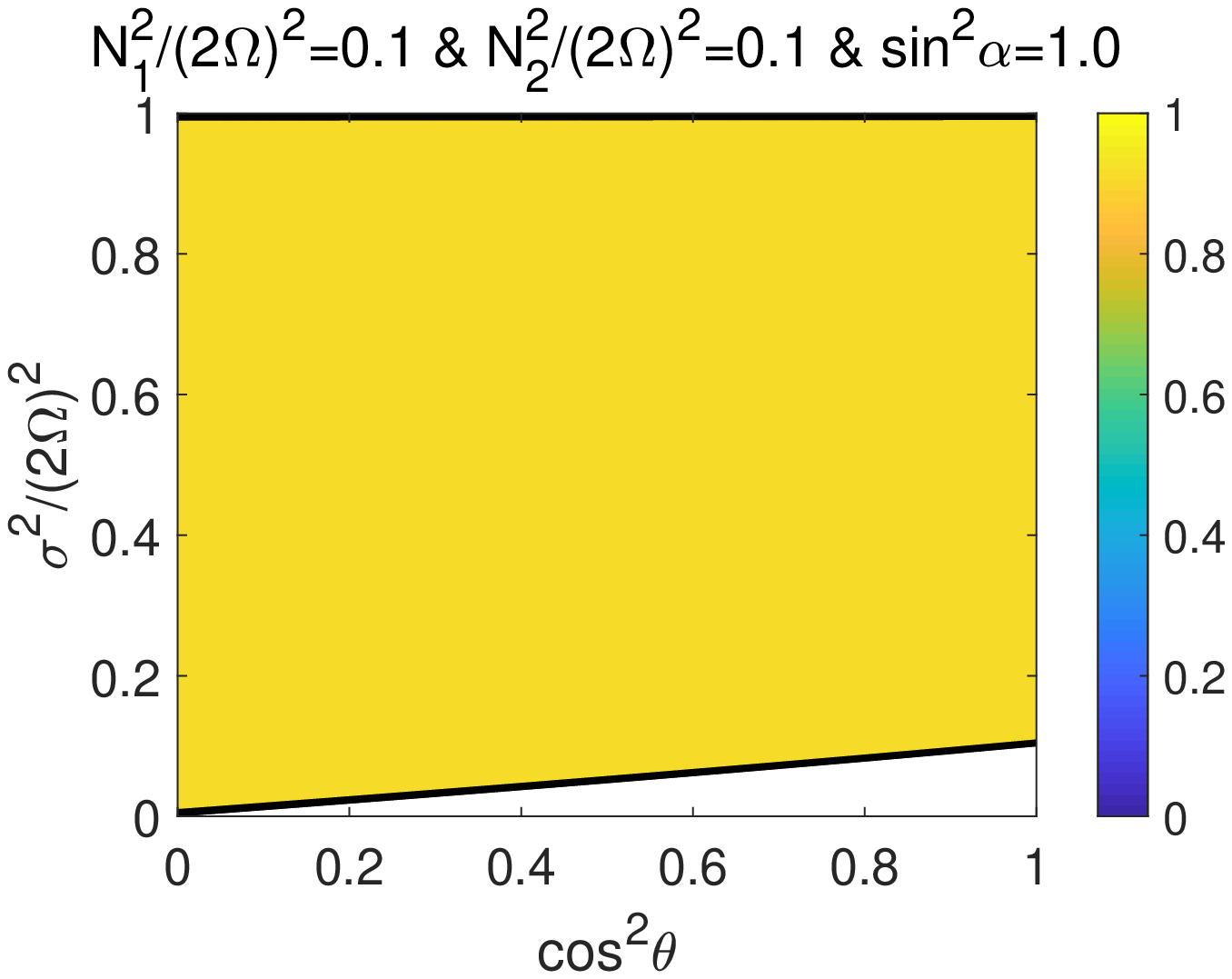}
\caption{}
\end{subfigure}

\caption{Transmission coefficient $\eta_{1}$ (the first part of (\ref{eq28})) for different $N_{1,2}^2/(2\Omega)^2$ and $\sin^2\alpha$ in a three-layer structure (one middle convective layer and two upper and lower neighbouring stable layers). The horizontal axis is $\cos^2\theta$, and the vertical axis is $\sigma^2/(2\Omega)^2$. From the left to the right panels, $\sin^2\alpha$ increases from 0.1 to 1.0. (a)-(c) Both stable layers are strongly stratified. (d)-(f) One stable layers are strongly stratified and the other is weakly stratified. (g)-(i) Both stable layers are weakly stratified. Wave propagation can only occur in colored regions. Regions are left white if wave propagation is prohibited. \label{fig:f3}}
\end{figure}

Fig.~\ref{fig:f3} shows the transmission ratio $\eta_{1}$ at three different combinations of $N_{1}^2$ and $N_{2}^2$. In the uppermost panel (figs.~\ref{fig:f3}(a-c)), both stable layers are strongly stratified, but $N_{1}^2/(2\Omega)^2$ is equal to $N_{2}^2/(2\Omega)^2$. It clearly shows that the transmission is enhanced. The middle panel (figs.~\ref{fig:f3}(d-f)) shows the result of the cases with one weakly and one strongly stratified layers. Apparently, the transmission is not as efficient as the cases shown in the uppermost panel. The lowermost panel (figs.~\ref{fig:f3}(g-i)) presents the result of the cases with two weakly stratified stable layers. The transmission is efficient because the rotational effect is important. In previous study on wave transmission in a two-layer structure \citep{cai20}, it has been shown that wave can be efficiently transmitted when the stable layer is weakly stratified. From fig.~\ref{fig:f3}, we also observe that $\sin^2\alpha$ has important effect on the frequency range. Frequency range increases with increasing $\sin^2\alpha$. When stable layers (or any of them) are strongly stratified, wave can only survive in a very thin region if $\sin^2\alpha$ is small. In the extreme case $\sin^2\alpha=0$, the surviving frequency range vanishes.

\begin{figure}
\centering

\begin{subfigure}{0.3\textwidth}
\includegraphics[width=\linewidth]{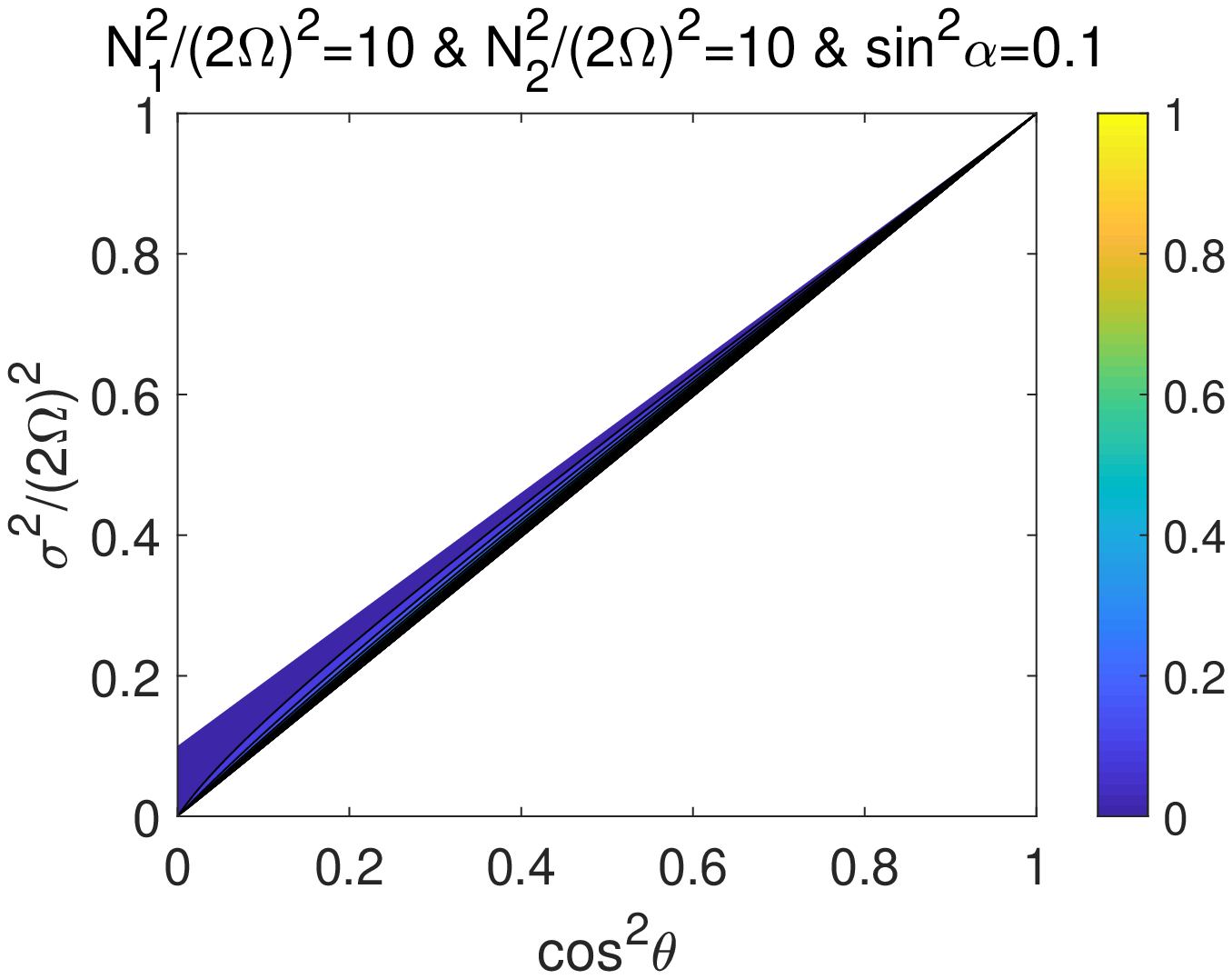}
\caption{}
\end{subfigure}
\begin{subfigure}{0.3\textwidth}
\includegraphics[width=\linewidth]{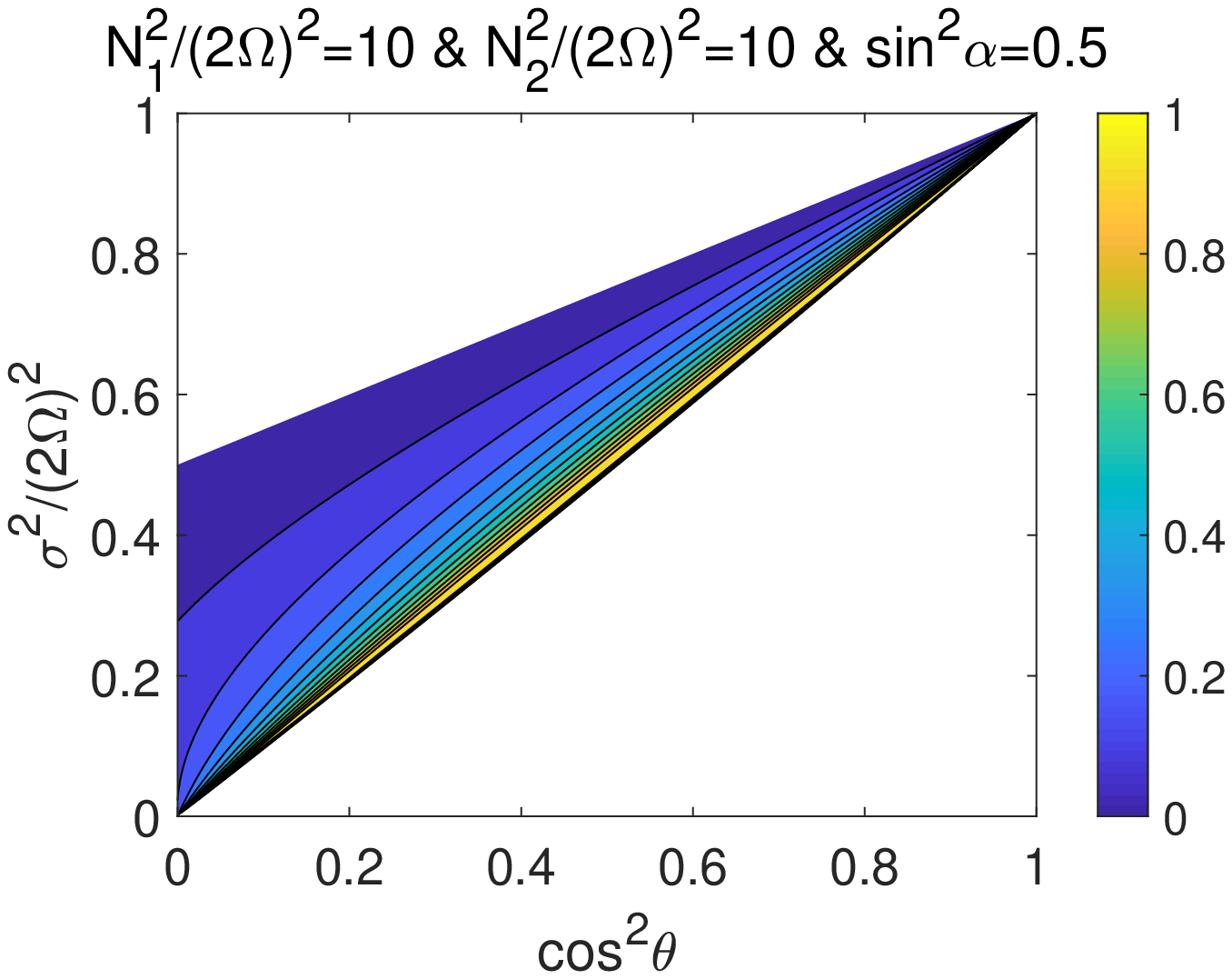}
\caption{}
\end{subfigure}
\begin{subfigure}{0.3\textwidth}
\includegraphics[width=\linewidth]{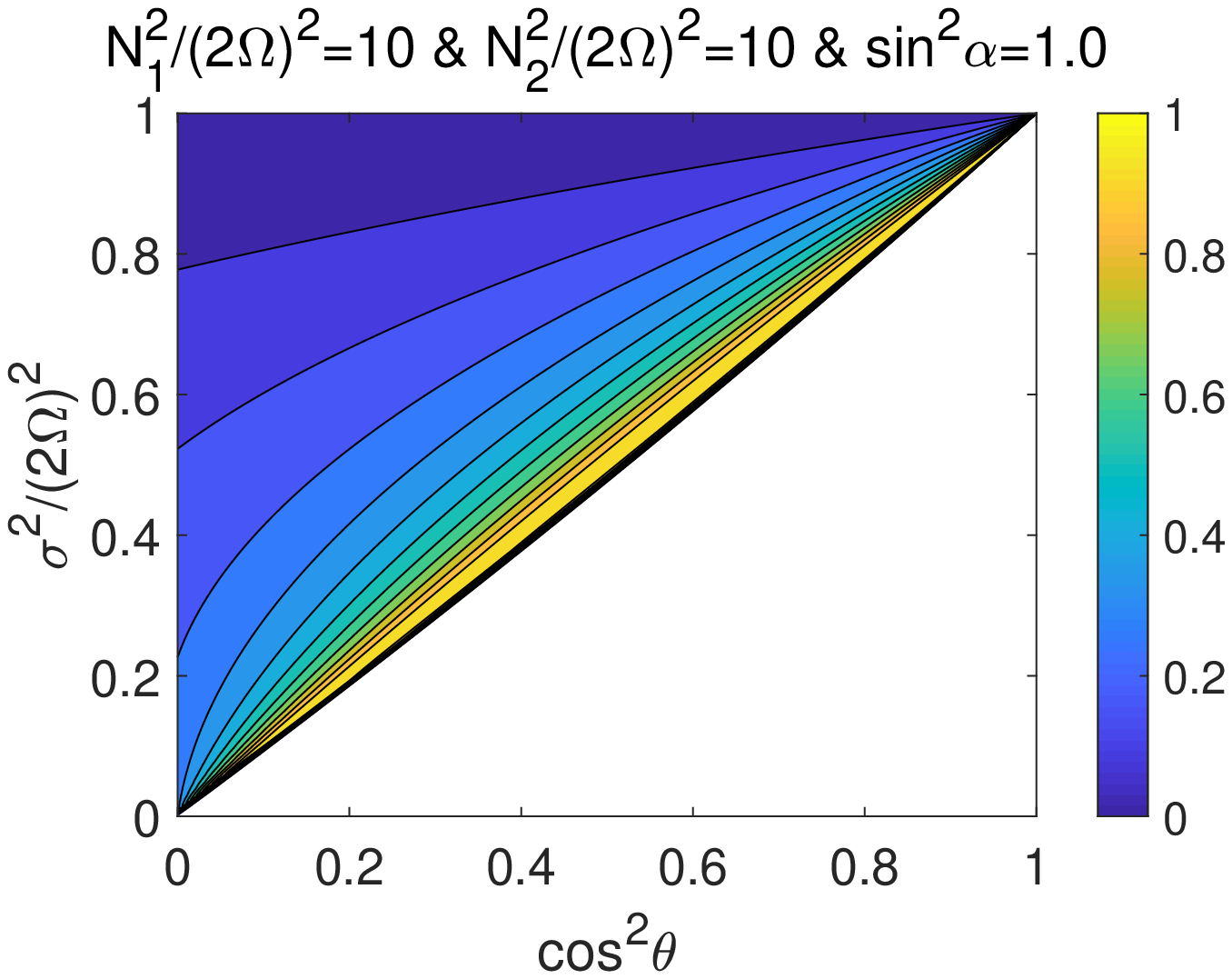}
\caption{}
\end{subfigure}

\medskip

\begin{subfigure}{0.3\textwidth}
\includegraphics[width=\linewidth]{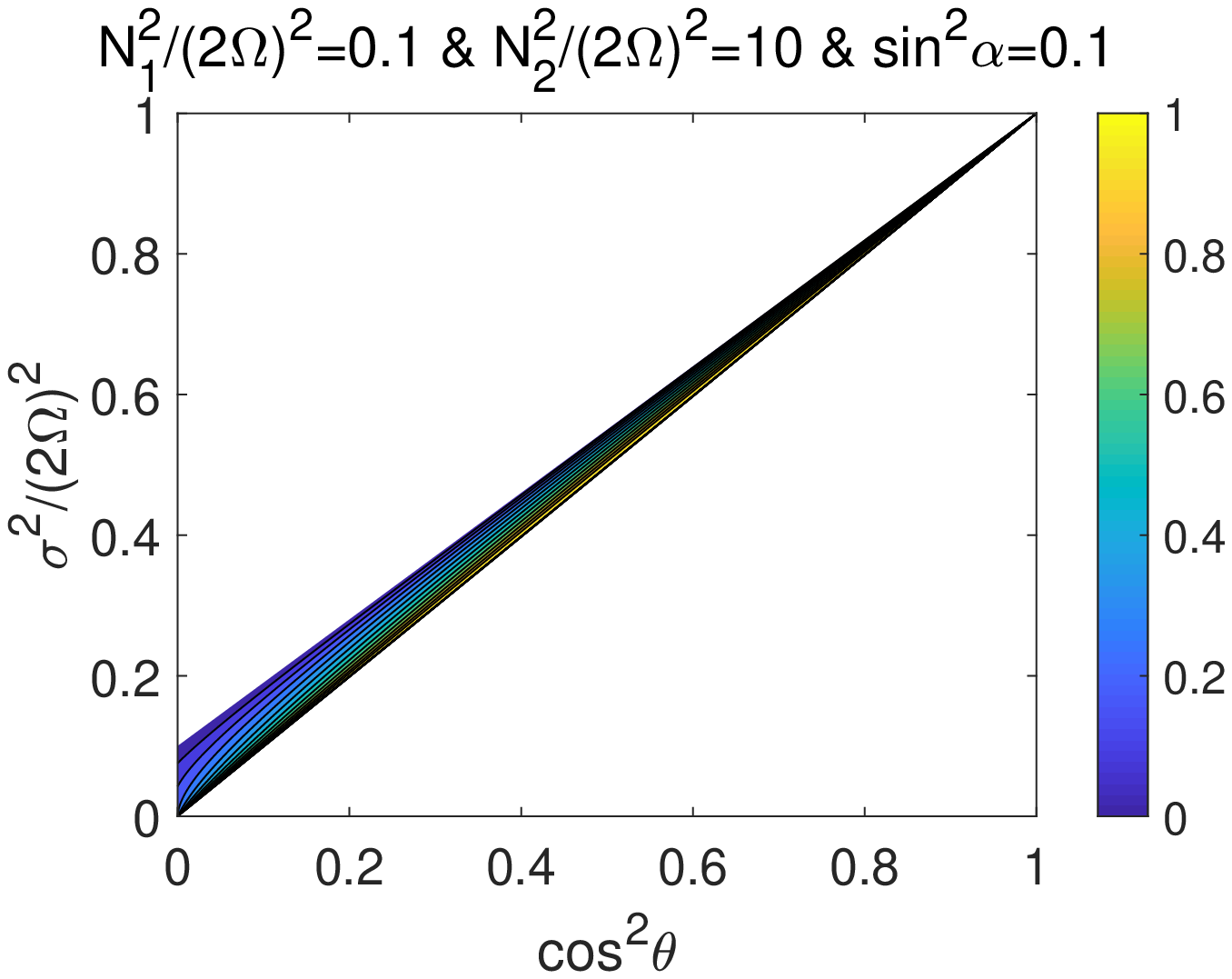}
\caption{}
\end{subfigure}
\begin{subfigure}{0.3\textwidth}
\includegraphics[width=\linewidth]{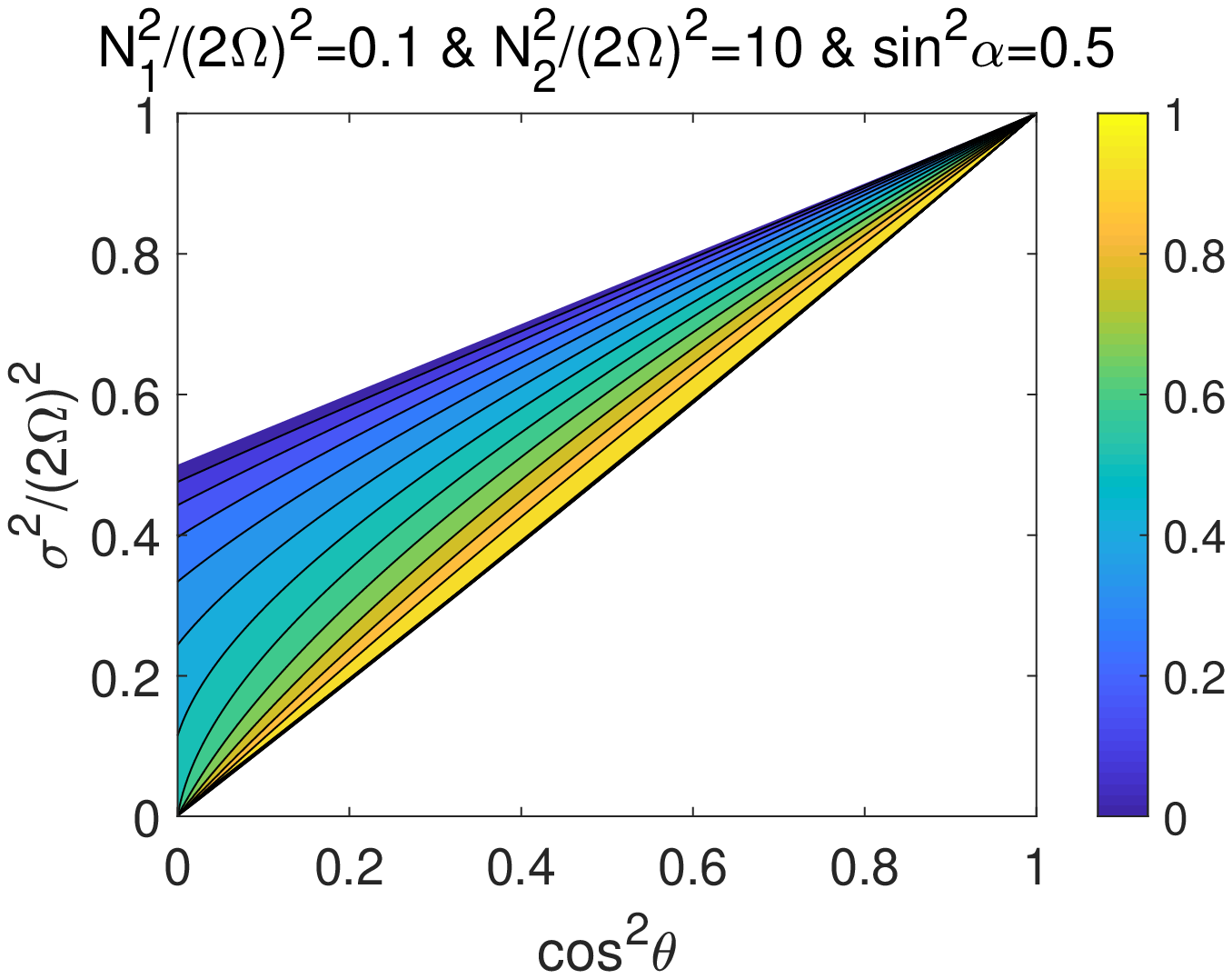}
\caption{}
\end{subfigure}
\begin{subfigure}{0.3\textwidth}
\includegraphics[width=\linewidth]{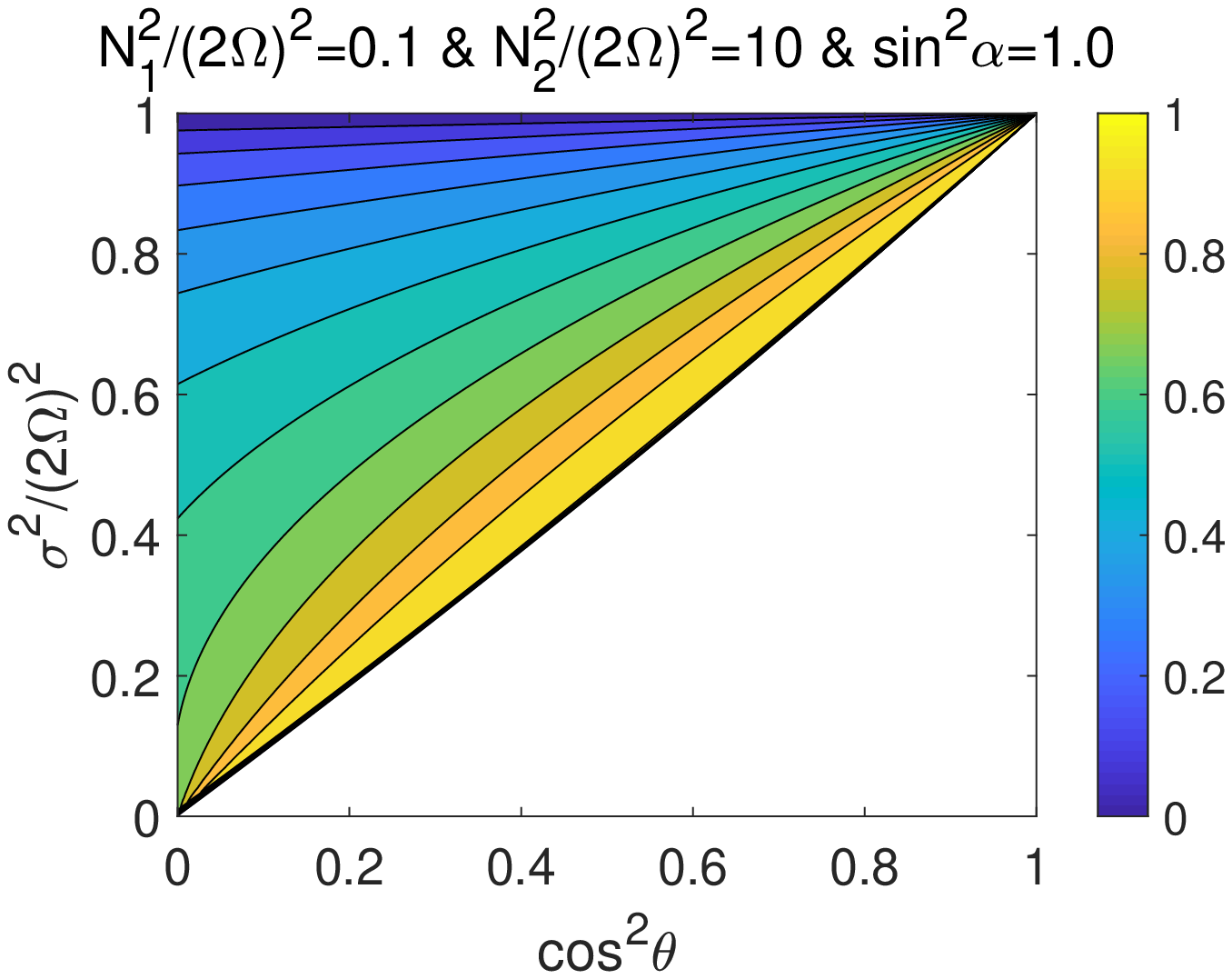}
\caption{}
\end{subfigure}

\medskip

\begin{subfigure}{0.3\textwidth}
\includegraphics[width=\linewidth]{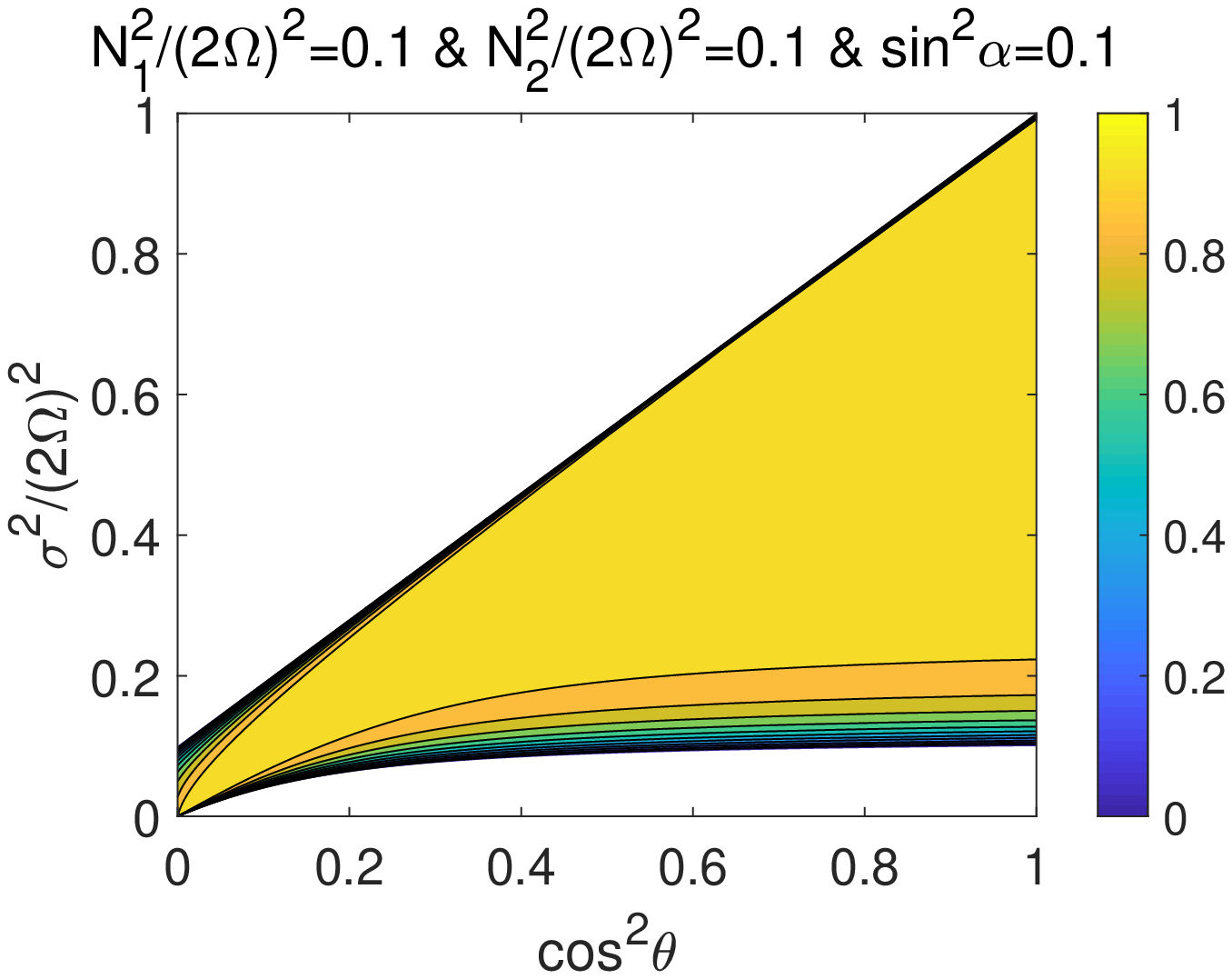}
\caption{}
\end{subfigure}
\begin{subfigure}{0.3\textwidth}
\includegraphics[width=\linewidth]{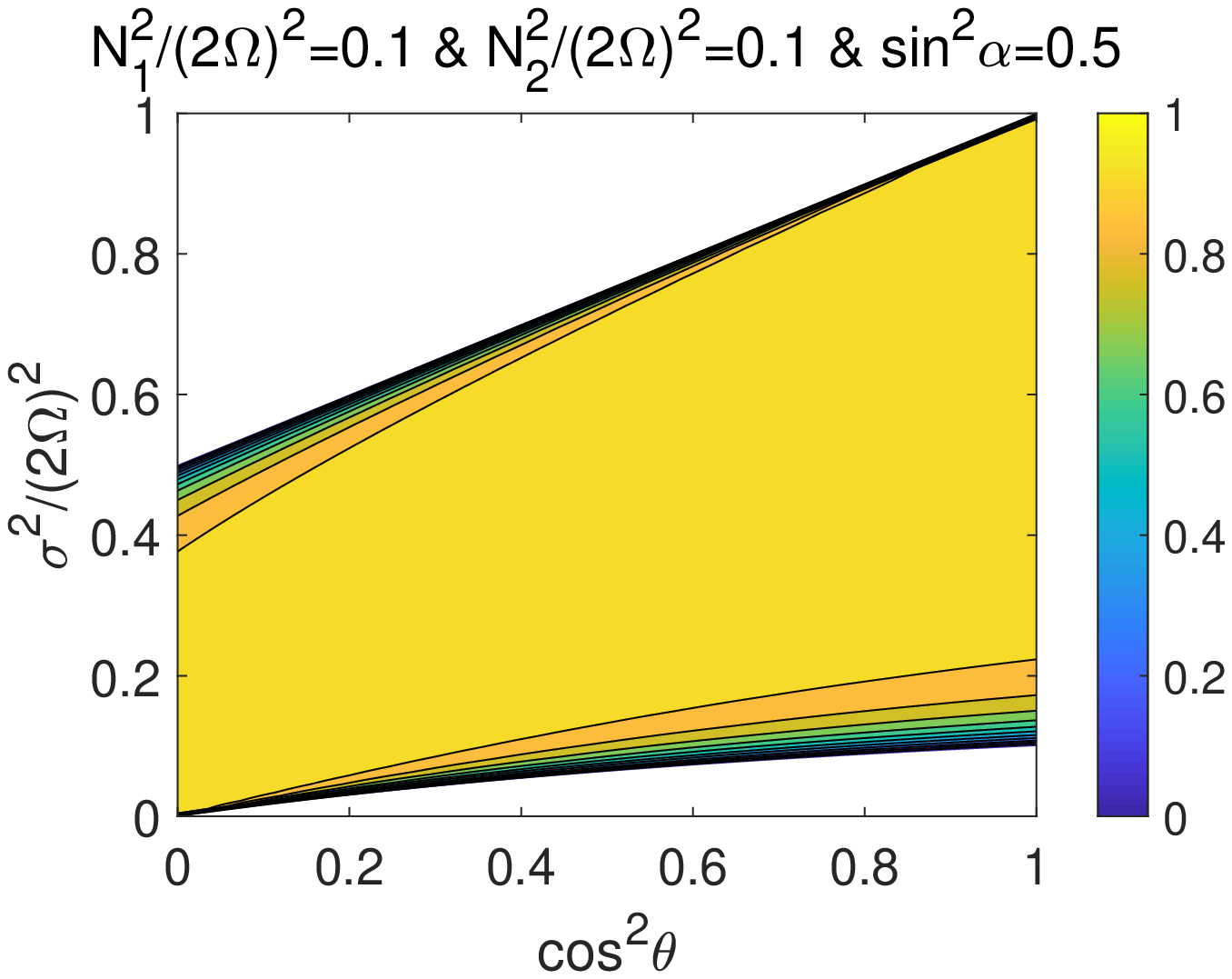}
\caption{}
\end{subfigure}
\begin{subfigure}{0.3\textwidth}
\includegraphics[width=\linewidth]{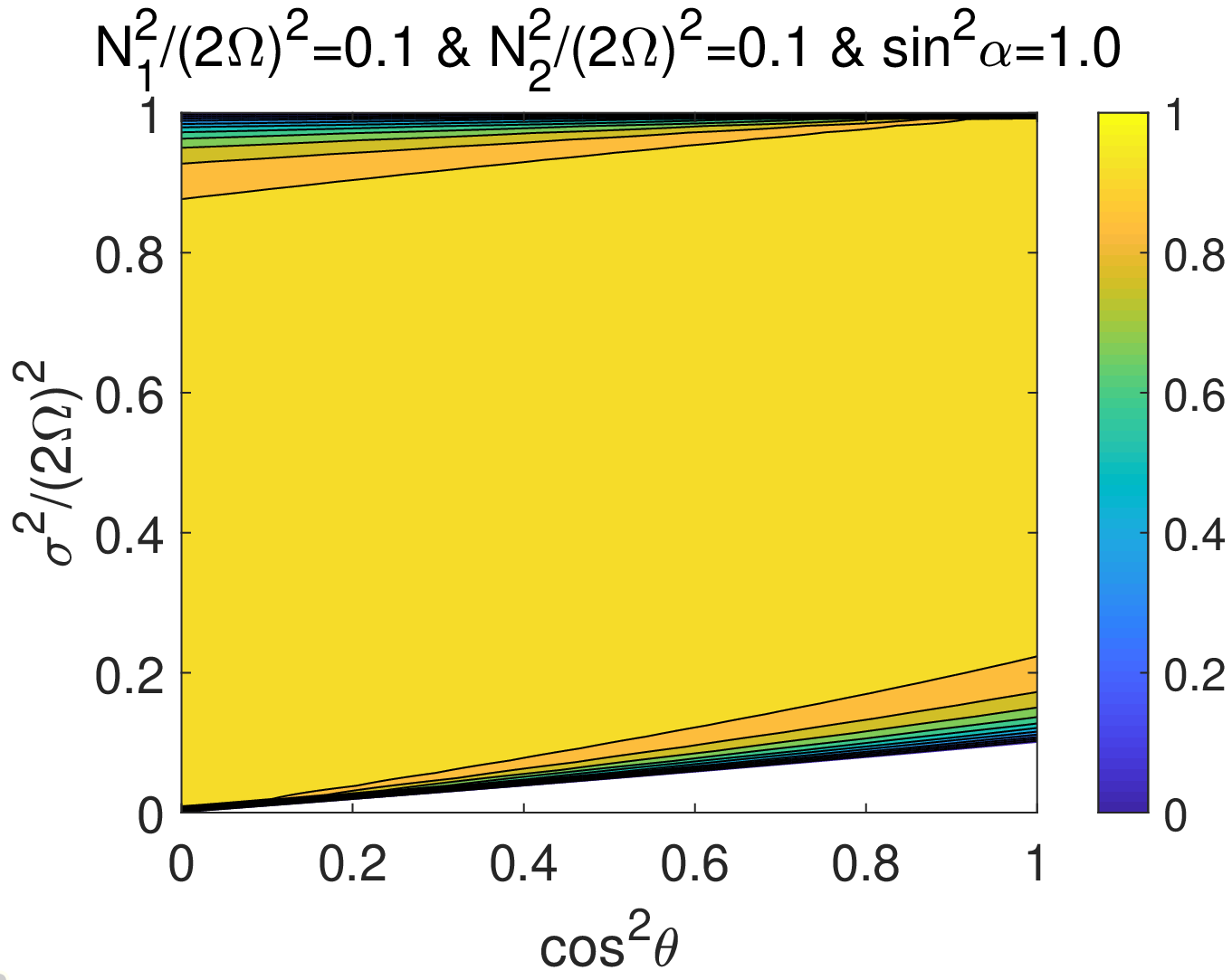}
\caption{}
\end{subfigure}

\caption{Transmission coefficient $\eta_{2}$ (the second part of (\ref{eq28})) for different $N_{1,2}^2/(2\Omega)^2$ and $\sin^2\alpha$ in a three-layer structure (one middle convective layer and two upper and lower neighbouring stable layers). The horizontal axis is $\cos^2\theta$, and the vertical axis is $\sigma^2/(2\Omega)^2$. From the left to the right panels, $\sin^2\alpha^2$ increases from 0.1 to 1.0. (a)-(c) Both stable layers are strongly stratified. (d)-(f) One stable layers are strongly stratified and the other is weakly stratified. (g)-(i) Both stable layers are weakly stratified. Wave propagation can only occur in colored regions. Regions are left white if wave propagation is prohibited.\label{fig:f4}}
\end{figure}

For the second part, the condition $\cos^2 (q_{1}\Delta z'_{1})=0$ is equivalent to $\Delta z'_{1}=(\ell+1/2) \pi/|q_{1}|$, which requires that the thickness of the middle convective layer is a multiple-and-a-half of the half wavelength of the propagating wave. In such case, the overall transmission ratio is
\begin{eqnarray}
\eta_{2}=\left[\frac{1}{4}\left(\sqrt{\frac{q_{1}^2}{s_{1}s_{2}}}+\sqrt{\frac{s_{1}s_{2}}{q_{1}^2}}\right)^2\right]^{-1}~.
\end{eqnarray}
Similarly, $\eta_{2}$ only depends on the wavenumber ratio $s_{1}s_{2}/q_{1}^2$. It decreases with $s_{1}s_{2}/q_{1}^2$ when $s_{1}s_{2}/q_{1}^2>1$ and increases with $s_{1}s_{2}/q_{1}^2$ when $s_{1}s_{2}/q_{1}^2<1$. Let us consider two cases $C>0$ and $C<0$. For the first case we have $|s_{1,2}|<|q_{1}|$, while for the second case we have $|s_{1,2}|>|q_{1}|$. Thus for $C>0$, we obtain $s_{1}s_{2}/q_{2}^2<1$ and $\eta_{2}$ always increases with $s_{1}s_{2}/q_{1}^2$. While for the other case $C<0$, we obtain $s_{1}s_{2}/q_{2}^2>1$ and $\eta_{2}$ always decreases with $s_{1}s_{2}/q_{1}^2$. Efficient transmission can occur if $s_{1}s_{2}/q_{1}^2 \rightarrow 1$, which basically requires both $N_{1}^2C/(B^2-A_{0}C)$ and $N_{2}^2C/(B^2-A_{0}C)$ to be small. It indicates that the transmission ratio will decrease if both $N_{1}^2/(2\Omega)^2$ and $N_{2}^2/(2\Omega)^2$ increase, no matter what the sign of $C$ is. Fig.~\ref{fig:f4} gives an example on such case. Apparently, it can be seen that the transmission ratio decreases when the stable layers are varied from weakly stratified (the lowermost panel) to strongly stratified (the uppermost panel). Also apparent is that transmission is efficient near the critical colatitudes, where $N_{1,2}^2C/(B^2-A_{0}C) \rightarrow 0$.

Although the deduced transmission ratio (\ref{eq28}) is for configuration 1, it can be generalized to other configurations. For configuration 2 (fig.\ref{fig:f2}(b)), the stable layer is embedded between two convective layers. We label the stable layer as $m+1/2$, and the neighbouring convective layers as $m$ and $m+1$. The neighbouring upper and lower interfaces of the stable layer $m+1/2$ is at $z_{m}$ and $z'_{m}$. By this setting, the transmission ratio can be deduced by simply interchanging $q_{m}$ with $s_{m}$ in (\ref{eq28}). Therefore we obtain the overall transmission ratio in configuration 2 is
\begin{eqnarray}
\eta=\left[\frac{1}{4}\left(\sqrt{\frac{q_{1}}{q_{2}}}+\sqrt{\frac{q_{2}}{q_{1}}}\right)^2\cos^2(s_{1}\Delta z'_{1}+\frac{1}{4}\left(\sqrt{\frac{s_{1}^2}{q_{1}q_{2}}}+\sqrt{\frac{q_{1}q_{2}}{s_{1}^2}}\right)^2\sin^2(s_{1}\Delta z'_{1})\right]^{-1}~.\label{eq31}
\end{eqnarray}
The wavenumbers in the convective layers are all the same, thus we have
\begin{eqnarray}
\eta=\left[\cos^2(s_{1}\Delta z'_{1})+\frac{1}{4}\left(|\frac{s_{1}}{q_{1}}|+|\frac{q_{1}}{s_{1}}|\right)^2\sin^2(s_{1}\Delta z'_{1})\right]^{-1}~. \label{eq32}
\end{eqnarray}
Again, we can show that wave transmission is enhanced when the thickness of the middle stable layer is a multiple of the half wavelength of the propagating wave.

\subsubsection{The Multiple-layer case}\label{sec2.1.2}
Now we consider the structure with more alternating layers. Here we first discuss the case of $2M+1$ alternating layers ($M+1$ stable layers and $M$ convective layers), and both the lowermost and uppermost layers are stable. From the recursive relation (\ref{eq15}), we have
\begin{eqnarray}
\left[
\begin{array}{c}
a_{1}\\
b_{1}
\end{array}
\right]=\prod_{m=1}^{M} \bm{T}_{m,m+1}
\left[
\begin{array}{c}
a_{M+1}\\
b_{M+1}
\end{array}
\right]=\left[
\begin{array}{cc}
T_{11} & T_{12}\\
T_{21} & T_{22}
\end{array}
\right]
\left[
\begin{array}{c}
a_{M+1}\\
b_{M+1}
\end{array}
\right].\label{eq33}
\end{eqnarray}
Let $a_{M+1}=0$, we obtain that the wave amplitude of the transmitted wave in the uppermost layer is $b_{M+1}=T_{22}^{-1}b_{1}$, and the transmission ratio is
\begin{eqnarray}
\eta=|\frac{s_{M+1}}{s_{1}}|\frac{1}{|T_{22}|^2}~.\label{eq34}
\end{eqnarray}
Now we discuss the case of $2M$ alternating layers ($M$ stable layers and $M$ convective layers), and the lowermost layer is stable and the uppermost layer is convective. Using the recursive relation (\ref{eq15}) and combining it with the boundary conditions at the uppermost interface, we have
\begin{eqnarray}
\left[
\begin{array}{c}
a_{1}\\
b_{1}
\end{array}
\right]=\left(\prod_{m=1}^{M-1} \bm{T}_{m,m+1}\right) S_{M}^{-1}\Lambda_{M,M}Q_{M,M}
\left[
\begin{array}{c}
c_{M}\\
d_{M}
\end{array}
\right]=\left[
\begin{array}{cc}
T'_{11} & T'_{12}\\
T'_{21} & T'_{22}
\end{array}
\right]
\left[
\begin{array}{c}
c_{M}\\
d_{M}
\end{array}
\right]~,\label{eq35}
\end{eqnarray}
where $T_{ij}',i,j\in\{1,2\}$ is defined by the matrix multiplications shown in the middle of (\ref{eq35}).
Let $c_{M}=0$, we obtain that the wave amplitude of the transmitted wave in the uppermost layer is $d_{M+1}=T'_{22}b_{1}$, and the transmission ratio is
\begin{eqnarray}
\eta=|\frac{q_{M}}{s_{1}}|\frac{1}{|T'_{22}|^2}~.\label{eq36}
\end{eqnarray}
Again, following similar procedures, we can deduce the transmission ratio of configuration 2 by interchanging $q_{m}$ with $s_{m}$.

We have shown that the transmission can be enhanced in the three-layer structure of configuration 1. Now we investigate whether the enhancement occurs in a structure with more layers. For the sake of simplicity, we assume that all convective layers have the same thickness $\Delta z_{c}$ and wavenumber $q$, and all stable layers have the same thickness $\Delta z_{s}$ and wavenumber $s$.

We first discuss the case when the number of layers ($2M+1$) is odd.
When $s_{m}=s$, $q_{m}=q$, and $\Delta z'_{m}=\Delta z_{c}$, the transfer matrix $T_{m,m+1}=T$ is
\begin{eqnarray}
&&\widehat{T}_{11}=\frac{1}{4}e^{-is\Delta z_{c}}\left\{[2+(\frac{q}{s}+\frac{s}{q})]e^{iq\Delta z_{c}}+[2-(\frac{q}{s}+\frac{s}{q})]e^{-iq\Delta z_{c}}\right\}~,\label{eq37}\\
&&\widehat{T}_{12}=\frac{1}{4}e^{i(sz_{m}+s z'_{m})}\left[(\frac{q}{s}-\frac{s}{q})e^{iq\Delta z_{c}}-(\frac{q}{s}-\frac{s}{q})e^{-iq\Delta z_{c}}\right]~,
\end{eqnarray}
with $\widehat{T}_{22}=\widehat{T}_{11}^{*}$ and $\widehat{T}_{21}=\widehat{T}_{12}^{*}$.
The eigenvalue satisfies the following equation
\begin{eqnarray}
(\widehat{T}_{11}-\lambda)(\widehat{T}_{22}-\lambda)=\widehat{T}_{12}\widehat{T}_{21}~.
\end{eqnarray}
After some manipulations, the equation can be written as
\begin{eqnarray}
\lambda^2-2\Re(\widehat{T}_{11})\lambda+1=0~,
\end{eqnarray}
or in an explicit form
\begin{eqnarray}
\lambda^2-2\left[\cos (q\Delta z_{c})\cos
(s\Delta z_{c})+\frac{1}{2}(\frac{q}{s}+\frac{s}{q})\sin(q\Delta z_{c})\sin(s\Delta z_{c})\right]\lambda+1=0~,
\end{eqnarray}
where $\Re$ denotes the real part of a complex number.
Let $\lambda_{1,2}$ be the two roots of the equation. Obviously we have $\lambda_{1}\lambda_{2}=1$. Let
\begin{eqnarray}
\Delta_{\lambda}&&=\left[\cos (q\Delta z_{c})\cos
(s\Delta z_{c})+\frac{1}{2}(\frac{q}{s}+\frac{s}{q})\sin(q\Delta z_{c})\sin(s\Delta z_{c})\right]^2-1\\
&&=\left\{\cos [(q-s)\Delta z_{c}]+\frac{1}{2}(\frac{q}{s}+\frac{s}{q}-2)\sin(q\Delta z_{c})\sin(s\Delta z_{c})\right\}^2-1~.
\end{eqnarray}
Then the eigenvalues $\lambda_{1,2}$ are real when $\Delta_{\lambda}>0$, and are complex when $\Delta_{\lambda}<0$. If $\lambda_{1,2}$ are real and $\lambda_{1}\neq\lambda_{2}$, then the maximum of $|\lambda_{1,2}|$ must be greater than one. For a multi-layer structure, the transfer matrix $T^{M} \propto \max(|\lambda_{1,2}|)^{M}$, which yields $\eta \propto \max(|\lambda_{1,2}|)^{-M}$. Since $\max(|\lambda_{1,2}|)$ is greater than one, the transmission ratio decays with the number of layers.

To ensure that the transmission does not decay, the solution $\lambda_{1,2}$ must be on the unit circle of the complex plane. This condition can be achieved when $\lambda_{1}=\lambda_{2}=\pm 1$, or $\lambda_{1,2}$ are a complex pair. Therefore a necessary condition (not sufficient) for efficient wave transmission is $\Delta_{\lambda}\leq 0$.

It should be emphasized that the condition $\Delta_{\lambda}\leq 0$ is not a sufficient condition. The transmission ratio is actually determined by $|T_{22}|$, which could possibly be much greater than one even though the eigenvalues $\lambda_{1,2}$ are on the unit circle. Here we take a further step to discuss when the value $|T_{22}|$ will be close to one, so as to ensure an efficient wave transmission.

Let us further define $z=0$ at the lowest interface, and $\alpha_{1}=\exp(-is\Delta z_{s})$ and $\alpha_{2}=\exp(-is\Delta z_{c})$. With such definitions, we have $z_{m}+z_{m}'=(2m-2)\Delta z_{s}+(2m-1)\Delta z_{c}$ and $\exp[-is(z_{m}+z_{m}')]=\alpha_{1}^{2m-2}\alpha_{2}^{2m-1}$. The transfer matrix can be rewritten as
\begin{equation}
\bm{T}_{m,m+1}=\left[
\begin{array}{cc}
\alpha_{2}\widetilde{T}_{11} & {\alpha_{1}^{*}}^{2m-2}{\alpha_{2}^{*}}^{2m-1}\widetilde{T}_{12}\\
{\alpha_{1}}^{2m-2}{\alpha_{2}}^{2m-1}\widetilde{T}_{12}^{*} & \alpha_{2}^{*}\widetilde{T}_{11}^{*}
\end{array}
\right]~,
\end{equation}
where
\begin{eqnarray}
&&\widetilde{T}_{11}=\frac{1}{4}\left\{[2+(\frac{q}{s}+\frac{s}{q})]e^{iq\Delta z_{c}}+[2-(\frac{q}{s}+\frac{s}{q})]e^{-iq\Delta z_{c}}\right\}~,\label{eq37}\\
&&\widetilde{T}_{12}=\frac{1}{4}\left[(\frac{q}{s}-\frac{s}{q})e^{iq\Delta z_{c}}-(\frac{q}{s}-\frac{s}{q})e^{-iq\Delta z_{c}}\right]~.
\end{eqnarray}
When $m=1$, we note that the transfer matrix $\bm{T}_{1,2}$ can be formulated as
\begin{equation}
\bm{T}_{1,2}=\left[
\begin{array}{cc}
\alpha_{2}\widetilde{T}_{11} & {\alpha_{2}^{*}}\widetilde{T}_{12}\\
{\alpha_{2}}\widetilde{T}_{12}^{*} & \alpha_{2}^{*}\widetilde{T}_{11}^{*}
\end{array}
\right]
=\bm{A}\left[
\begin{array}{cc}
\alpha_{2} & 0\\
0 & \alpha_{2}^{*}
\end{array}
\right]~,
\end{equation}
where
\begin{equation}
\bm{A}=\left[
\begin{array}{cc}
\widetilde{T}_{11} & \widetilde{T}_{12}\\
\widetilde{T}_{12}^{*} & \widetilde{T}_{11}^{*}
\end{array}
\right]~.
\end{equation}
Here we consider a special case with $\alpha_{1}^2={\alpha_{1}^{*}}^2=1$, which can be achieved by letting $|s|\Delta z_{s}=\ell' \pi$, where $\ell'$ is a non-negative integer. For this special case, it can be proved that
\begin{eqnarray}
\prod_{m=1}^{M} \bm{T}_{m,m+1}
=\bm{A}^{M}\left[
\begin{array}{cc}
{\alpha_{2}}^M & 0\\
0 & {{\alpha_{2}}^{*}}^{M}
\end{array}
\right]~.
\end{eqnarray}
Now we try to derive the explicit form of $\bm{A}^{M}$. It is obvious that $\widetilde{T}_{12}$ is a pure imaginary number, thus we can write $\bm{A}$ as
\begin{eqnarray}
\bm{A}=\left[
\begin{array}{cc}
\Re(\widetilde{T}_{11})+i\Im(\widetilde{T}_{11}) & i\Im(\widetilde{T}_{12})\\
-i\Im(\widetilde{T}_{12}) & \Re(\widetilde{T}_{11})-i\Im(\widetilde{T}_{11})
\end{array}
\right]=\cos(q \Delta z_{c})\bm{I}+i\bm{U}~,
\end{eqnarray}
where
\begin{equation}
\bm{U}=
\left[
\begin{array}{cc}
\Im(\widetilde{T}_{11}) & \Im(\widetilde{T}_{12})\\
-\Im(\widetilde{T}_{12}) & -\Im(\widetilde{T}_{11})
\end{array}
\right]~,
\end{equation}
and it is easy to verify
\begin{equation}
\bm{U^2}=\sin^2(q\Delta z_{c})\bm{I}~.
\end{equation}
If $\sin(q\Delta z_{c})=0$, we can show $\bm{A}=\cos(q\Delta z_{c})\bm{I}$ and $\bm{A}^{M}=\cos^{M}(q\Delta z_{c})\bm{I}$. If $\sin(q\Delta z_{c})\neq 0$, then we obtain
\begin{eqnarray}
\bm{A}^{M}&=&\left(\cos(q \Delta z_{c})\bm{I}+i\bm{U}\right)^{M}\\
&=&\bm{I}\sum_{k\in even}C_{k}^{M}\cos^{M-k}(q\Delta z_{c})i^{k}\sin^{k}(q\Delta z_{c})\nonumber\\
  &&+\bm{U}[\sin(q\Delta z_{c})]^{-1}\sum_{k\in odd}C_{k}^{M}\cos^{M-k}(q\Delta z_{c})i^{k}\sin^{k}(q\Delta z_{c})\\
&=& \cos(Mq\Delta z_{c})\bm{I}+i\sin(Mq\Delta z_{c})[\sin(q\Delta z_{c})]^{-1}\bm{U}~,
\end{eqnarray}
where $C_{k}^{M}=M!/((M-k)!k!)$ is the combination function.
From the above calculation, the analytical solution of transmission ratio can be obtained. If $\sin(q\Delta z_{c})=0$, we have
\begin{eqnarray}
T_{22}={\alpha_{2}^{*}}^{M}\cos^{M}(q\Delta z_{c})~,
\end{eqnarray}
and the transmission ratio is
\begin{eqnarray}
\eta=1~. \label{eq63}
\end{eqnarray}
If $\sin(q\Delta z_{c})\neq 0$, we have
\begin{eqnarray}
T_{22}={\alpha_{2}^{*}}^{M}\left[\cos(Mq\Delta z_{c})-i\frac{1}{2}(\frac{q}{s}+\frac{s}{q})\sin(Mq\Delta z_{c})\right]~,
\end{eqnarray}
and the transmission ratio is
\begin{eqnarray}
\eta=\left[\cos^2 (Mq\Delta z_{c})+\frac{1}{4}(\frac{q}{s}+\frac{s}{q})^2\sin^2(Mq\Delta z_{c})\right]^{-1}~. \label{eq65}
\end{eqnarray}
(\ref{eq63}) can be synthesized into (\ref{eq65}), since $\sin(q\Delta z_{c})=0$ implies $\sin^2(Mq\Delta z_{c})=0$ and $\cos^2(Mq\Delta z_{c})=1$. Therefore we conclude that, under the condition $|s|\Delta z_{s}=\ell'\pi$, the wave transmission ratio can be described by (\ref{eq65}). Comparing with the result of three-layer structure case, we see that (\ref{eq28}) is just a special case of (\ref{eq65}) when $M=1$. The discussion on efficiency of wave transmission based on (\ref{eq65}) is similar to that in three-layer structure case, and here we will not repeat it. From the analytical solution, it is clear that wave will be totally transmitted when $M|q|\Delta z_{c}=\ell \pi$ and $|s|\Delta z_{s}=\ell'\pi$.

Analytical solution on the general cases of $|s|\Delta z_{s} \neq \ell'\pi$ is more difficult, but some insights can be provided from the discussion of eigenvalues of the transmission matrix. It is worth mentioning for the special case when $\sin q\Delta z_{c}\rightarrow 0$ and $\sin s\Delta z_{c}\rightarrow 0$, the eigenvalues $|\lambda_{1,2}|\rightarrow 1$. In this limit, it can be shown that
\begin{eqnarray}
\Delta_{\lambda}\sim \frac{[mod((q-s)\Delta z_{c},\pi)]^4}{12}>0~,
\end{eqnarray}
where $mod$ is the modulo function. The eigenvalues are real numbers, and one of $|\lambda_{1,2}|$ is slightly greater than 1. The transmission decays slowly as the wave crosses each layer. The wave transmission can be efficient when the number of layers is not too large. The eigenvalues can be estimated as
\begin{eqnarray}
\lambda_{1,2}=\sqrt{1+\Delta_{\lambda}}\pm \sqrt{\Delta_{\lambda}}\sim 1\pm \frac{[mod((q-s)\Delta z_{c},\pi)]^2}{3\sqrt{2}}~.
\end{eqnarray}
Thus the decay rate of transmission ratio is approximately $\lambda_{2}^{-M} \sim \{1-12^{-1/2}M [mod((q-s)\Delta z_{c},\pi)]^2\}^{-1}$. The transmission can be efficient when $M\ll M_{c}=[mod((q-s)\Delta z_{c},\pi)]^{-2}$. When $mod((q-s)\Delta z_{c},\pi)\rightarrow 0$, the critical value $M_{c}$ is very large. As a result, the transmission can be approximately efficient in this case. From the above discussion, we infer that the transmission can be efficient when $\sin (q\Delta z_{c})\ll 1$ or $\sin(s\Delta z_{c})\ll 1$. This conclusion is useful when embedded convective layers are very thin.

When the total number of layers is even ($2M$), we can consider it as $(2M-1)$ layers plus an addition layer. The best scenario on transmission for $(2M-1)$ layers is that the incident wave is totally transmitted to the $(2M-1)$-th layer. Now the wave is incident from the $(2M-1)$-th layer to the $(2M)$-th layer, and it can be considered as a two-layer problem. For a two-layer problem, wave transmission is generally not efficient when the stable layer is strongly stratified \citep{wei20,cai20}.

\begin{figure}
\centering

\begin{subfigure}{0.3\textwidth}
\includegraphics[width=\linewidth]{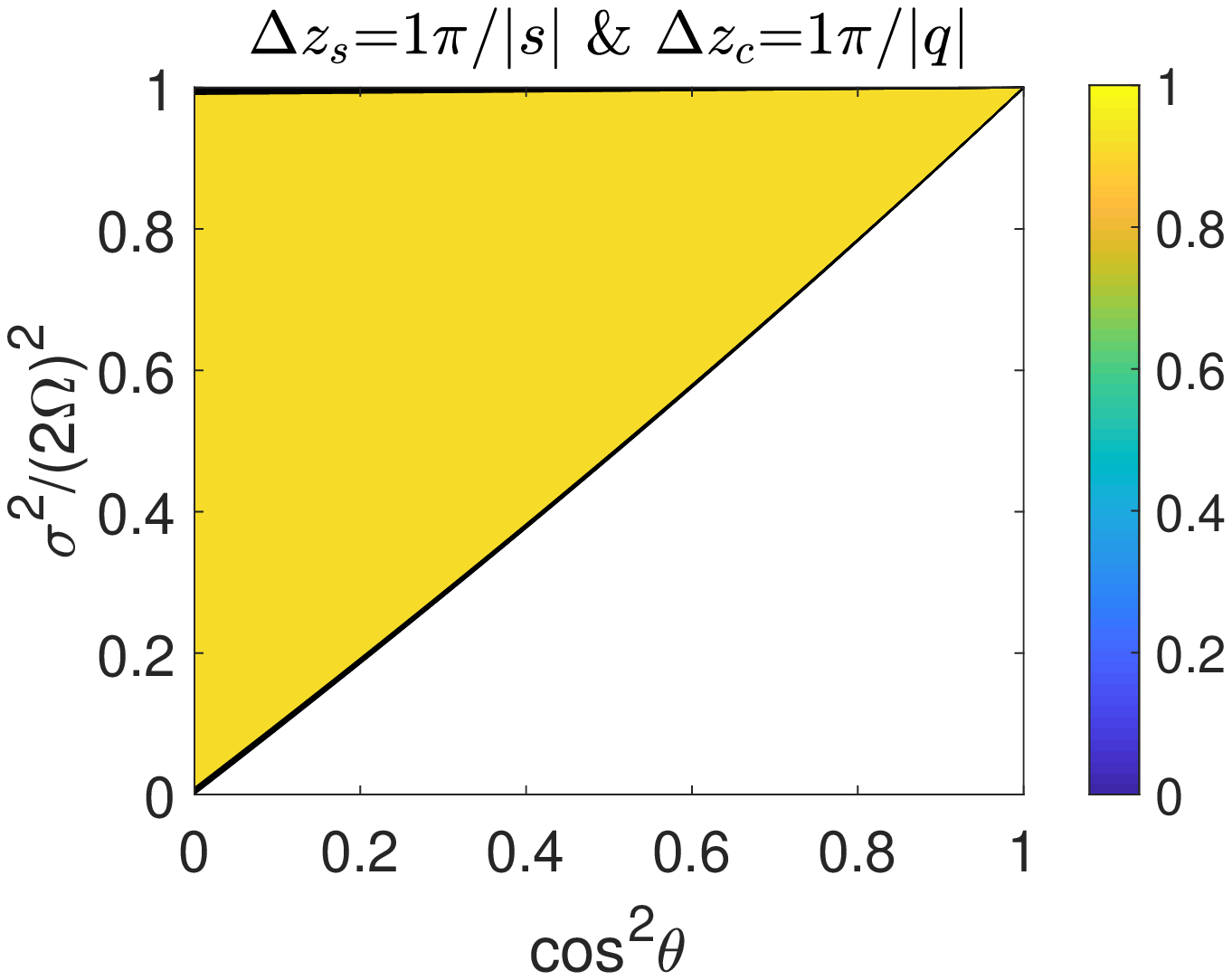}
\caption{}
\end{subfigure}
\begin{subfigure}{0.3\textwidth}
\includegraphics[width=\linewidth]{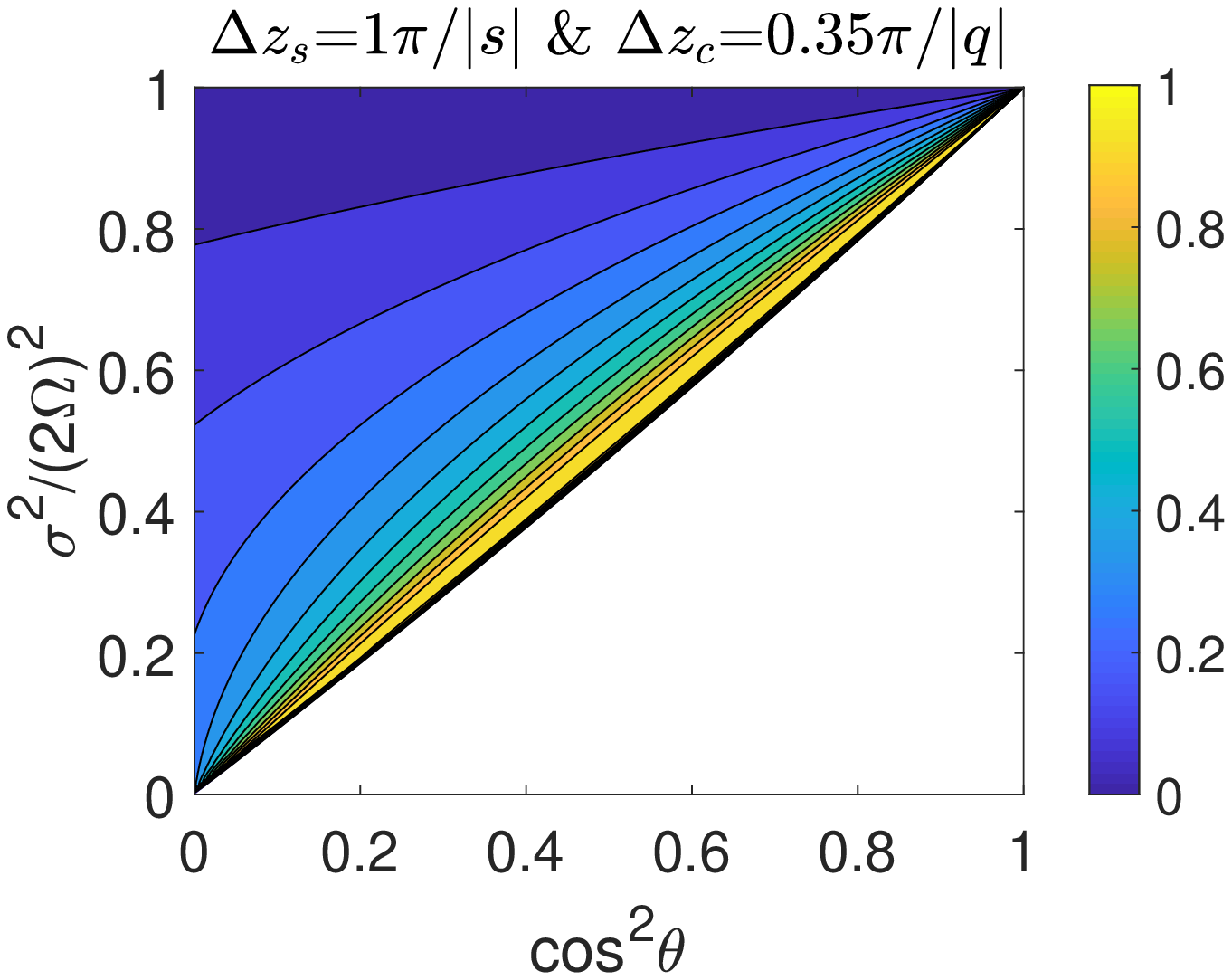}
\caption{}
\end{subfigure}
\begin{subfigure}{0.3\textwidth}
\includegraphics[width=\linewidth]{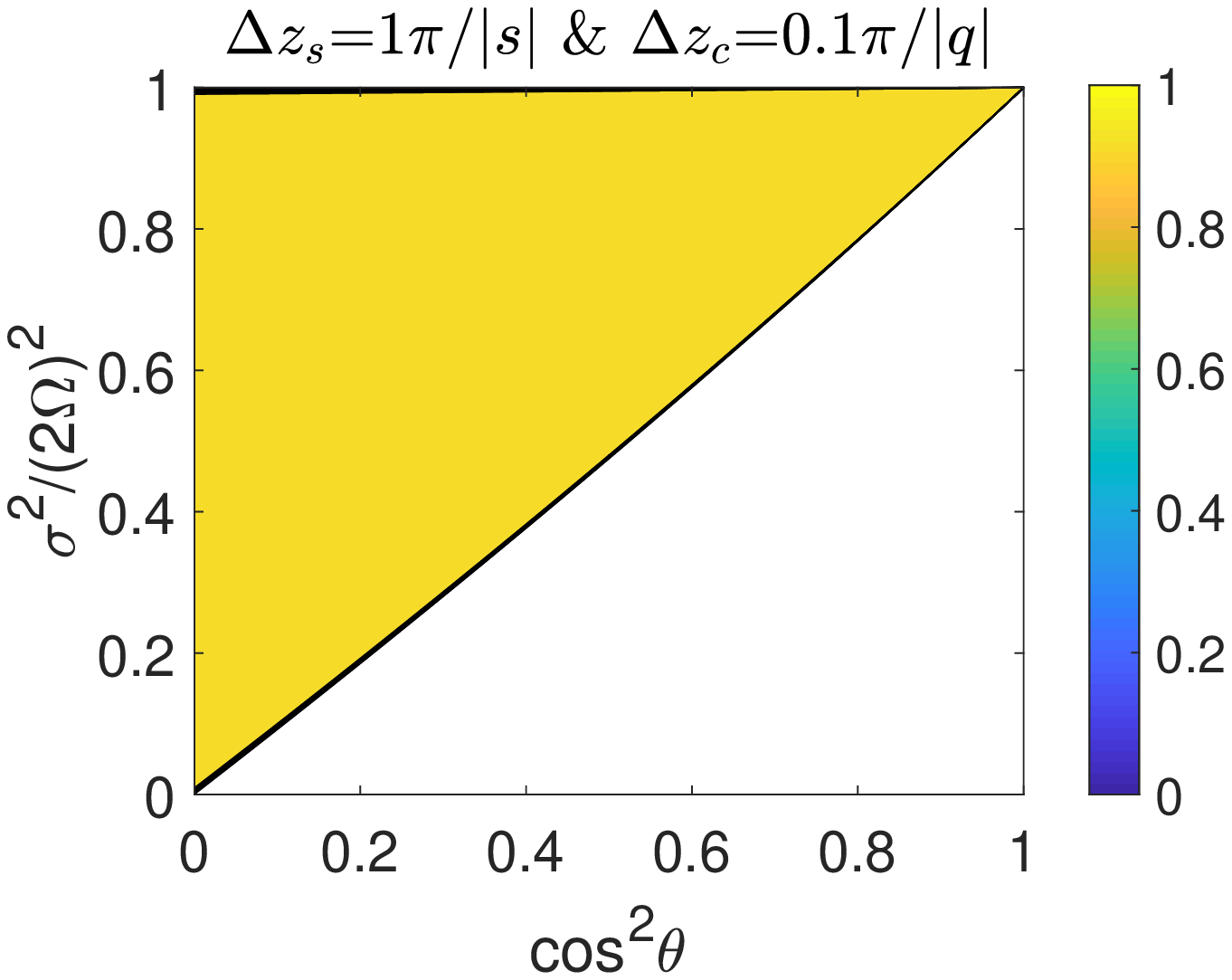}
\caption{}
\end{subfigure}

\medskip

\begin{subfigure}{0.3\textwidth}
\includegraphics[width=\linewidth]{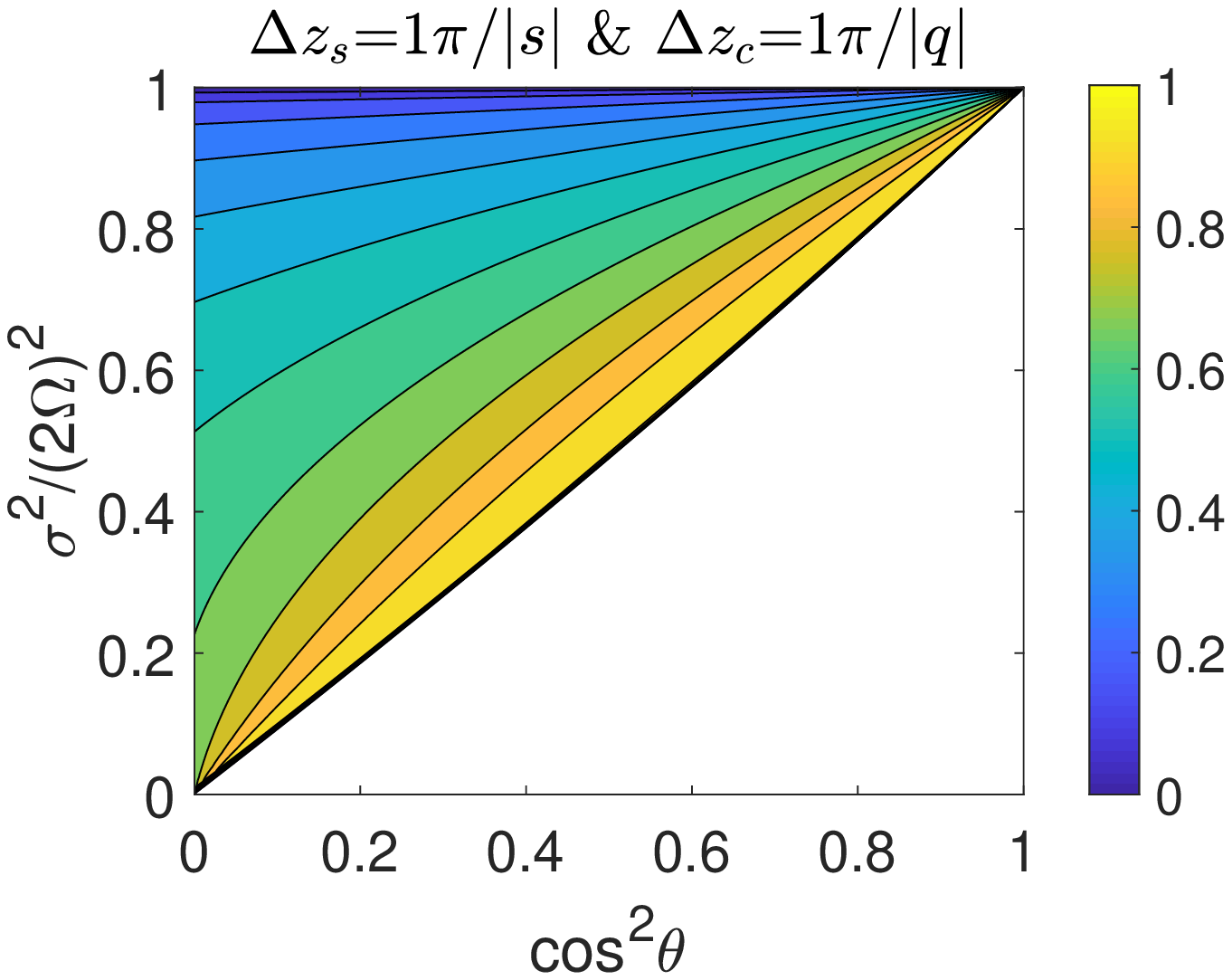}
\caption{}
\end{subfigure}
\begin{subfigure}{0.3\textwidth}
\includegraphics[width=\linewidth]{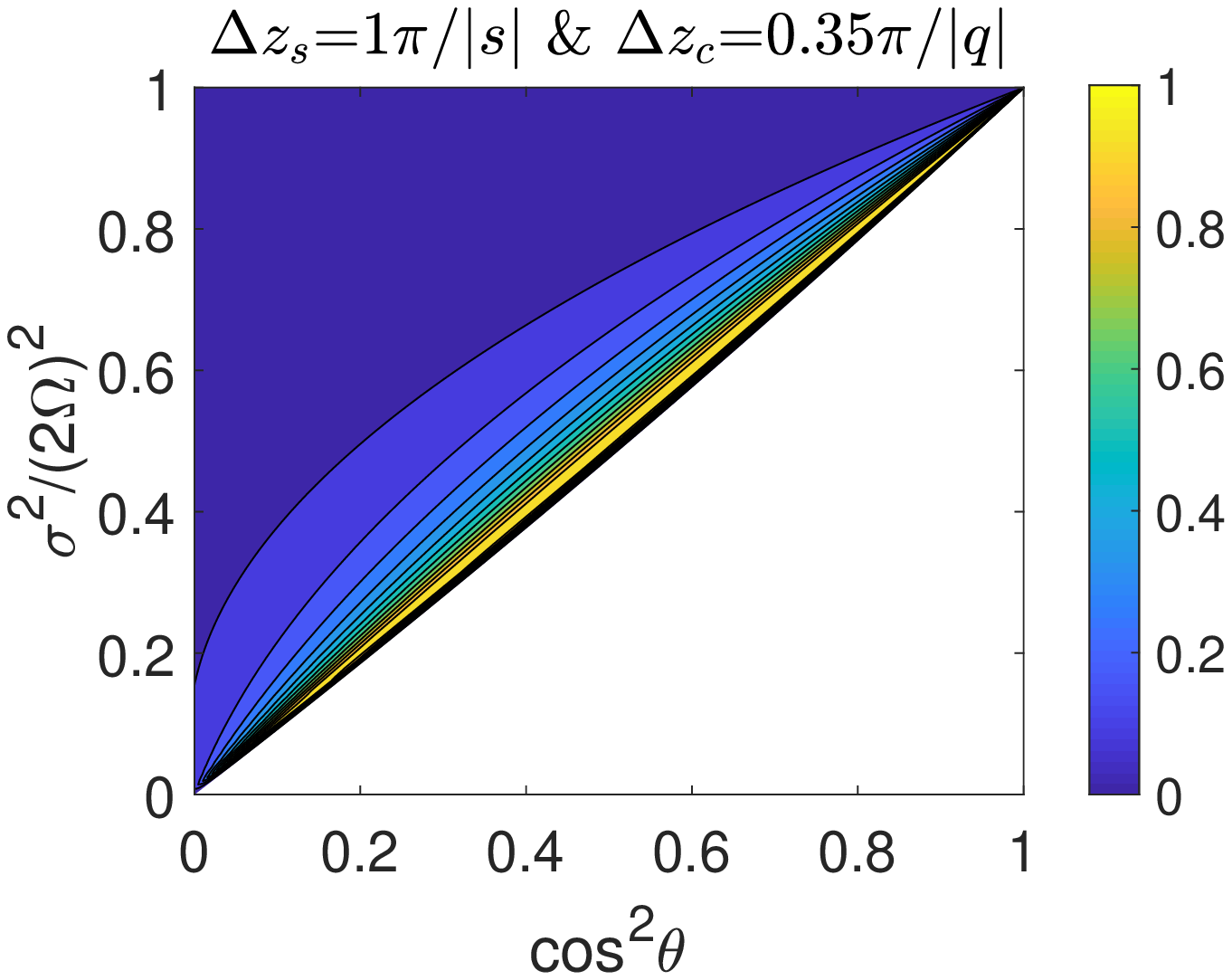}
\caption{}
\end{subfigure}
\begin{subfigure}{0.3\textwidth}
\includegraphics[width=\linewidth]{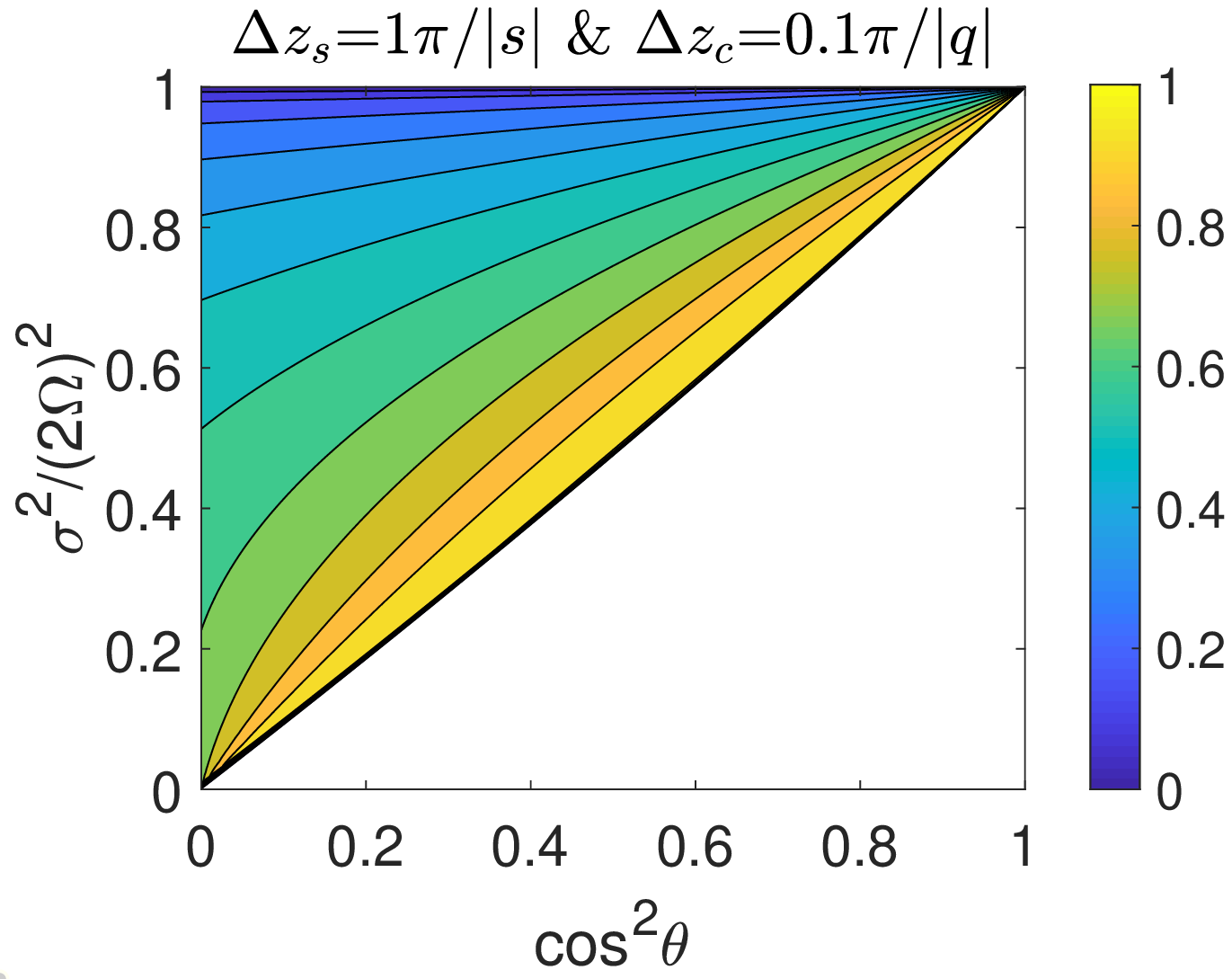}
\caption{}
\end{subfigure}

\medskip

\begin{subfigure}{0.3\textwidth}
\includegraphics[width=\linewidth]{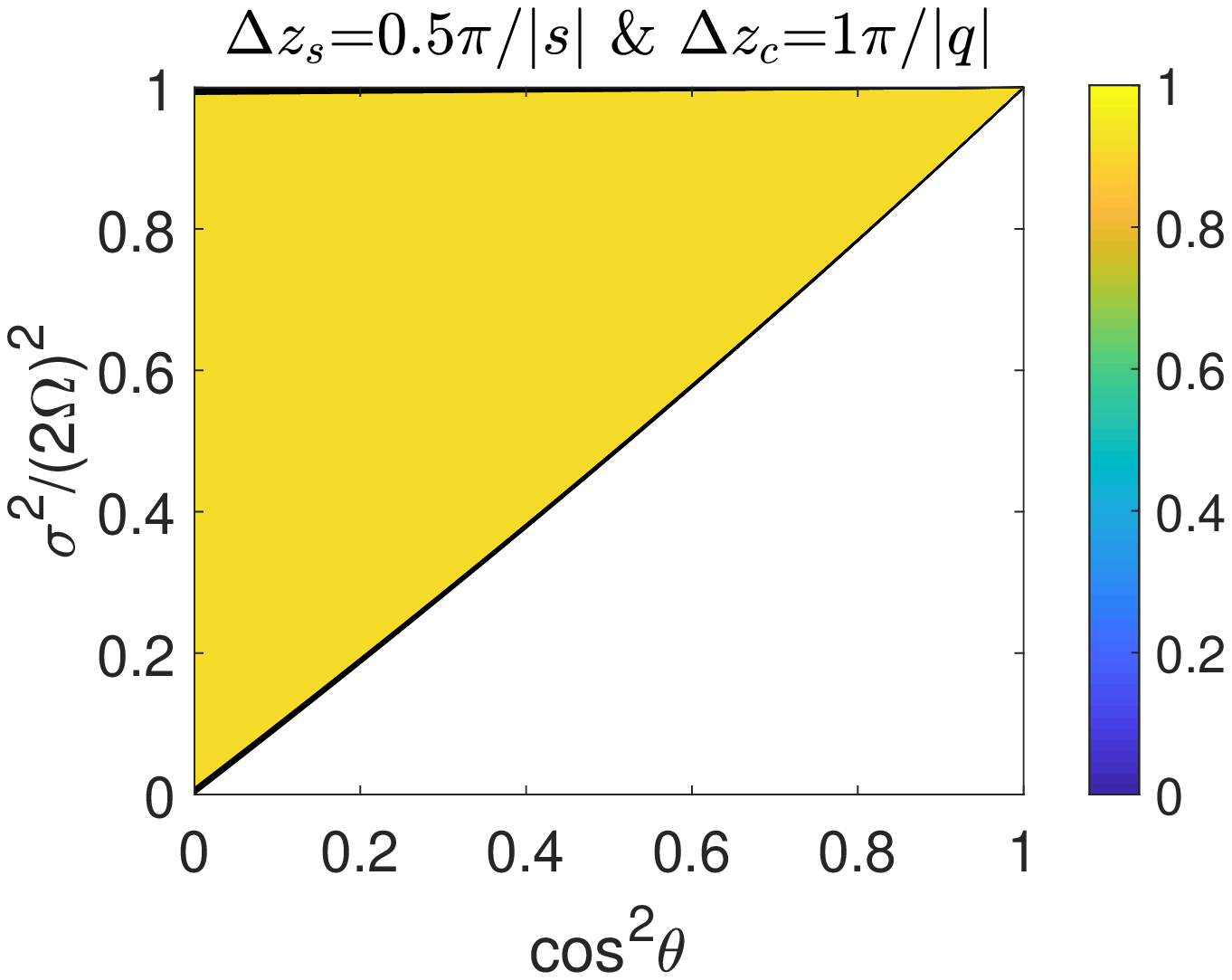}
\caption{}
\end{subfigure}
\begin{subfigure}{0.3\textwidth}
\includegraphics[width=\linewidth]{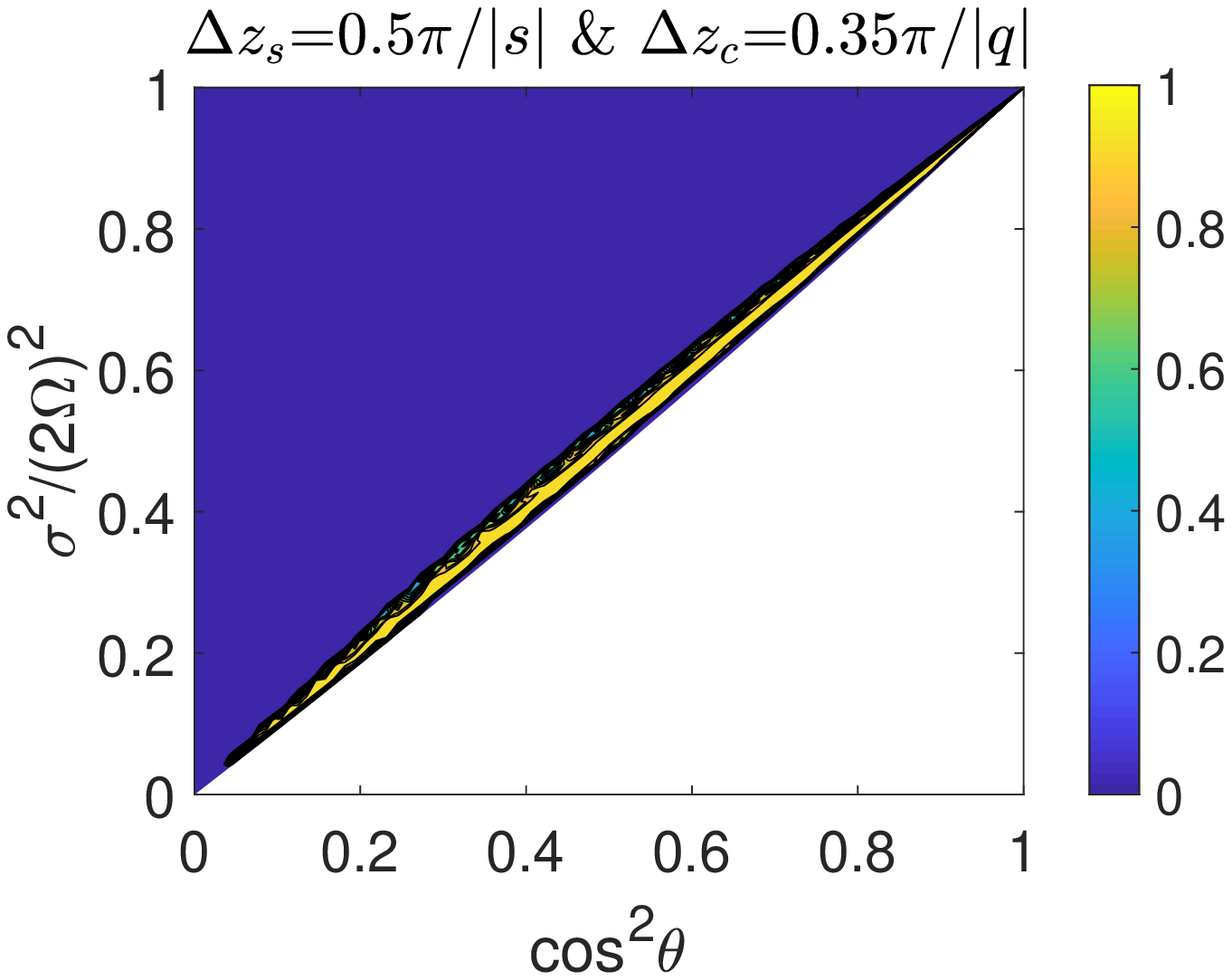}
\caption{}
\end{subfigure}
\begin{subfigure}{0.3\textwidth}
\includegraphics[width=\linewidth]{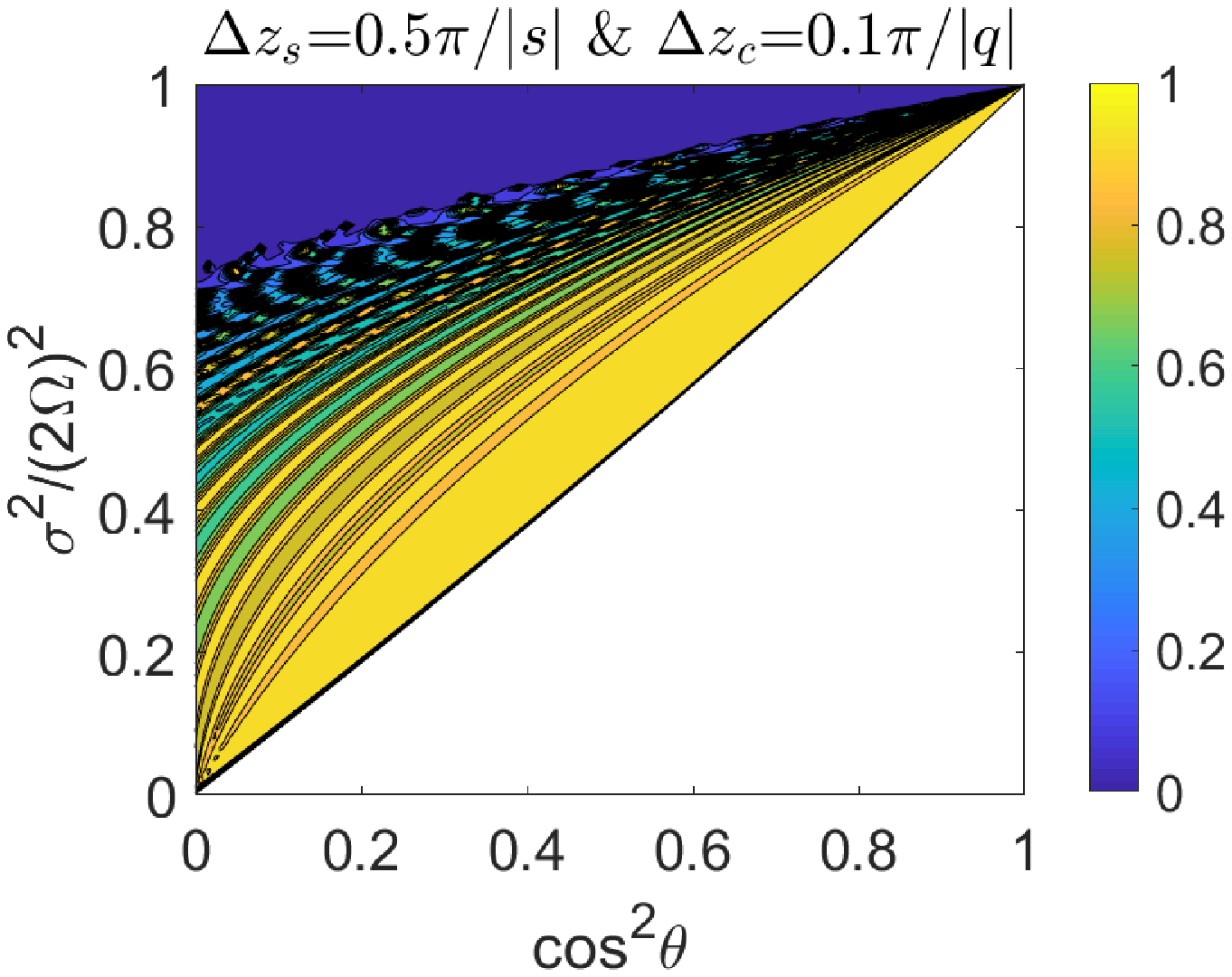}
\caption{}
\end{subfigure}

\caption{Transmission ratios in multi-layer structures. For all the cases, $N_{m}^2/(2\Omega)^2=10$ and $\sin^2\alpha=1$, and the lowermost layer is stable. $\Delta z_{s}=\pi/|s|$ for (a-f) but $\Delta z_{s}=0.5\pi/|s|$ for (g-i). In each case, all the convective layers have the same thickness $\Delta z_{c}$ and wavenumber $q$, and all the stable layers have the same thickness $\Delta z_{s}$ and wave number $s$. (a-c)The transmission ratio for a 101-layer structure with $\Delta z_{c}=(1,0.35,0.1)\pi/|q|$. (d-f) The transmission ratio for a 102-layer structure with $\Delta z_{c}=(1,0.35,0.1)\pi/|q|$. (g-i) Similar to (a-c) but with different $\Delta z_{s}$. Wave propagation can only occur in colored regions. Regions are left white if wave propagation is prohibited. \label{fig:f5}}
\end{figure}

Fig.~\ref{fig:f5} plots the transmission ratios in a 101-layer structure (the upper panel) and a 102-layer structure (the middle panel). In all the cases, we choose $N_{m}^2/(2\Omega)^2=10$, $\sin^2\alpha=1$ and the lowermost layer is stable. Thus in all cases, the stable layers are strongly stratified. It clearly shows that the transmission is enhanced in fig.~\ref{fig:f5}(a) and (c), where the total number of layers is odd and $\Delta z_{c}$ and $\Delta z_{s}$ satisfy the conditions $M|q|\Delta z_{c}=\ell \pi$ and $|s|\Delta z_{s}=\ell'\pi$. The transmission in figs.~\ref{fig:f5}(b) is not enhanced because $M|q|\Delta z_{c} \neq \ell \pi$. For the case of 102-layer structure (figs.~\ref{fig:f5}(d) and (f)), the transmission is not enhanced even when the above conditions are satisfied. For these two cases, wave is almost totally transmitted from the 1st layer to 101st layer (see figs.~\ref{fig:f5}(a) and (c)). Wave transmission from the 101st layer to 102nd layer can be viewed as a two-layer problem, and it is generally not efficient when the stable layer is strongly stratified. Therefore, if the stable layer is strongly stratified, the enhancement of transmission only takes place when the number of alternating layers is odd. In other words, enhanced wave transmission only occurs in a multi-layer structure with stable layers embedded within convective layers, or convective layers embedded within stable layers. The enhancement of transmission also depends on the thicknesses of embedded layers $\Delta z_{s}$. Figs.~\ref{fig:f5}({g-i}) present results of three cases with $|s|\Delta z_{s}=0.5\pi$ and different values of $|q|\Delta z_{c}$. It can be seen that the transmission of the case $|q|\Delta z_{c}=\pi$ is enhanced, while the transmission of the case $|q|\Delta z_{c}=0.1\pi$ is only partially enhanced.

The result obtained in the configuration 1 is also true in the configuration 2. Here we do not repeat the discussion on configuration 2.

\citet{andre17} have also observed that wave transmission can be enhanced in a multi-layer structure. They provided a physical explanation for the enhancement in wave transmission when the incident wave is resonant with waves in adjacent layers with half-wavelengths equal to the layer depth. Our analysis verifies this phenomenon from a mathematical point of view.

\subsection{Wave tunneling}
\begin{figure}
\centering
\begin{subfigure}{0.45\textwidth}
\includegraphics[width=\linewidth]{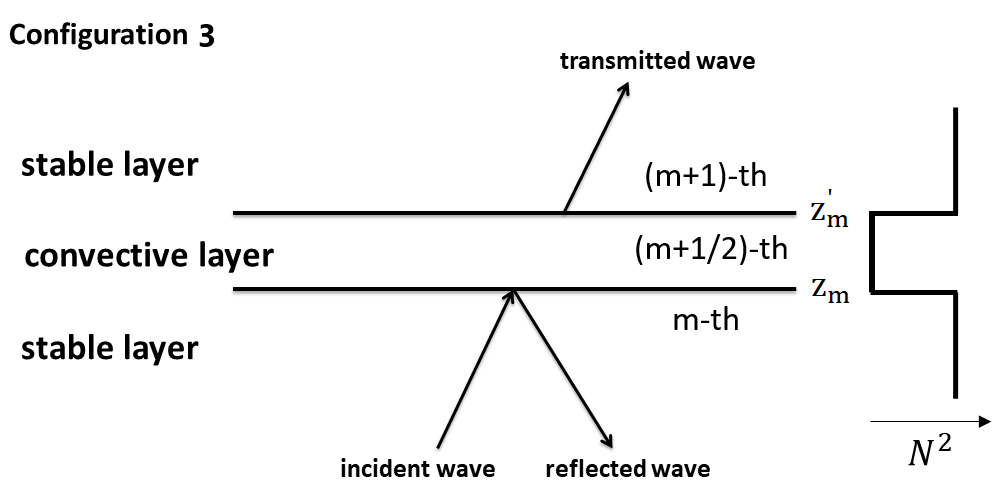}
\caption{}
\end{subfigure}
\begin{subfigure}{0.45\textwidth}
\includegraphics[width=\linewidth]{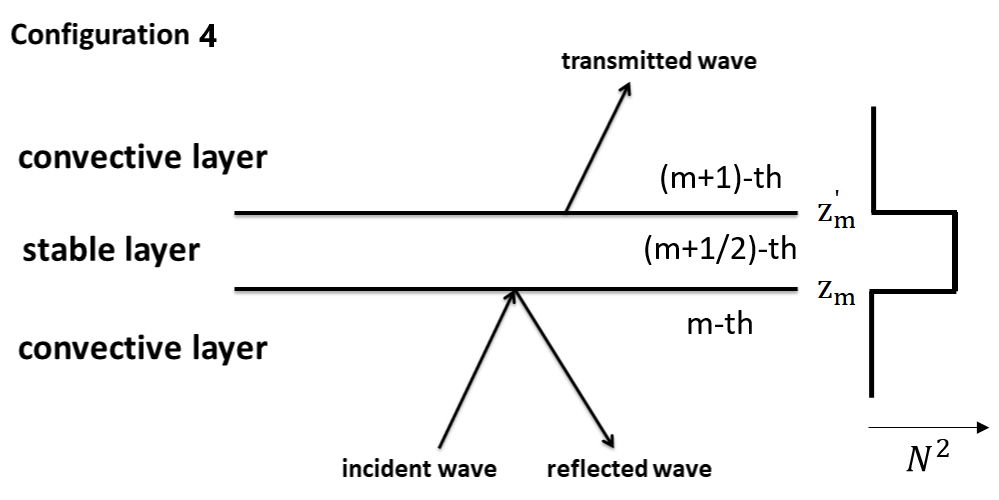}
\caption{}
\end{subfigure}
\caption{Two configurations of tunneling of gravity and inertial waves. (a)Tunneling of gravity wave. Both top and bottom layers are stable. Gravity wave can propagate in the stable layer, but wave cannot propagate in the convective layer. (b)Tunneling of inertial wave. Both top and bottom layers are convective. Gravity wave can propagate in the convective layer, but wave cannot propagate in the stable layer. \label{fig:f6}}
\end{figure}

In the previous section, we have considered wave transmissions in multiple convective and radiative (stable) layers when the condition $B^2-AC>0$ is satisfied everywhere in the domain. Here we consider wave transmissions in two other configurations: (1)wave solution exists in stable layers but not in convective layers; (2)wave solution exists in convective layers but not in stable layers. For the first configuration, gravity wave can propagate in stable layers; and for the second configuration, inertial wave can propagate in convective layers. Fig.~\ref{fig:f6} shows the sketch plots of a three-layer structure for the two configurations. In fig.~\ref{fig:f6}(a), the convective layer is embedded within two stable layers. In fig.~\ref{fig:f6}(b), the stable layer is embedded within two convective layers. For multiple layer structures in fig.~\ref{fig:f6}, although wave cannot propagate in the whole domain, it still can transmit through a tunneling process \citep{mihalas2013foundations,sutherland2004internal}. In the followings, we will discuss the tunneling of gravity and inertial waves, respectively.

\subsubsection{Tunneling of gravity waves}
For configuration 3 in fig.~\ref{fig:f6}(a), wave can propagate in the stable layer but cannot propagate in the convective layer. Thus the wave frequency must be in the range $\sigma_{4}^2<\sigma^2<\sigma_{5}^2$.
The width of the frequency domain can be written as
\begin{eqnarray}
\sigma_{5}^2-\sigma_{4}^2=\frac{1}{2}[(N_{min}^2 -f^2-\tilde{f}_{s}^2)+\sqrt{(N_{min}^2 -f^2-\tilde{f}_{s}^2)^2+4N_{min}^2 \tilde{f}_{s}^2}]~. \label{eq68}
\end{eqnarray}
For the sake of convenience, we only discuss waves at the northern atmosphere ($\theta\in[0,\pi/2]$). The result at the southern atmosphere can be inferred from symmetry property. From (\ref{eq68}), it can be proved that $\sigma_{5}^2-\sigma_{4}^2$ always increases with $\theta$, $\sin^2\alpha$, and $N_{min}^2/(2\Omega)^2$ (see Appendix \ref{appendixb}). Therefore, the frequency domain is wider at equatorial regions than polar regions. Also, it is wider when the meridional wavenumber dominates the zonal wavenumber, and it is wider when the degree of stratification is stronger.

Now we discuss the wave tunneling in configuration 3. We use the same setting as that discussed in wave propagation. The derivation of wave transmission of tunneling is similar to that of wave propagation. The only difference is that now wave cannot propagate in the convective layer, and thus $q_{m}=i\hat{q}_{m}$ is pure imaginary number. Then the transfer relation from $m$-th layer to $(m+1)$-th layer can be easily obtained by replacing $q_{m}$ with $i\hat{q}_{m}$ in (\ref{eq15}).

\begin{figure}
\centering
\begin{subfigure}{0.3\textwidth}
\includegraphics[width=\linewidth]{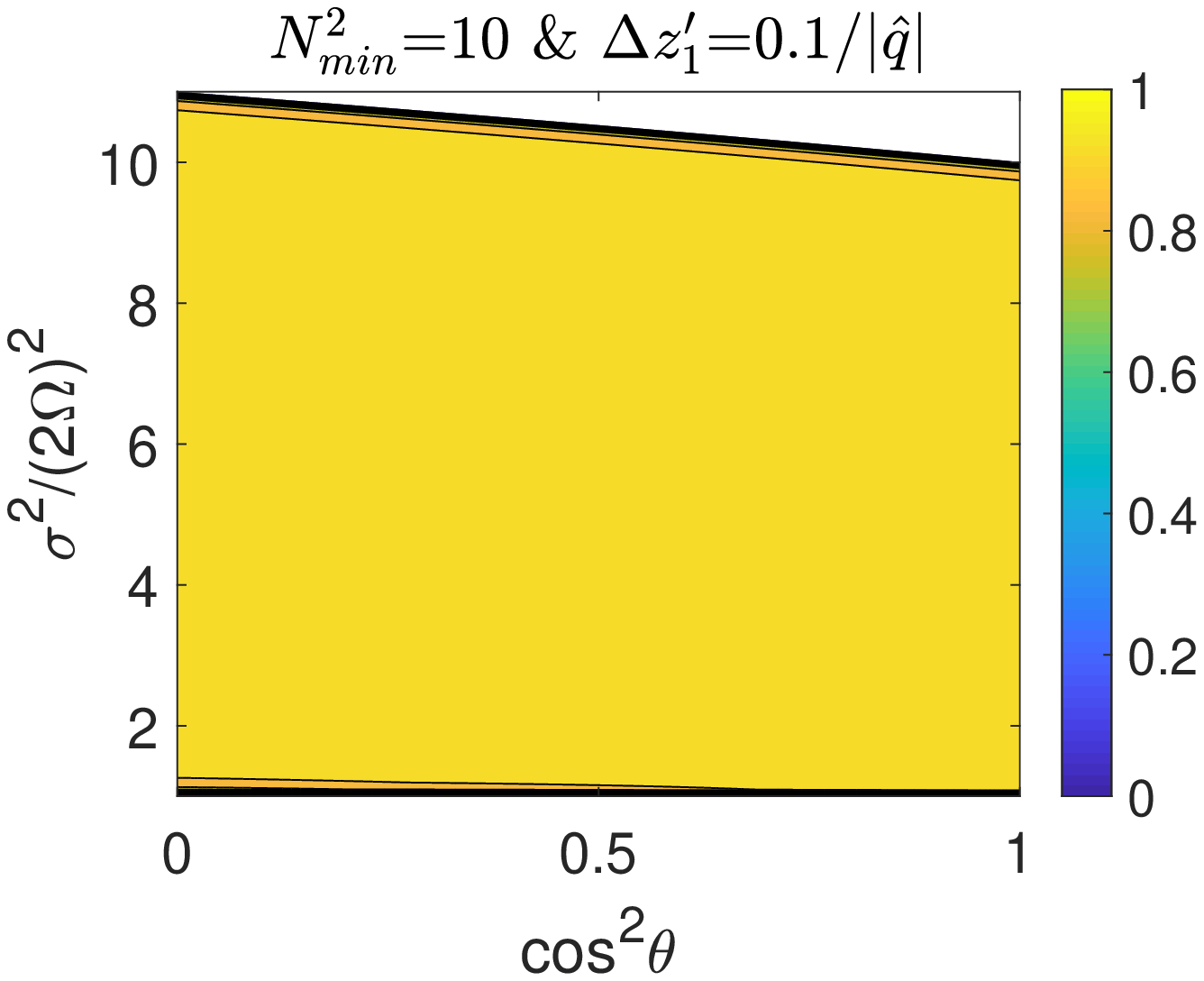}
\caption{}
\end{subfigure}
\begin{subfigure}{0.3\textwidth}
\includegraphics[width=\linewidth]{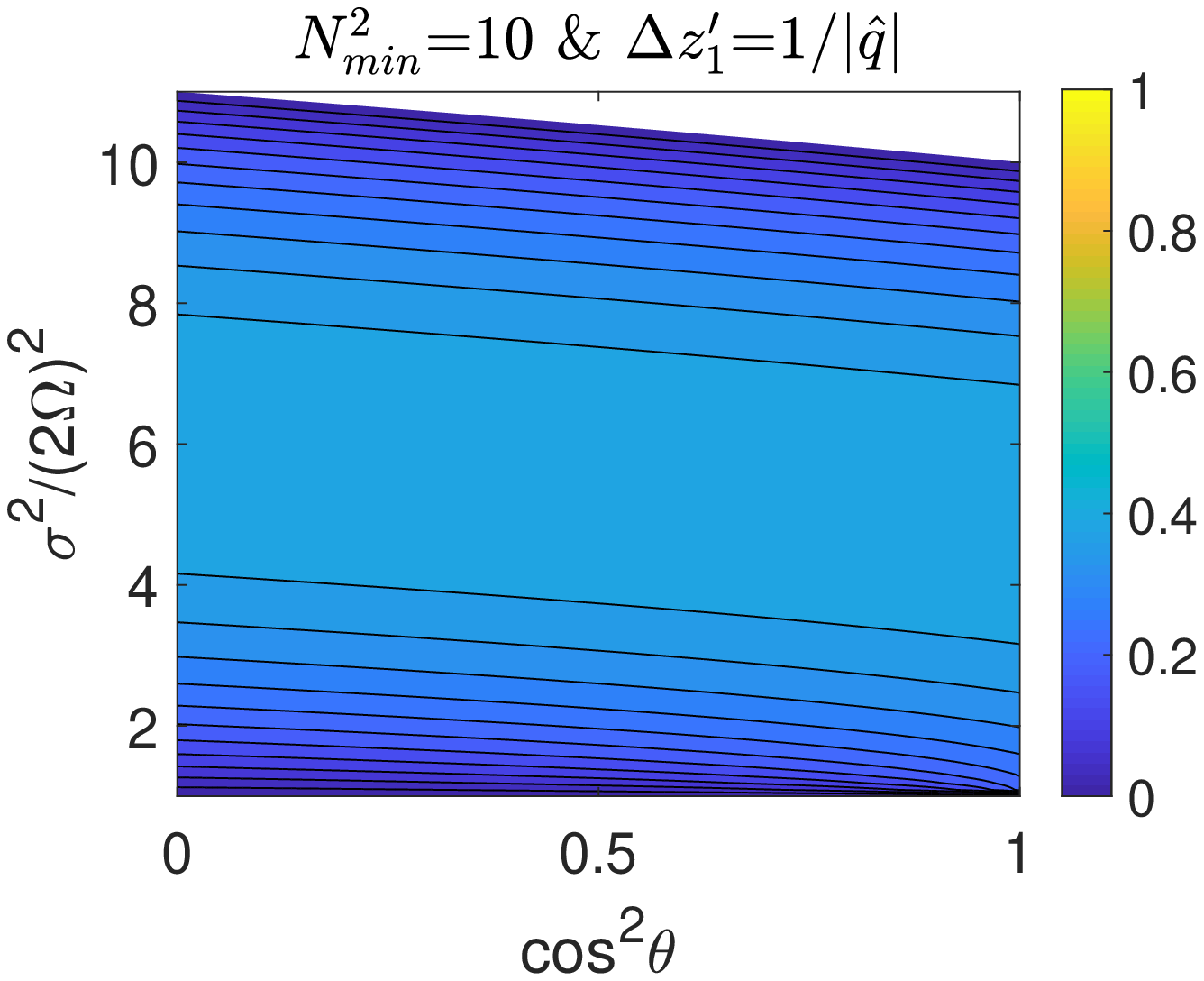}
\caption{}
\end{subfigure}
\begin{subfigure}{0.3\textwidth}
\includegraphics[width=\linewidth]{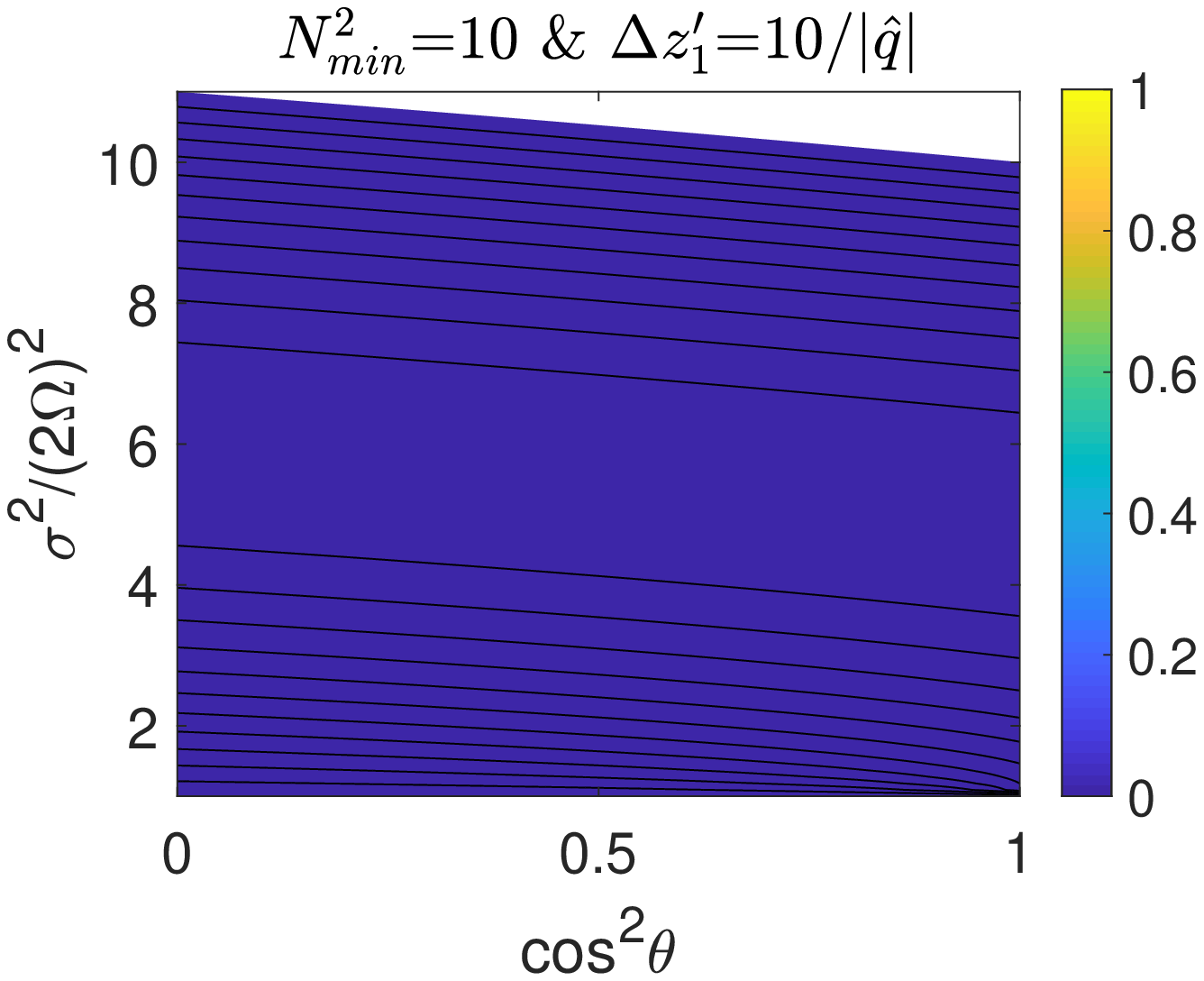}
\caption{}
\end{subfigure}

\medskip

\begin{subfigure}{0.3\textwidth}
\includegraphics[width=\linewidth]{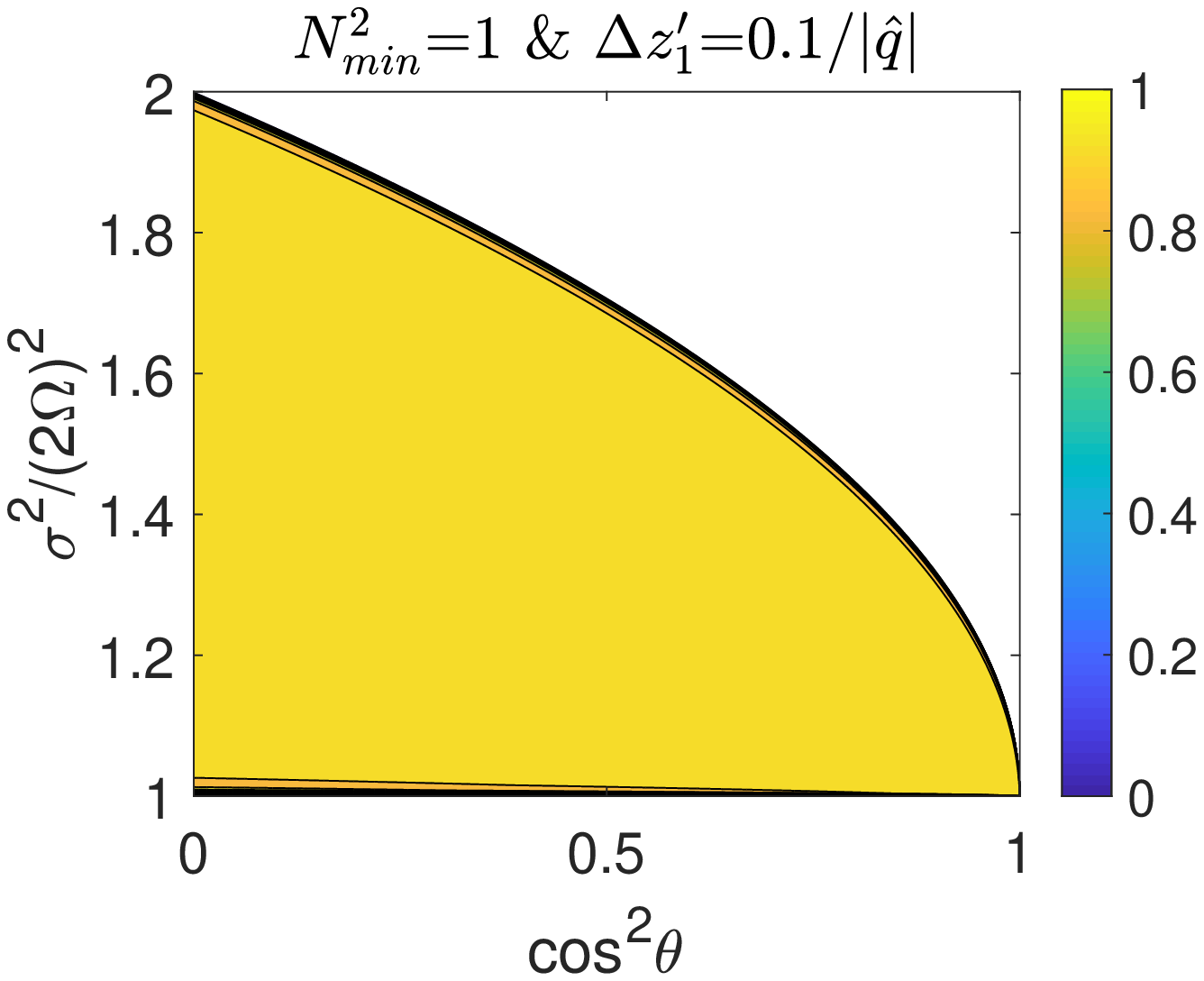}
\caption{}
\end{subfigure}
\begin{subfigure}{0.3\textwidth}
\includegraphics[width=\linewidth]{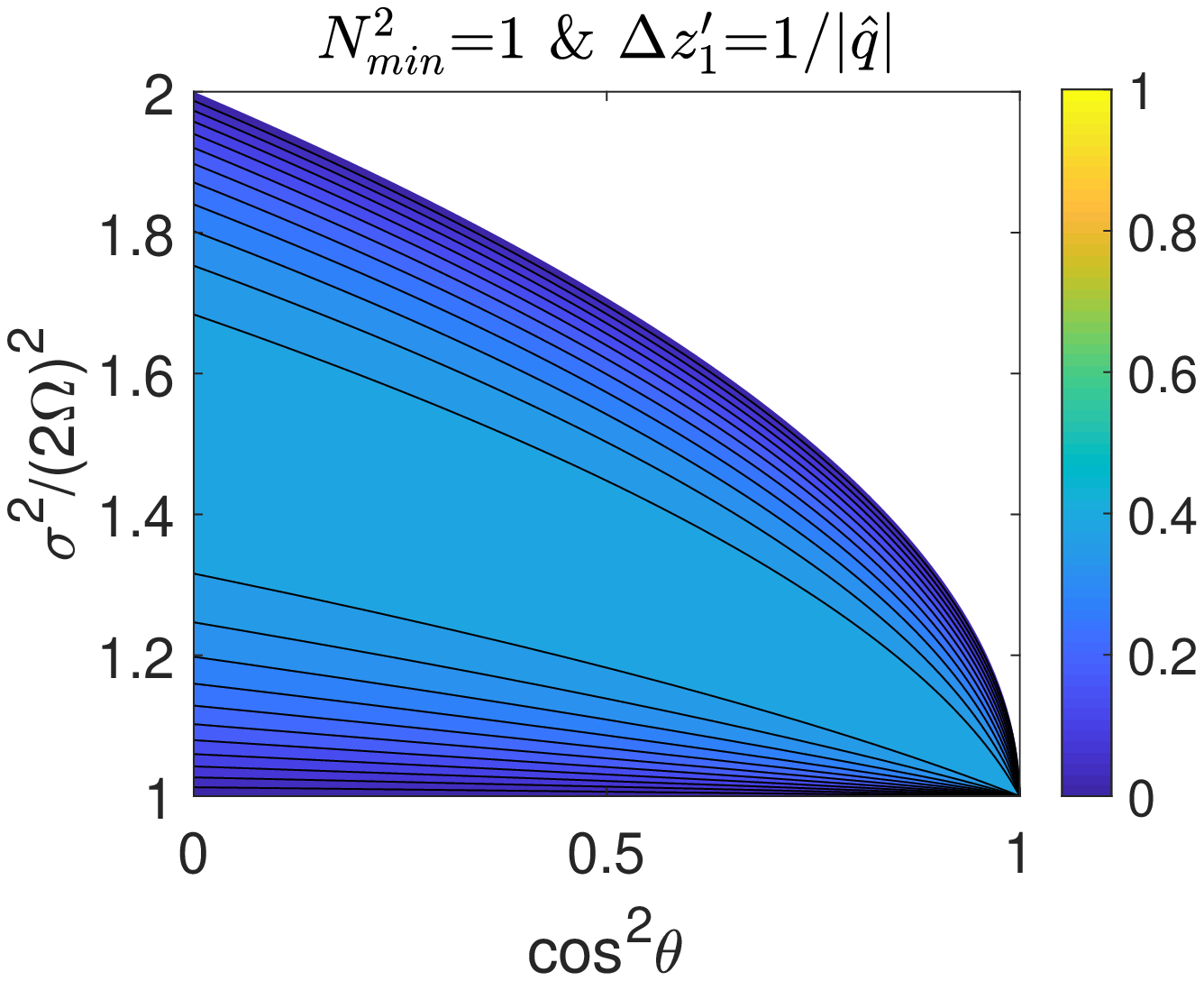}
\caption{}
\end{subfigure}
\begin{subfigure}{0.3\textwidth}
\includegraphics[width=\linewidth]{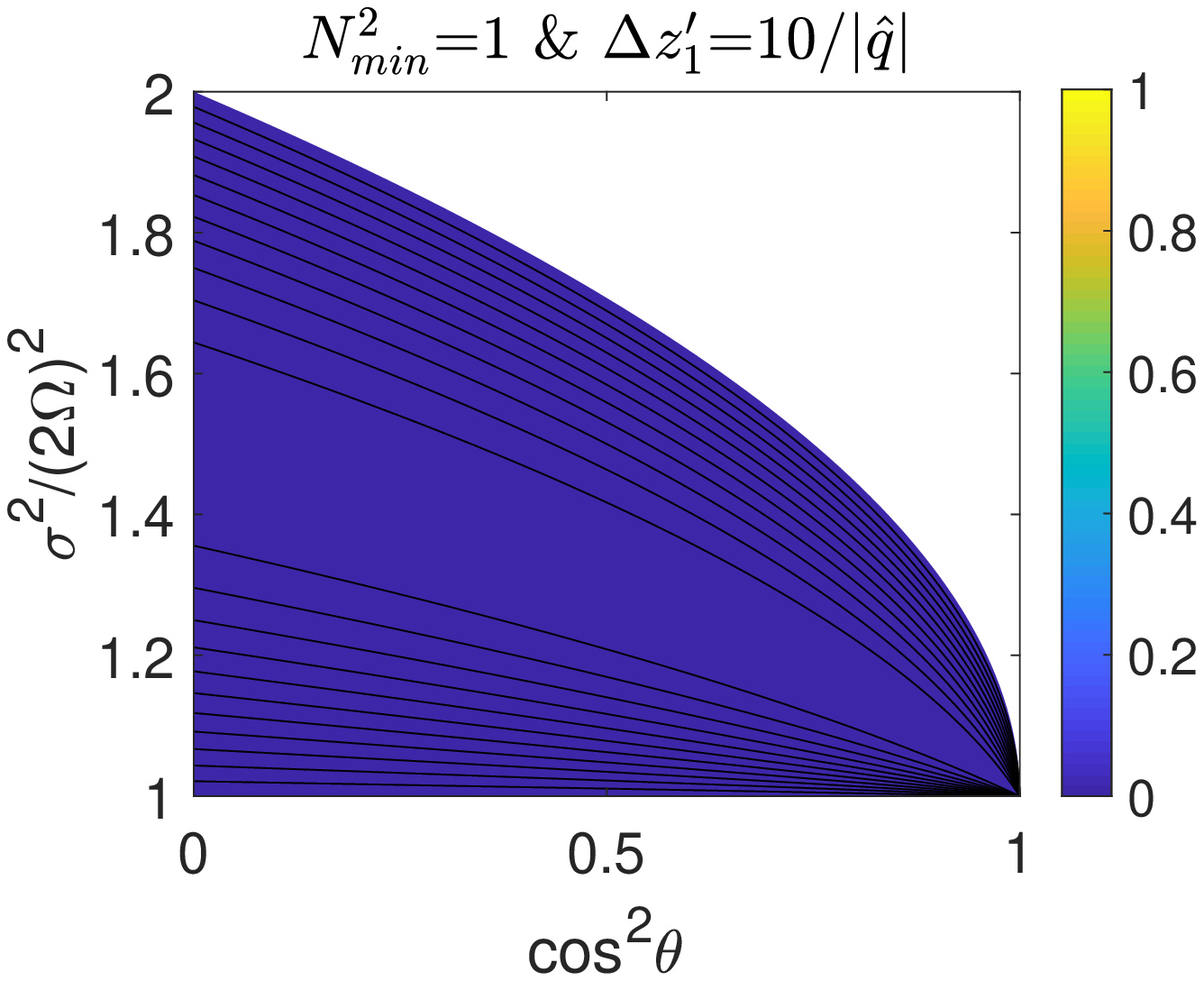}
\caption{}
\end{subfigure}

\medskip

\begin{subfigure}{0.3\textwidth}
\includegraphics[width=\linewidth]{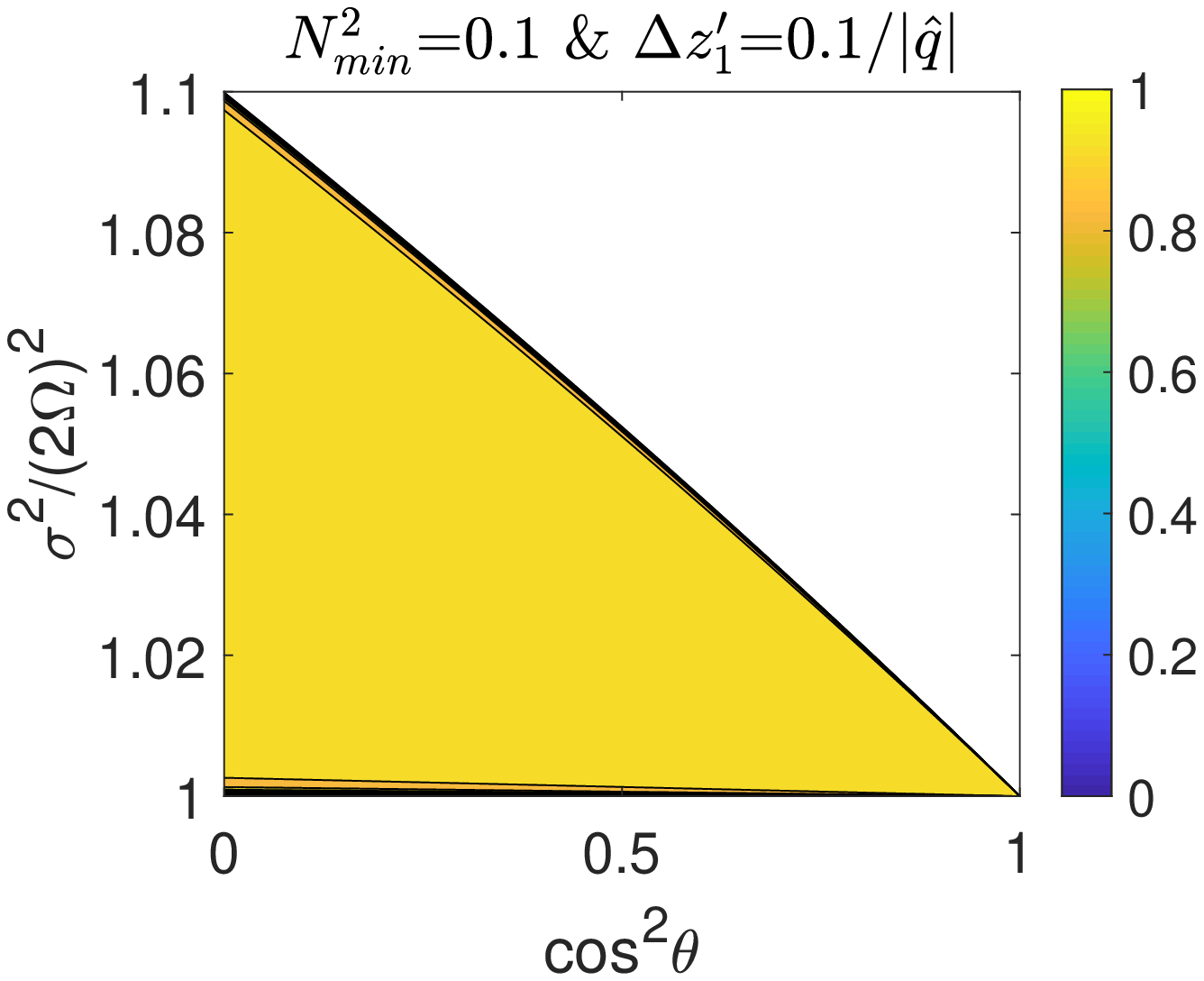}
\caption{}
\end{subfigure}
\begin{subfigure}{0.3\textwidth}
\includegraphics[width=\linewidth]{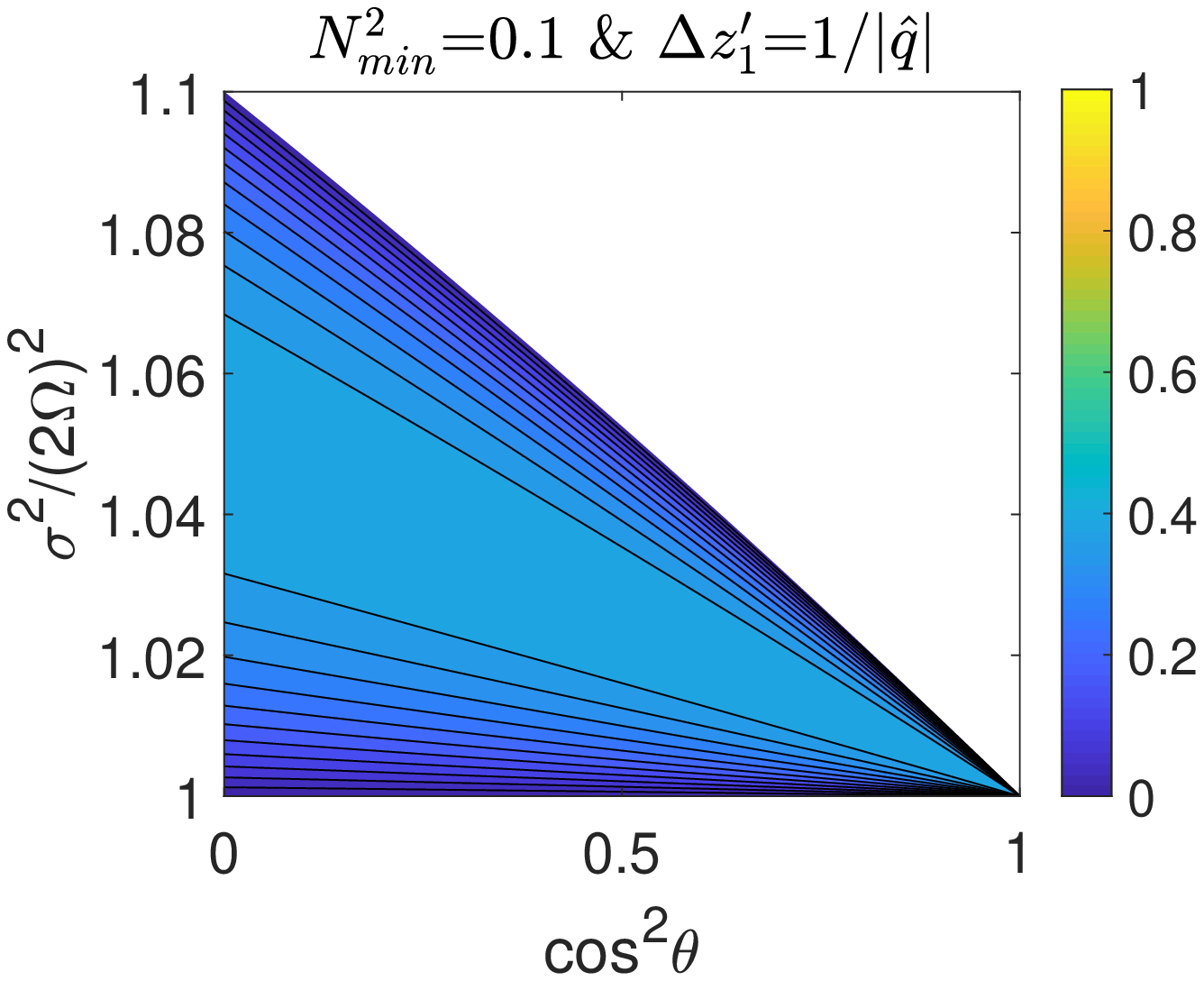}
\caption{}
\end{subfigure}
\begin{subfigure}{0.3\textwidth}
\includegraphics[width=\linewidth]{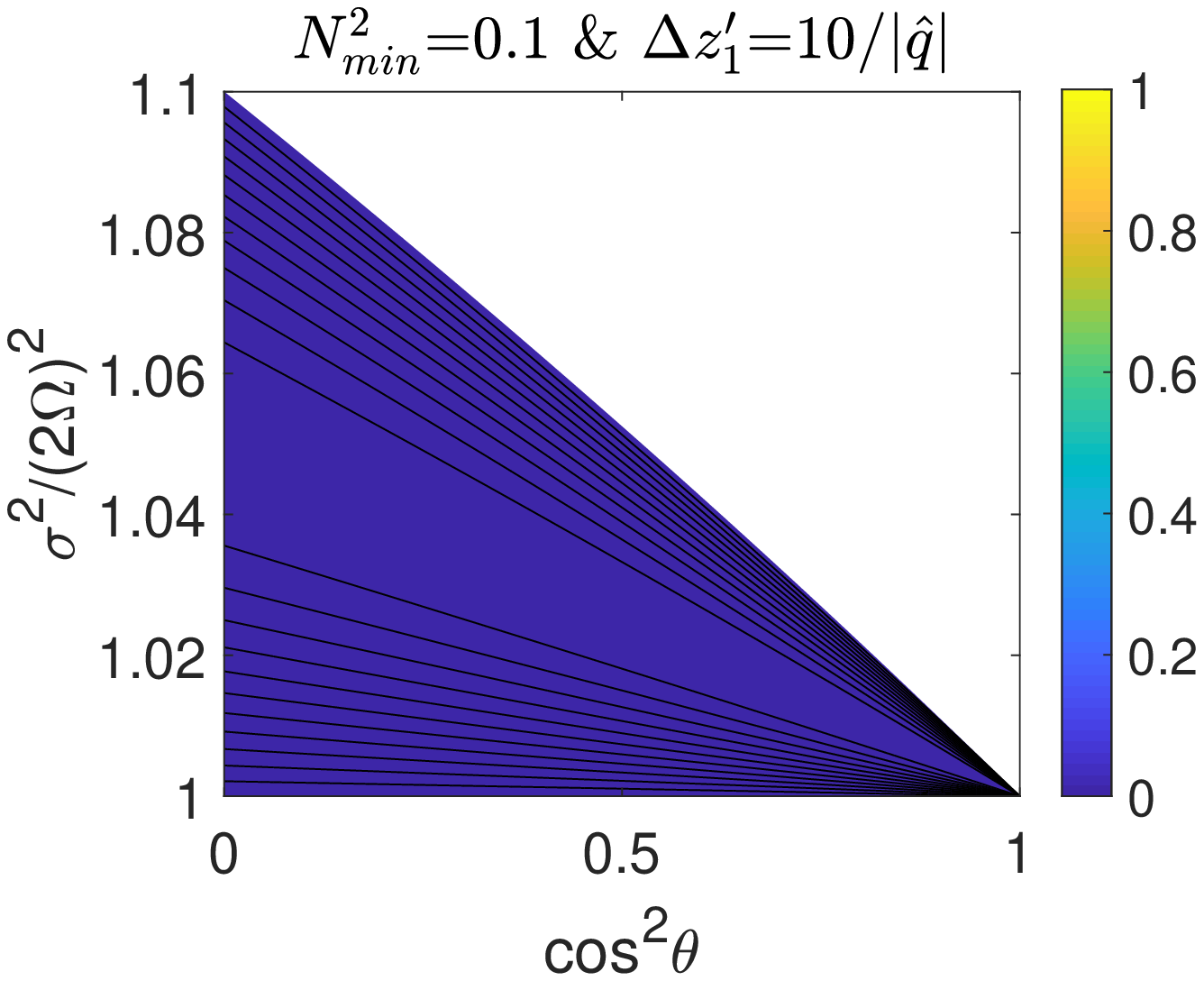}
\caption{}
\end{subfigure}

\caption{Contour plots of transmission ratios for the tunneling of gravity wave in a three-layer structure. (a-c)Transmission ratios when the buoyancy frequency $N_{min}^2=10$, and $\Delta z'_{1}=(0.1,1,10)/|\hat{q}_{1}|$. (d-f)Transmission ratios when the buoyancy frequency $N_{min}^2=1$, and $\Delta z'_{1}=(0.1,1,10)/|\hat{q}_{1}|$. (g-i)Transmission ratios when the buoyancy frequency $N_{min}^2=0.1$, and $\Delta z'_{1}=(0.1,1,10)/|\hat{q}_{1}|$. Tunneling of gravity waves can only occur in colored regions. Regions are left white if tunneling of gravity waves is prohibited.\label{fig:f7}}
\end{figure}

Replacing $q_{m}$ with $i\hat{q}_{m}$ in (\ref{eq15}), we can obtain the transfer matrix.
In this configuration, we always have $C=f^2-\sigma^2<0$ (since $\sigma_{4}^2>f^2$). Hence for an outgoing transmitted wave, we have $b_{M+1}=0$ (modified phase velocity has an opposite sign as the group velocity). Again, let us first discuss the transmission ratio for a three-layer structure. In such case, we have $a_{1}=T_{11}a_{2}$ and
\begin{eqnarray}
&&T_{11}=\frac{1}{4}e^{-i(s_{1}z_{1}-s_{2}z'_{1})}[(1+\frac{i\hat{q}_{1}}{s_{1}}+\frac{s_{2}}{s_{1}}+\frac{s_{2}}{i\hat{q}_{1}})e^{-\hat{q}_{1}\Delta z'_{1}}+(1-\frac{i\hat{q}_{1}}{s_{1}}+\frac{s_{2}}{s_{1}}-\frac{s_{2}}{i\hat{q}_{1}})e^{\hat{q}_{1}\Delta z'_{1}}]~.
\end{eqnarray}
Thus the wave transmission ratio is
\begin{eqnarray}
\eta&&=|\frac{s_{2}}{s_{1}}|\frac{1}{|T_{11}|^2}\\
&&=16\left[(\sqrt{\frac{s_{1}}{s_{2}}}+\sqrt{\frac{s_{2}}{s_{1}}})^2(e^{-\hat{q}_{1}\Delta z'_{1}}+e^{\hat{q}_{1}\Delta z'_{1}})^{2}+(\sqrt{\frac{\hat{q}_{1}^2}{s_{1}s_{2}}}
-\sqrt{\frac{s_{1}s_{2}}{\hat{q}_{1}^2}})^{2}(e^{-\hat{q}_{1}\Delta z'_{1}}-e^{\hat{q}_{1}\Delta z'_{1}})^{2}\right]^{-1}\\
&&=16\left\{(\sqrt{\frac{s_{1}}{s_{2}}}+\sqrt{\frac{s_{2}}{s_{1}}})^2(e^{-\hat{q}_{1}\Delta z'_{1}}+e^{\hat{q}_{1}\Delta z'_{1}})^{2}+\left[(\sqrt{\frac{\hat{q}_{1}^2}{s_{1}s_{2}}}
+\sqrt{\frac{s_{1}s_{2}}{\hat{q}_{1}^2}})^{2}-4\right]\left[(e^{-\hat{q}_{1}\Delta z'_{1}}+e^{\hat{q}_{1}\Delta z'_{1}})^{2}-4\right]\right\}^{-1}
\end{eqnarray}
Obviously, $\eta$ depends on the values of $s_{1}/s_{2}$, $\hat{q}_{1}^2/(s_{1}s_{2})$, and $|\hat{q}_{1}|\Delta z'_{1}$. It increases with $s_{1}/s_{2}$ when $s_{1}/s_{2}<1$, and decreases with $s_{1}/s_{2}$ when $s_{1}/s_{2}>1$. Similarly, we see that it increases with $\hat{q}_{1}^2/(s_{1}s_{2})$ when $\hat{q}_{1}^2/(s_{1}s_{2})<1$, and decreases with $\hat{q}_{1}^2/(s_{1}s_{2})$ when $\hat{q}_{1}^2/(s_{1}s_{2})>1$. It is also noted that it decreases with $|\hat{q}_{1}|\Delta z'_{1}$. Thus efficient transmission ratio can be achieved when $s_{1}\rightarrow s_{2}$, $\hat{q}_{1}^2 \rightarrow s_{1}s_{2}$, and $|\hat{q_{1}}|\Delta z'_{1}\rightarrow 0$. The condition $\hat{q}_{1}^2 \rightarrow s_{1}s_{2}$ can be relaxed if $|\hat{q_{1}}|\Delta z'_{1}\rightarrow 0$. Therefore, efficient transmission requires that the wavenumbers ($s_{1}$ and $s_{2}$) in the stable layers are similar in magnitudes, and the thickness of the convective layer ($\Delta z'_{1}$) is much smaller than the e-folding decay distance ($1/|\hat{q}_{1}|$). Fig.~\ref{fig:f7} shows the contour plots of the transmission ratios for different $N_{min}^2$ and $\Delta z'_{1}$. In all the cases, we set $s_{1}=s_{2}$ and $\sin^2\alpha=1$. First, the figure clearly shows that the frequency domain increases with $N_{min}^2$ and $\theta$. It is consistent with the previous analysis on the width of the frequency domain. Second, we see that the transmission ratio is mainly affected by the thickness of the convective layer $\Delta z'_{1}$. The shallower the convective layer is, the higher the transmission ratio is. We also note that the transmission ratio is insensitive to $N_{min}^2$. The effect of degree of stratification on the transmission ratio is insignificant.

Now we consider the wave transmission in configuration 3 with more layers. Again, we assume that all of the stable layers have the same degree of stratification ($N_{1}^2$) and thickness ($\Delta z_{s}$). Similarly, we assume that all of the convective layers have the same thickness $\Delta z_{c}$. By these settings, $s_{m}=s$ and $q_{m}=i\hat{q}$ are constants in the stable and convective layers, respectively. When $s_{m}=s$, $q_{m}=i\hat{q}$, and $\Delta z'_{m}=\Delta z_{c}$, the transfer matrix $T_{m,m+1}$ in (\ref{eq15}) can be written as
\begin{eqnarray}
&&\widehat{T}_{11}=\frac{1}{4}e^{-is\Delta z_{c}}\left\{\left[2+i(\frac{\hat{q}}{s}-\frac{s}{\hat{q}})\right]e^{-\hat{q}\Delta z_{c}}+\left[2-i(\frac{\hat{q}}{s}-\frac{s}{\hat{q}})\right]e^{\hat{q}\Delta z_{c}}\right\}~,\label{eq55}\\
&&\widehat{T}_{12}=\frac{1}{4}e^{is(z_{m}+z_{m}')}\left\{\left[(\frac{\hat{q}}{s}+\frac{s}{\hat{q}})\right]e^{-\hat{q}\Delta z_{c}}-\left[(\frac{\hat{q}}{s}+\frac{s}{\hat{q}})\right]e^{\hat{q}\Delta z_{c}}\right\}~,
\end{eqnarray}
with $\widehat{T}_{22}^{*}=\widehat{T}_{11}$ and $\widehat{T}_{21}=\widehat{T}_{12}^{*}$.
The eigenvalues of $\bm{T}_{m,m+1}$ are the roots of the following equation
\begin{eqnarray}
\lambda^2-2\Re(\widehat{T}_{11})\lambda+1=0~.
\end{eqnarray}
Similarly, we can obtain a sufficient condition for efficient wave tunneling if
\begin{eqnarray}
\Delta_{\lambda}=\left[\frac{1}{2}(e^{-\hat{q}\Delta z_{c}}+e^{\hat{q}\Delta z_{c}})\cos(s\Delta z_{c})+\frac{1}{4}\left(\frac{\hat{q}}{s}-\frac{s}{\hat{q}}\right)(e^{-\hat{q}\Delta z_{c}}-e^{\hat{q}\Delta z_{c}})\sin(s\Delta z_{c})\right]^2-1~,
\end{eqnarray}
and $\Delta_{\lambda}\leq 0$.
It should be noted that $\Delta_{\lambda}\leq 0$ is only a necessary but not sufficient condition for the efficient transmission. If $|\hat{q}|\Delta z_{c} \gg 0$, then $\Delta_{\lambda}$ is likely to be greater than zero. Therefore the probability is higher at small $|\hat{q}|\Delta z_{c}$ for the efficient transmission to take place.

By using similar technique as mentioned in section \ref{sec2.1.2}, we can find analytical solution of transmission ratio for the special case $|s|\Delta z_{s}=(\ell'+1/2) \pi$ (so that $|s|\Delta z_{s}-\pi/2=\ell'\pi$ and $\hat{T}_{12}/\exp(\pi/2)$ is pure imaginary number, and the problem is analogous to that in section \ref{sec2.1.2}). Here we only give the result without showing the details. Under the condition $|s|\Delta z_{s}=(\ell'+1/2) \pi$, the transmission ratio in the multi-layer structure for tunneling of gravity wave is
\begin{eqnarray}
\eta=\left\{\cos^2 M\beta+\frac{1}{4}\left[(\frac{\hat{q}}{s}-\frac{s}{\hat{q}})(e^{-\hat{q}\Delta z_{c}}+e^{\hat{q}\Delta z_{c}})\right]^2 \left[\left(\frac{\hat{q}}{s}+\frac{s}{\hat{q}}\right)^2 - (e^{-\hat{q}\Delta z_{c}}+e^{\hat{q}\Delta z_{c}})^2\right]^{-1}\sin^2 M\beta\right\}^{-1}~,\label{eq79}
\end{eqnarray}
where
\begin{eqnarray}
\beta=\arg\left((e^{-\hat{q}\Delta z_{c}}+e^{\hat{q}\Delta z_{c}})+i\left[\left(\frac{\hat{q}}{s}+\frac{s}{\hat{q}}\right)^2 - (e^{-\hat{q}\Delta z_{c}}+e^{\hat{q}\Delta z_{c}})^2\right]^{1/2}\right),
\end{eqnarray}
and $\arg$ is the argument function operating on complex numbers. Note that $\sin^2 M\beta \propto ({\hat{q}}/{s}+{s}/{\hat{q}})^2 - (e^{-\hat{q}\Delta z_{c}}+e^{\hat{q}\Delta z_{c}})^2$, and there is no singularity problem in (\ref{eq79}). From (\ref{eq79}), we see that efficient tunneling of gravity waves generally requires $|\hat{q}|\Delta z_{c}$ to be small. In the limit $|\hat{q}|\Delta z_{c}\rightarrow 0$, we find $\eta\rightarrow 1$. Thus enhanced transmission of wave tunneling can occur at $|\hat{q}|\Delta z_{c}\rightarrow 0$ and $|s|\Delta z_{s}=(\ell'+1/2) \pi$. We call this phenomenon `resonant tunneling'.

\begin{figure}
\centering
\begin{subfigure}{0.3\textwidth}
\includegraphics[width=\linewidth]{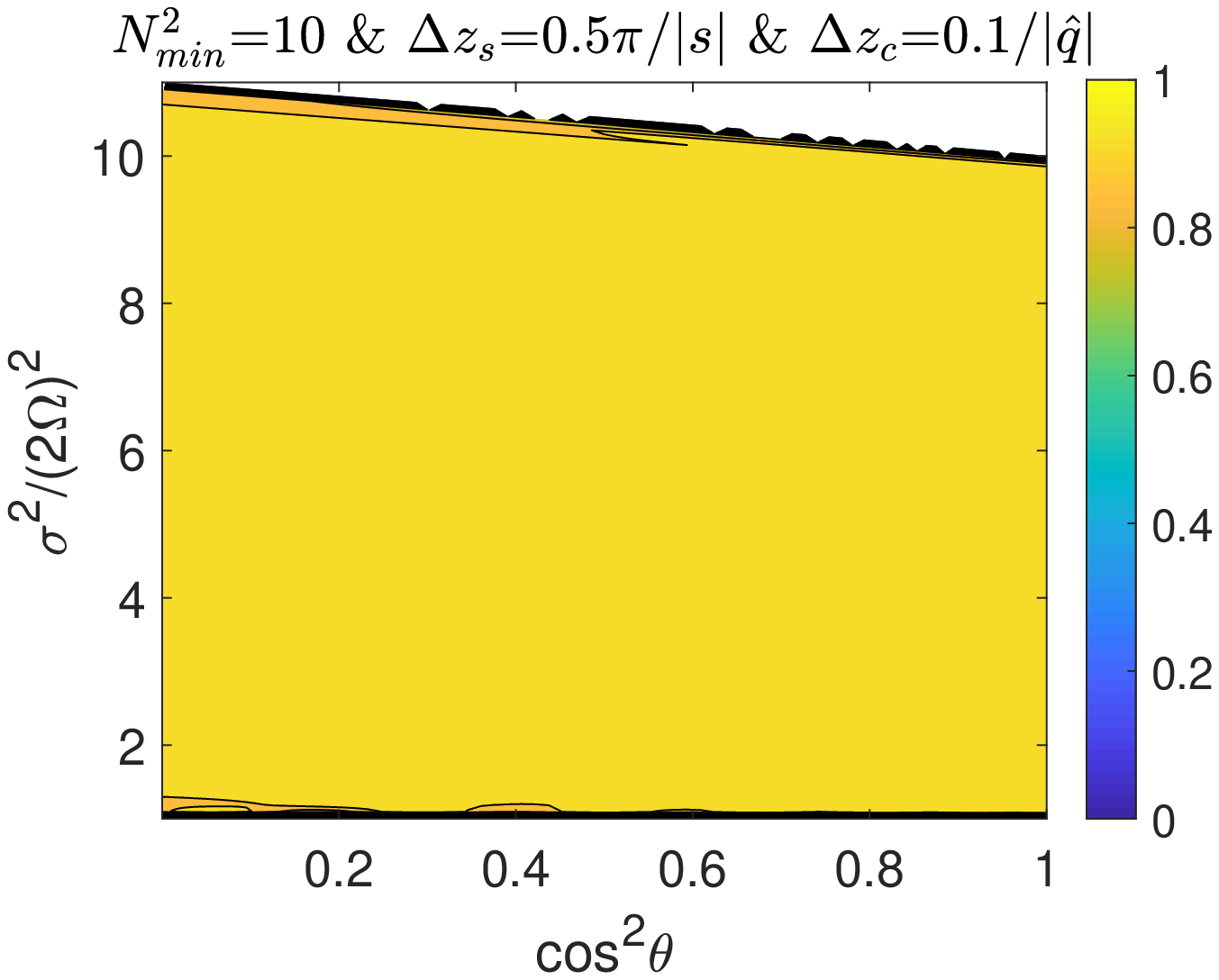}
\caption{}
\end{subfigure}
\begin{subfigure}{0.3\textwidth}
\includegraphics[width=\linewidth]{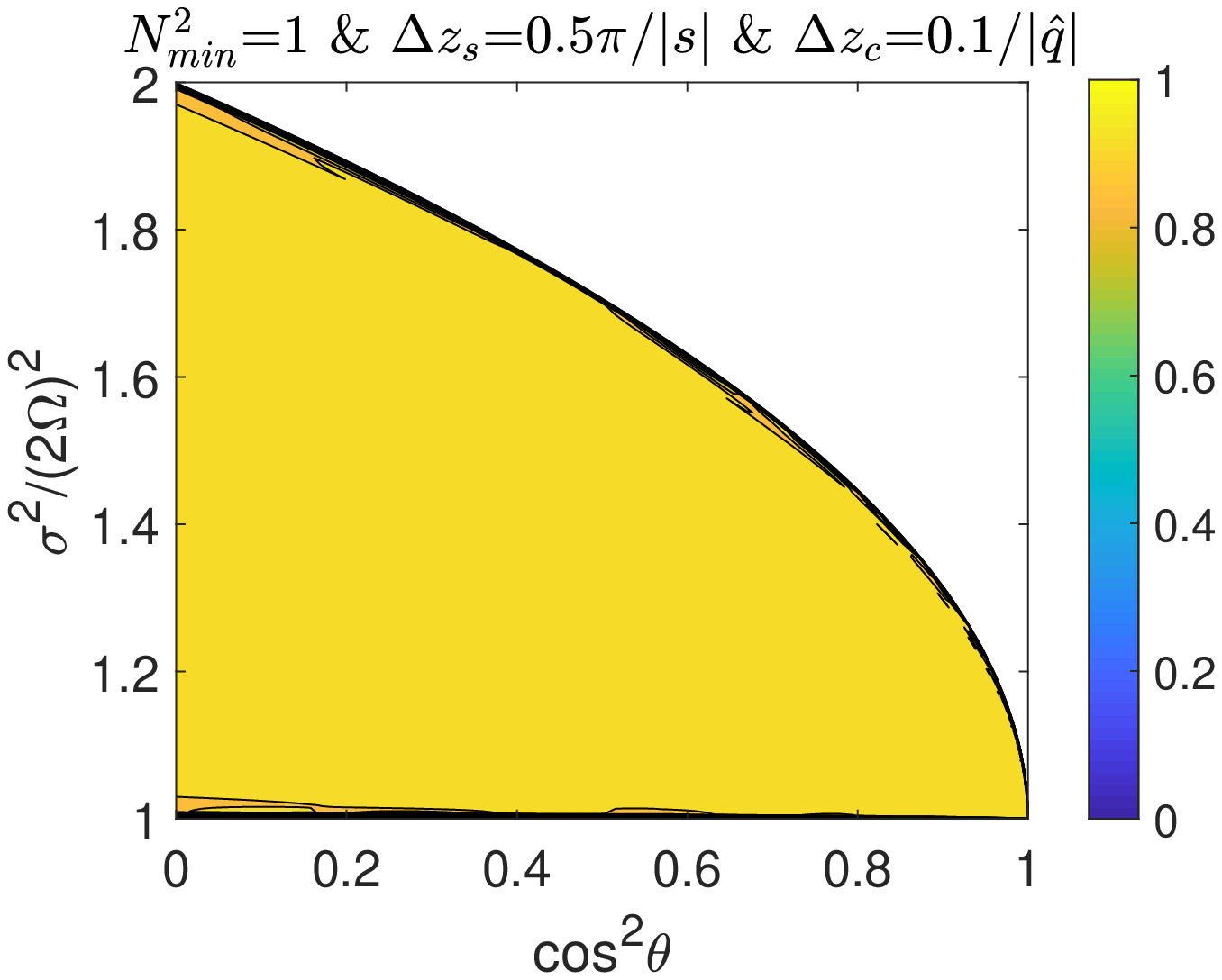}
\caption{}
\end{subfigure}
\begin{subfigure}{0.3\textwidth}
\includegraphics[width=\linewidth]{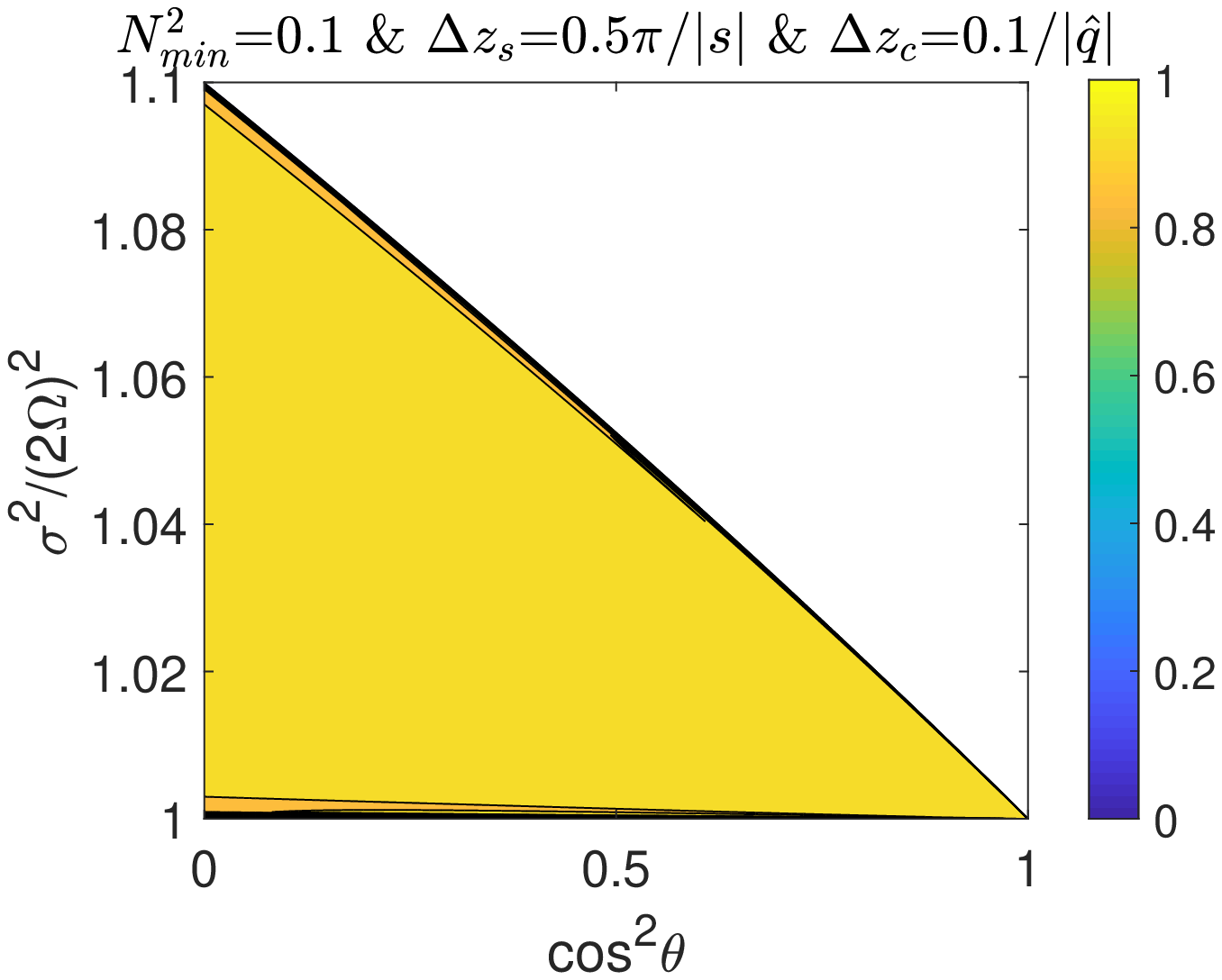}
\caption{}
\end{subfigure}

\medskip

\begin{subfigure}{0.3\textwidth}
\includegraphics[width=\linewidth]{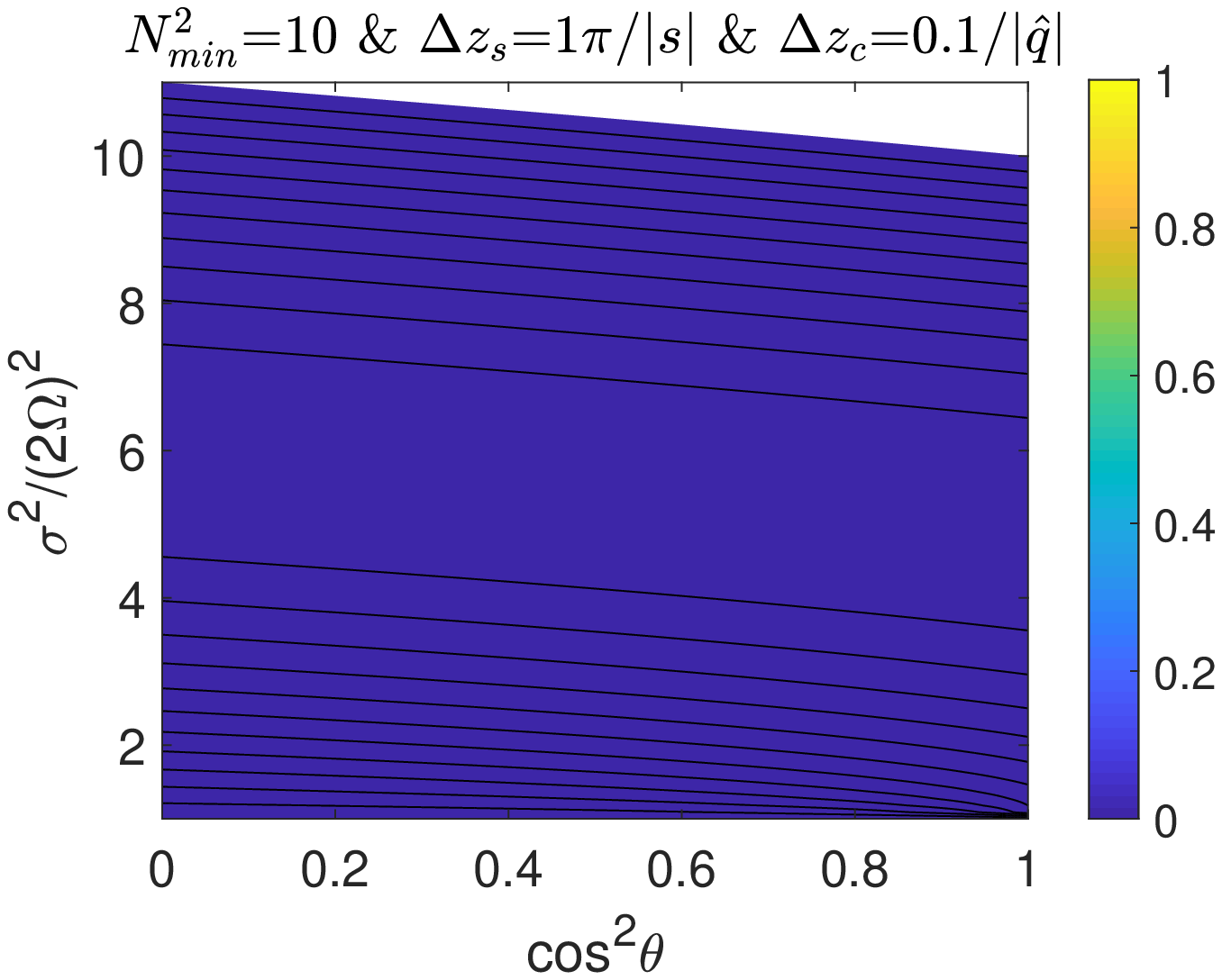}
\caption{}
\end{subfigure}
\begin{subfigure}{0.3\textwidth}
\includegraphics[width=\linewidth]{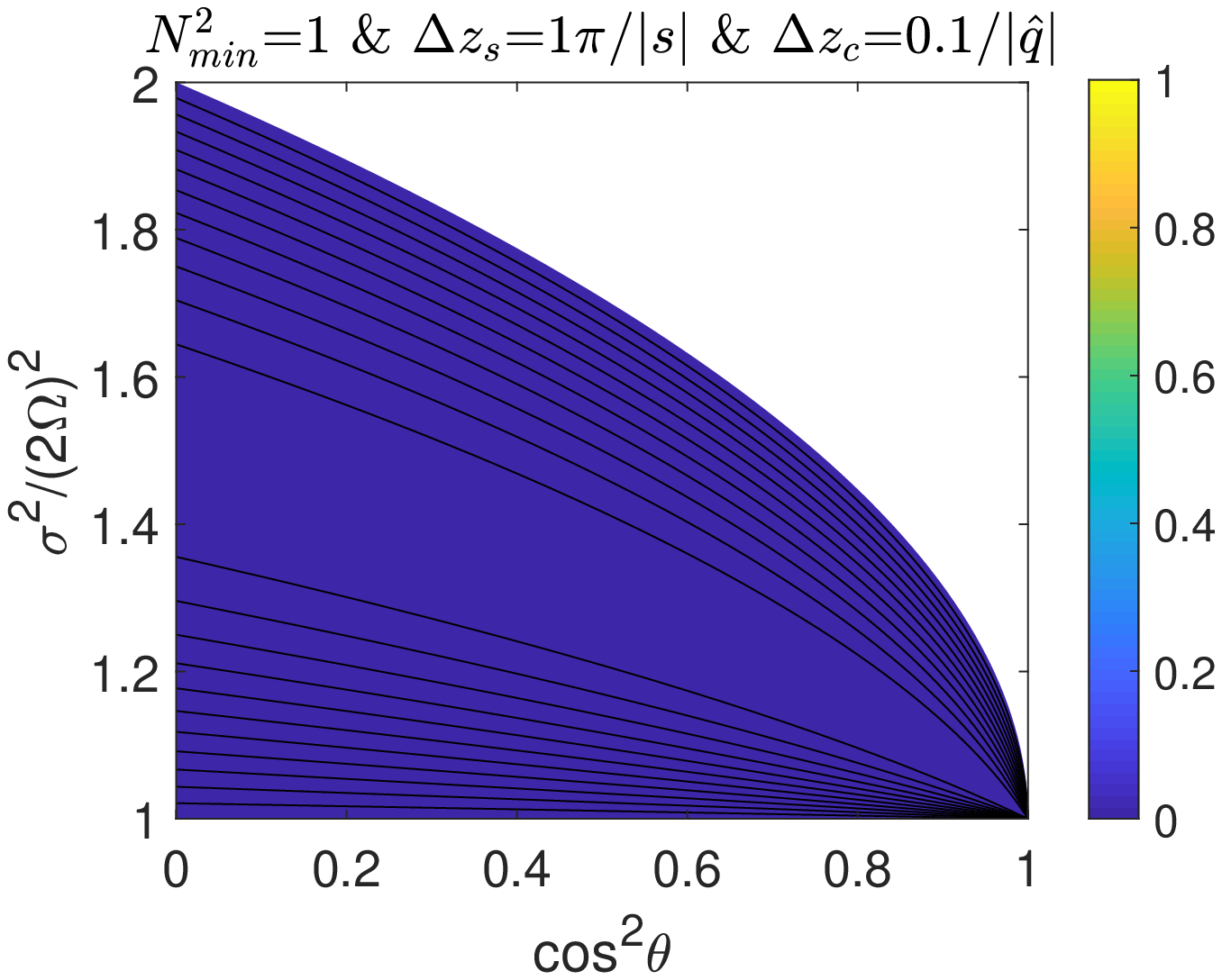}
\caption{}
\end{subfigure}
\begin{subfigure}{0.3\textwidth}
\includegraphics[width=\linewidth]{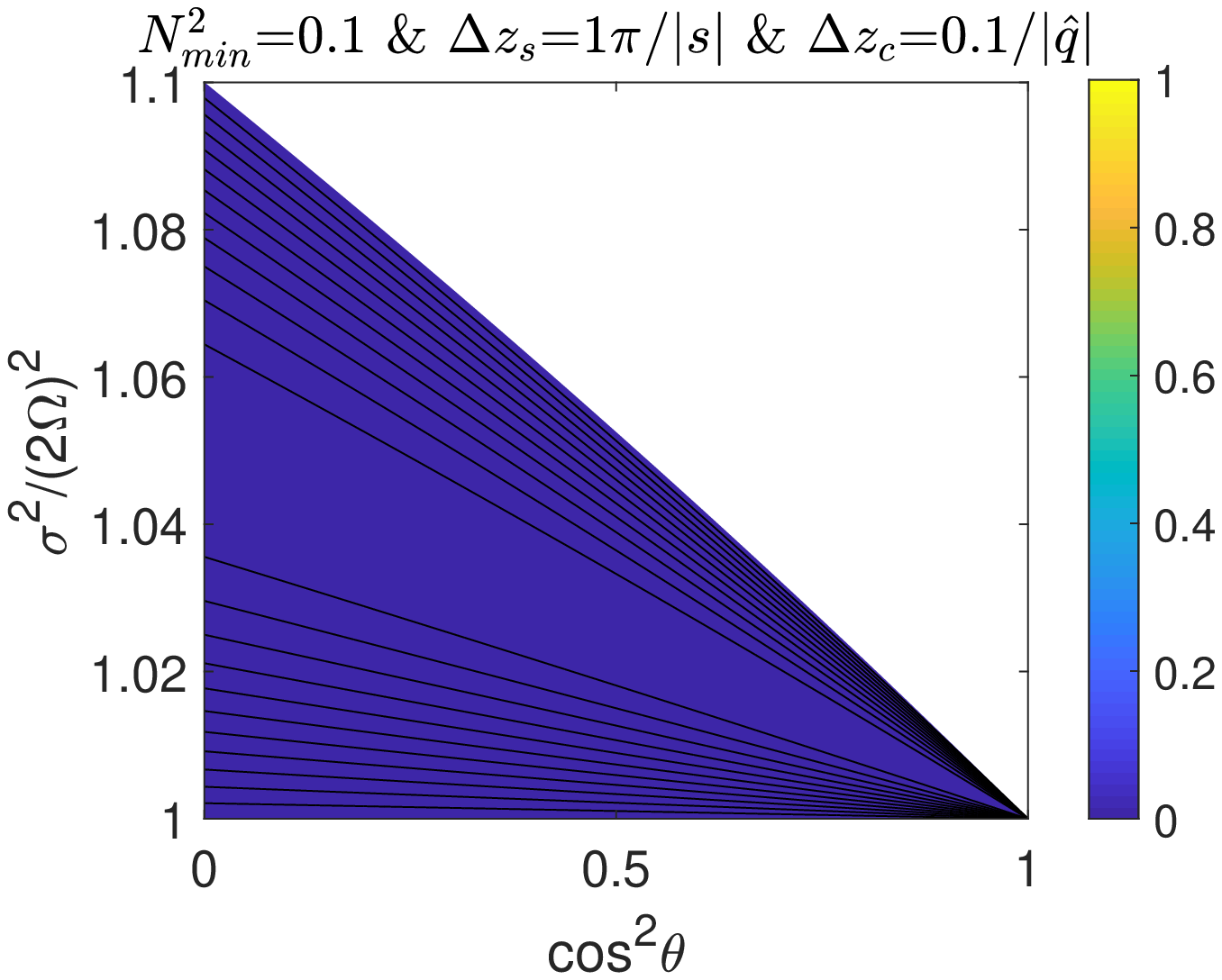}
\caption{}
\end{subfigure}

\caption{Contour plots of transmission ratios for the tunneling of gravity wave in a multi-layer structure. The buoyancy frequency and thicknesses of the convective layers are all the same with the values of $N_{min}^2$ and $\Delta z_{c}$, and the thicknesses of the stable layers are all the same with the value of $\Delta z_{s}$. (a-c)Transmission ratios when the buoyancy frequency $N_{min}^2=10,1,0.1$, $\Delta z_{c}=0.1/|\hat{q}|$, and $\Delta z_{s}=0.5\pi/|\hat{s}|$. (d-f)Transmission ratios when the buoyancy frequency $N_{min}^2=10,1,0.1$, $\Delta z_{c}=0.1/|\hat{q}|$, and $\Delta z_{s}=\pi/|\hat{s}|$. Tunneling of gravity waves can only occur in colored regions. Regions are left white if tunneling of gravity waves is prohibited.. \label{fig:f8}}
\end{figure}

Fig.~\ref{fig:f8} shows the contour plots of the transmission ratios for different $N_{min}^2$ in a 101-layer structure (M=50). In this calculation, all of the stable layers are assumed to have the same degree of stratification ($N_{min}^2$) and thickness ($\Delta z_{s}$), and the convective layers are assumed to have the same thickness ($\Delta z_{c}$). The analysis of the three-layer structure shows that efficient transmission only occurs when $\Delta z_{c}$ is small. The mathematical analysis on the eigenvalues of transfer matrices also indicates that efficient transmission is more likely to take place when $\Delta z_{c}$ is small. For this reason, we set $\Delta z_{c}=0.1/|\hat{q}|$ for all the computed cases. From the figure, we see that the transmission ratio is insensitive to the degree of stratification. Instead, the thickness of $\Delta z_{s}$ is more important. In our calculations, we find that the transmission is efficient when $\Delta z_{s}=(\ell+0.5) \pi/|s|$, and inefficient when $z_{s}=\ell \pi/|s|$, where $\ell$ is an integer. Therefore, the tunneling of gravity wave is efficient when each convective layer $\Delta z_{c}$ is much shallower than the e-folding decay distance $1/|\hat{q}|$ and the thickness of each stable layer is close to a multiple-and-a-half of the half wavelength.

\subsubsection{Tunneling of inertial waves}
In this section, we discuss the tunneling of inertial waves. Similarly, we consider a $(2M+1)$-layer structure with alternating $M+1$ convective layers and $M$ stable layers. The sketch plot is shown as configuration 4 in fig.~\ref{fig:f6}(b). For this configuration, inertial wave can propagate in convective layers but no wave could propagate in stable layers, and thus the frequency range is $\sigma_{1}^2<\sigma^2<\sigma_{2}^2$.
The width of the frequency domain is
\begin{eqnarray}
\sigma_{2}^2-\sigma_{1}^2=\frac{1}{2}[(f^2+\tilde{f}_{s}^2+N_{min}^2)- \sqrt{(f^2+\tilde{f}_{s}^2+N_{min}^2)^2-4N_{min}^2 f^2}]~.
\end{eqnarray}
Analysis shows that $\sigma_{2}^2-\sigma_{1}^2$ decreases with $\theta$ and $\sin^2\alpha$, and increases with $N_{min}^2/(2\Omega)^2$ (see Appendix \ref{appendixc}). Therefore the frequency domain is wider at polar regions than equatorial regions. Also it is wider when the zonal wavenumber dominates the meridional wavenumber, and it is wider when the degree of stratification is stronger.

\begin{figure}
\centering
\begin{subfigure}{0.3\textwidth}
\includegraphics[width=\linewidth]{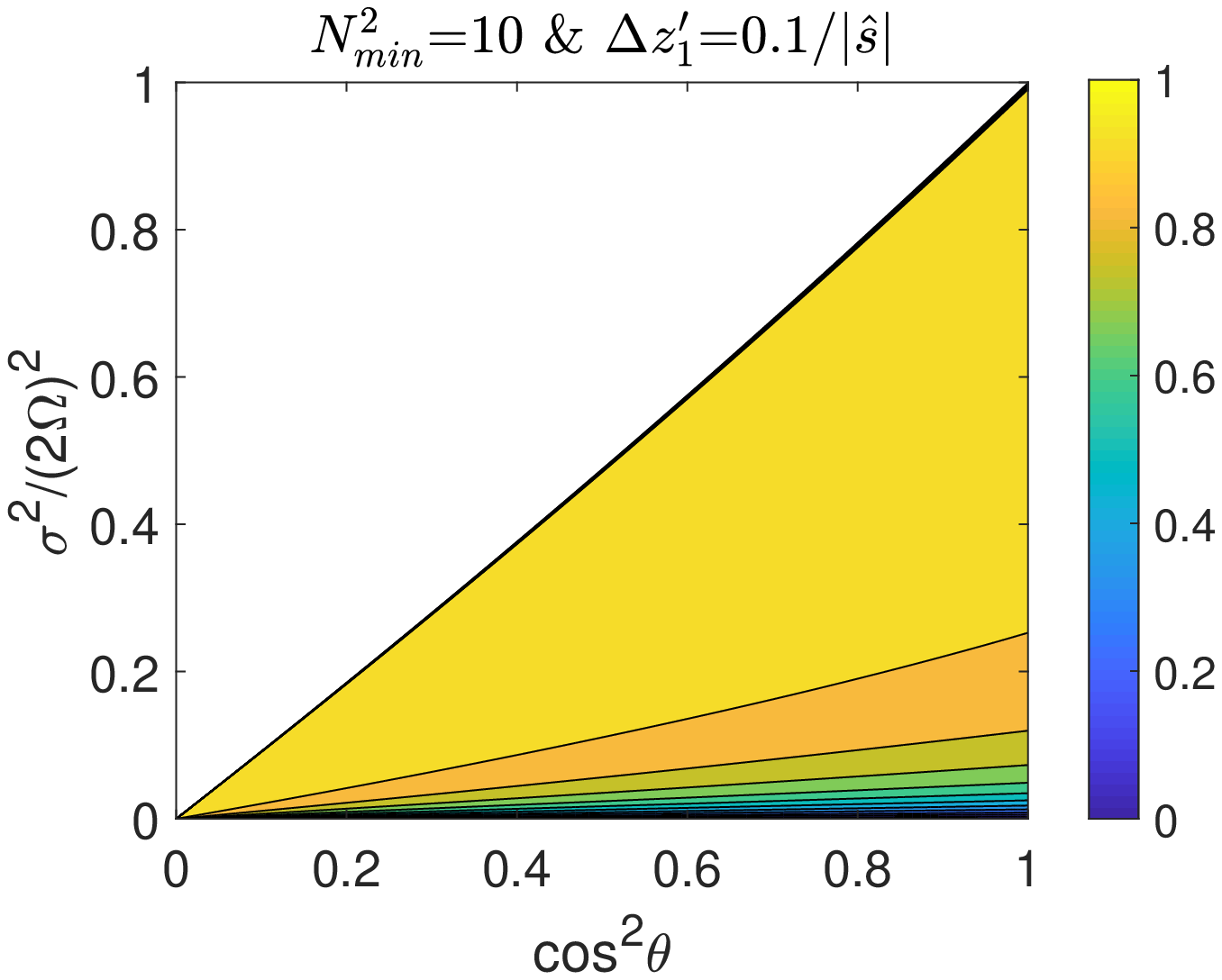}
\caption{}
\end{subfigure}
\begin{subfigure}{0.3\textwidth}
\includegraphics[width=\linewidth]{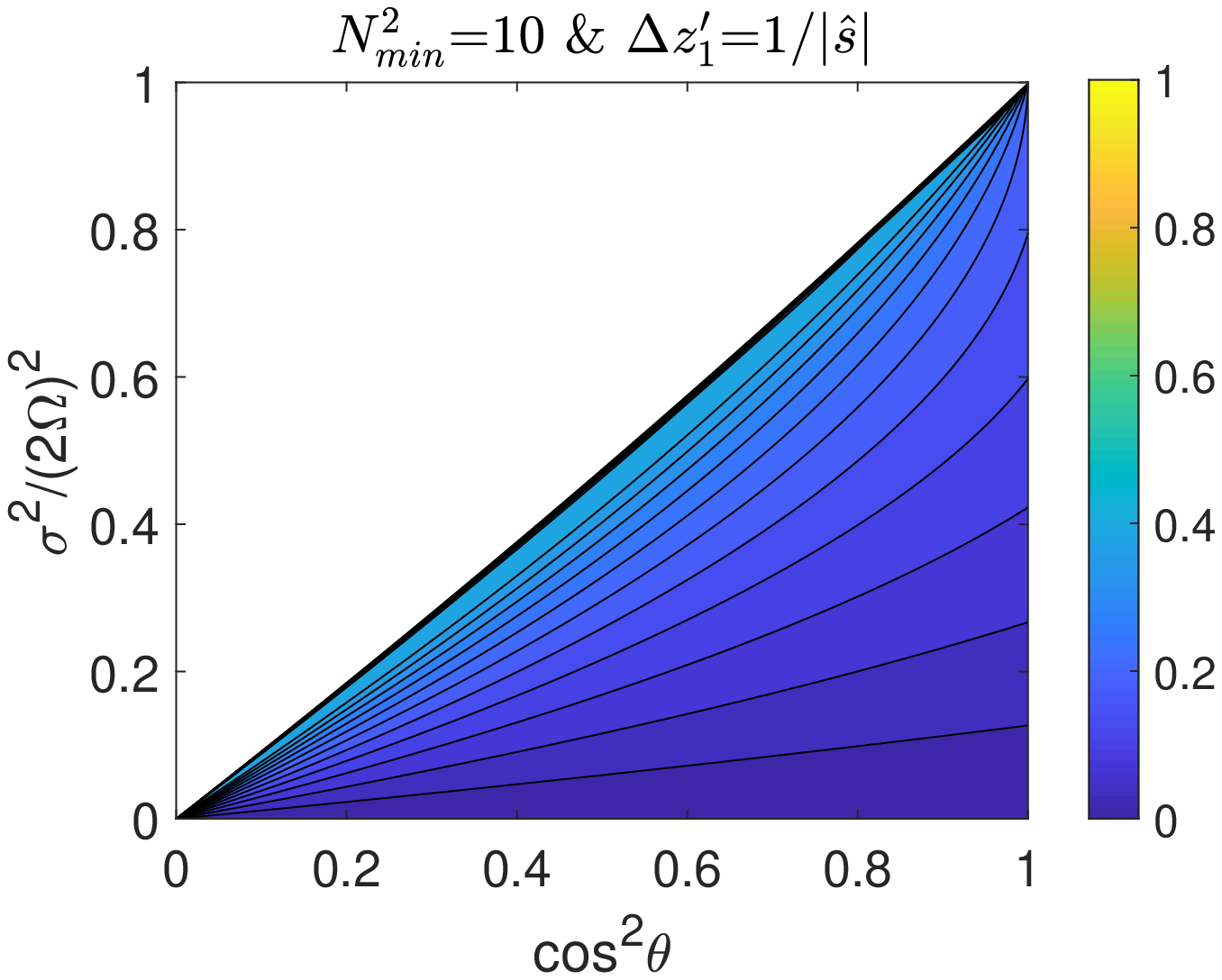}
\caption{}
\end{subfigure}
\begin{subfigure}{0.3\textwidth}
\includegraphics[width=\linewidth]{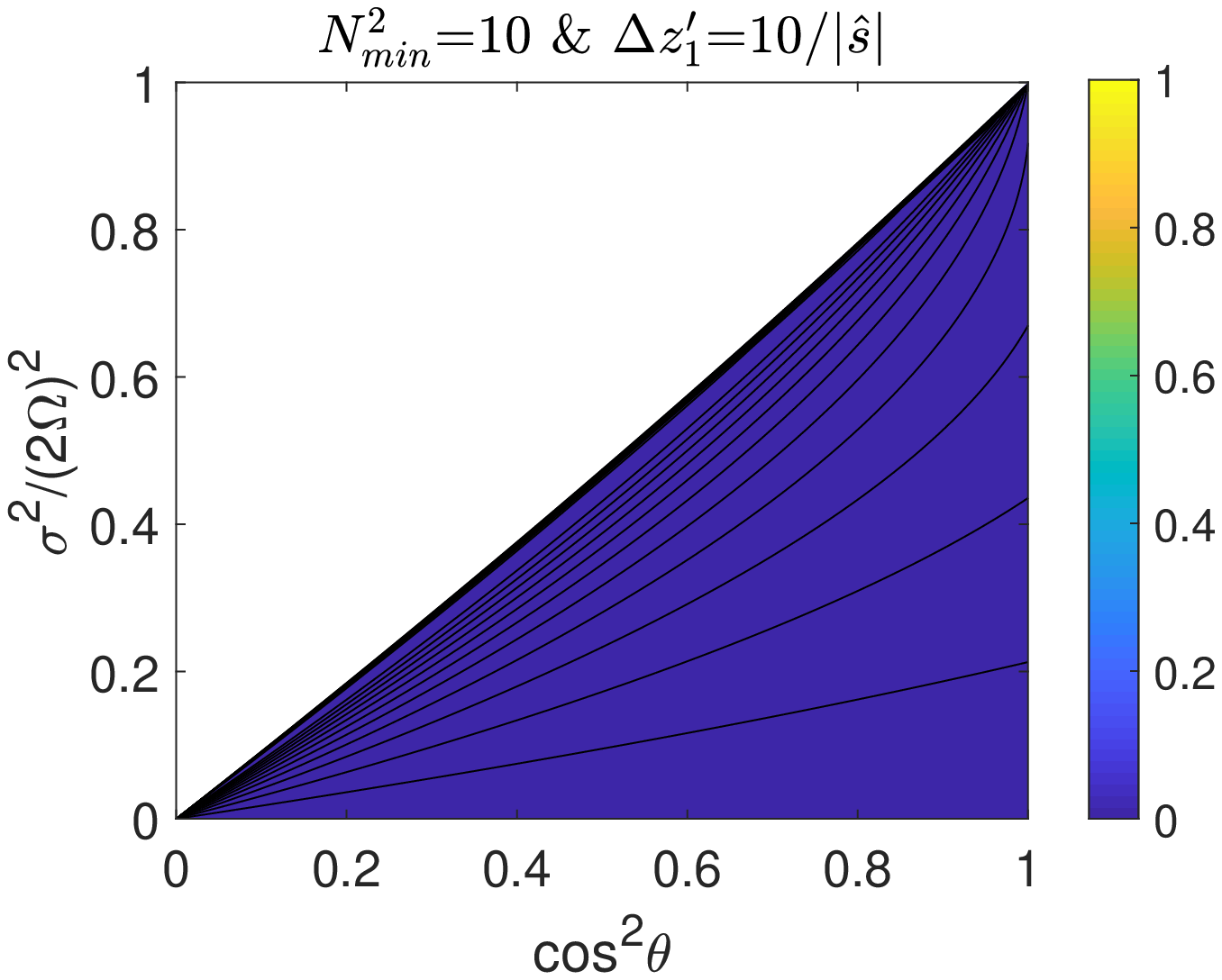}
\caption{}
\end{subfigure}

\medskip

\begin{subfigure}{0.3\textwidth}
\includegraphics[width=\linewidth]{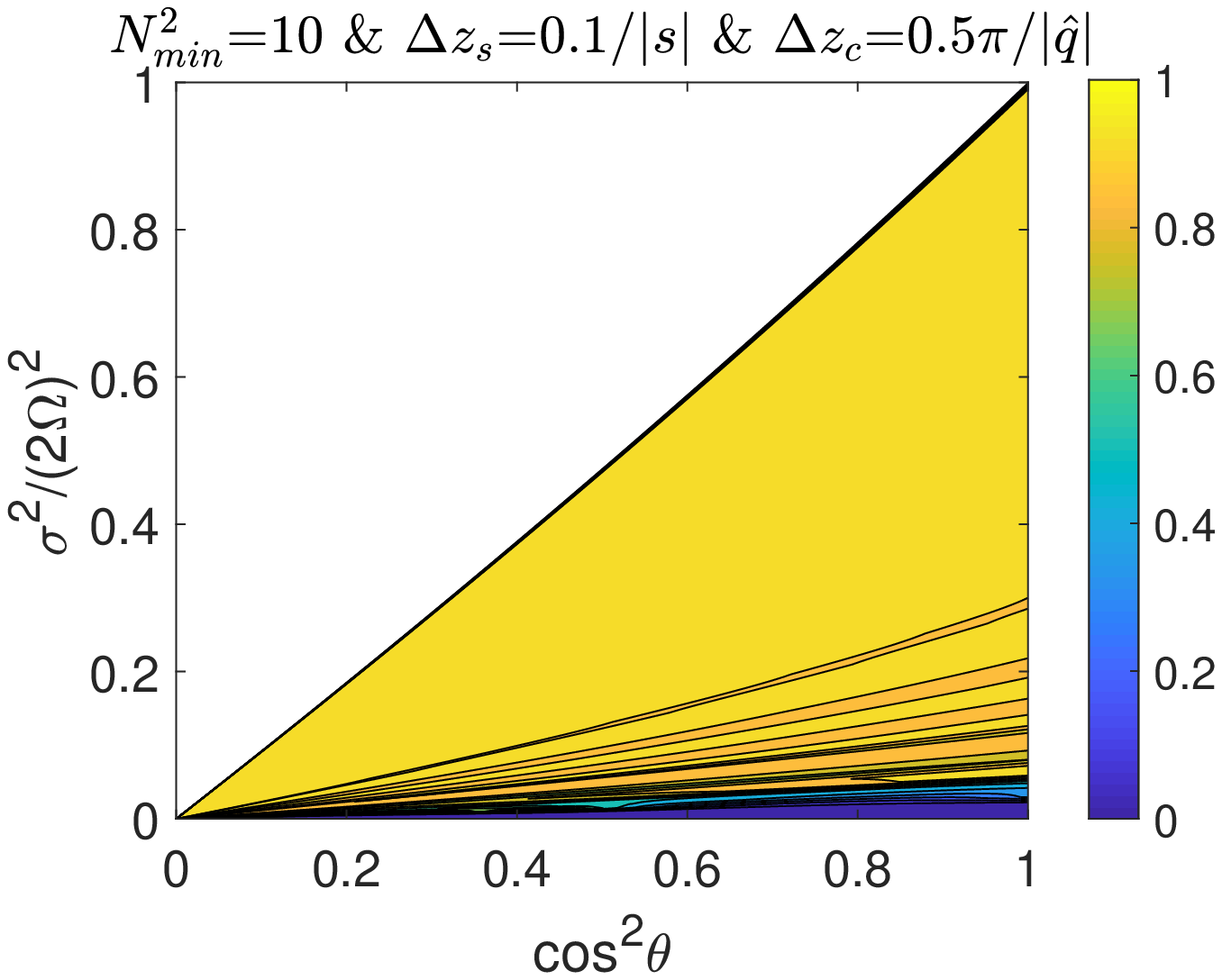}
\caption{}
\end{subfigure}
\begin{subfigure}{0.3\textwidth}
\includegraphics[width=\linewidth]{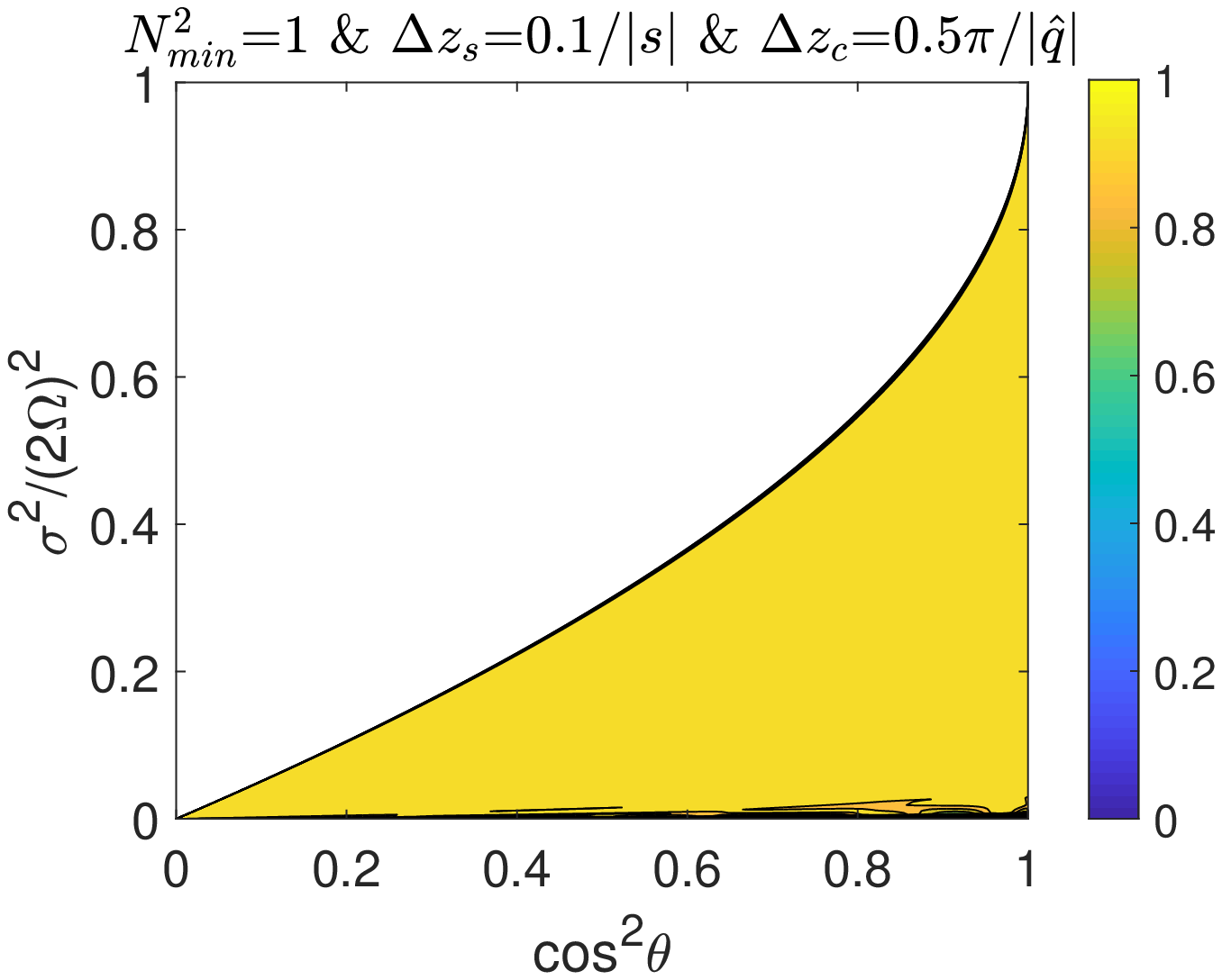}
\caption{}
\end{subfigure}
\begin{subfigure}{0.3\textwidth}
\includegraphics[width=\linewidth]{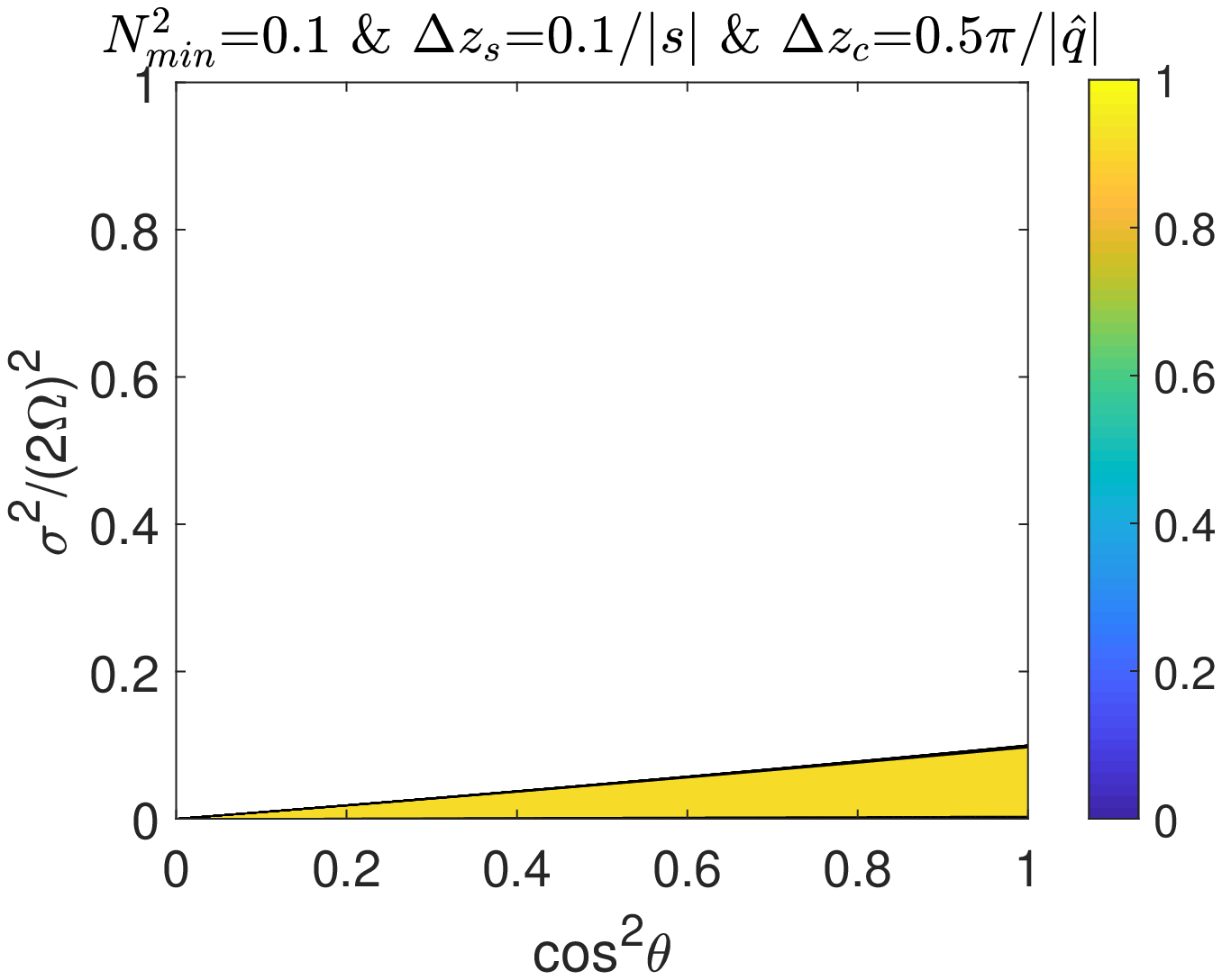}
\caption{}
\end{subfigure}

\medskip

\begin{subfigure}{0.3\textwidth}
\includegraphics[width=\linewidth]{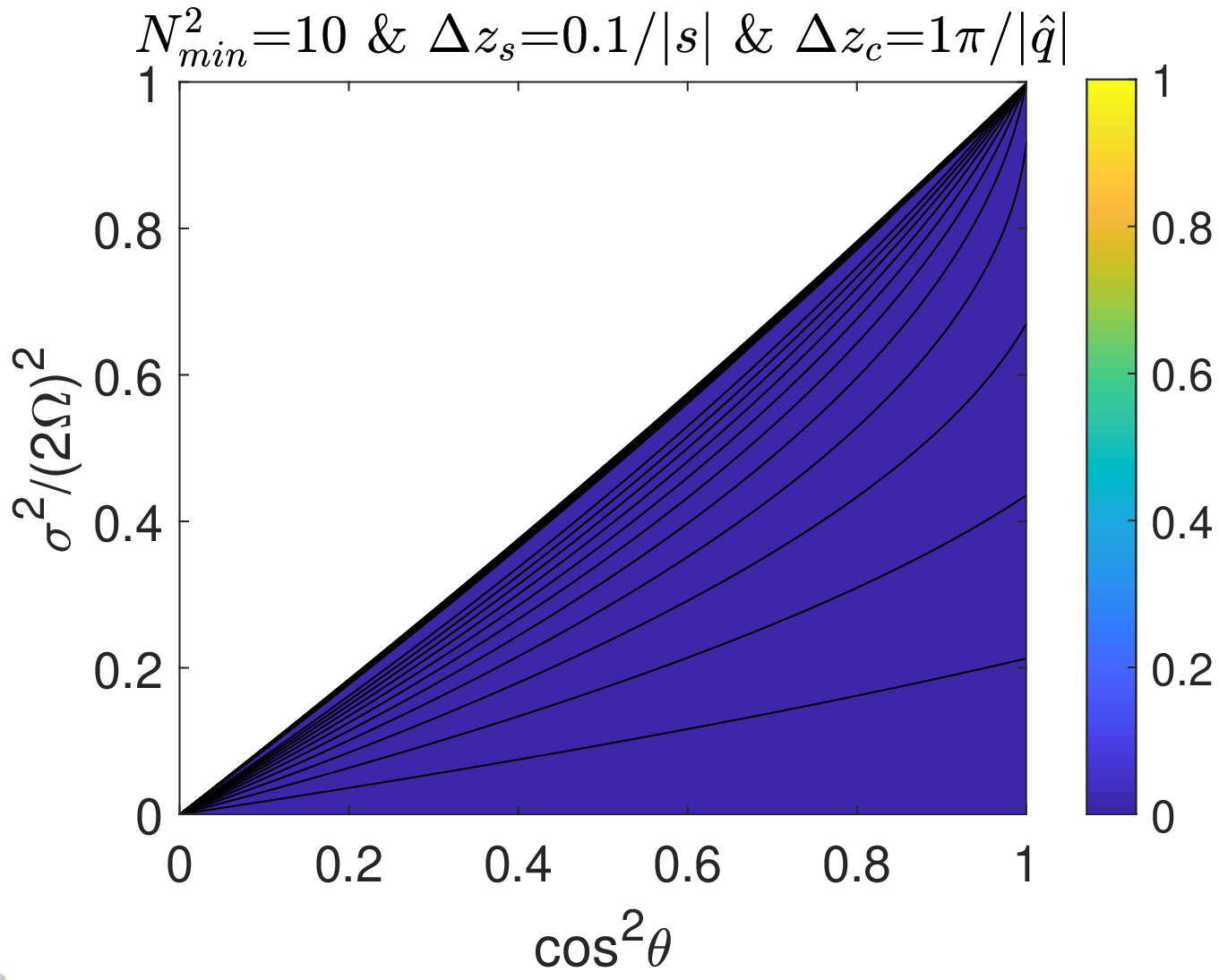}
\caption{}
\end{subfigure}
\begin{subfigure}{0.3\textwidth}
\includegraphics[width=\linewidth]{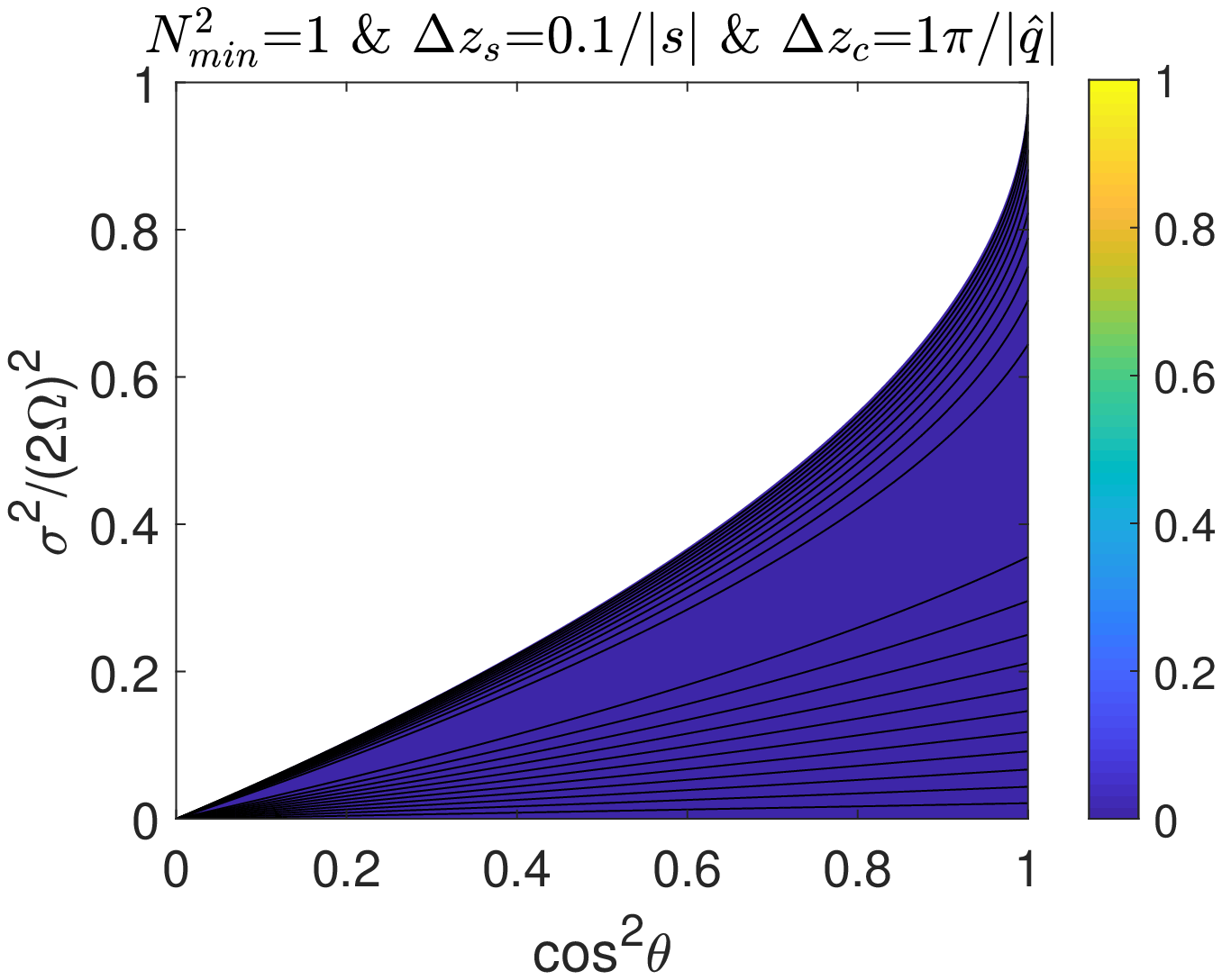}
\caption{}
\end{subfigure}
\begin{subfigure}{0.3\textwidth}
\includegraphics[width=\linewidth]{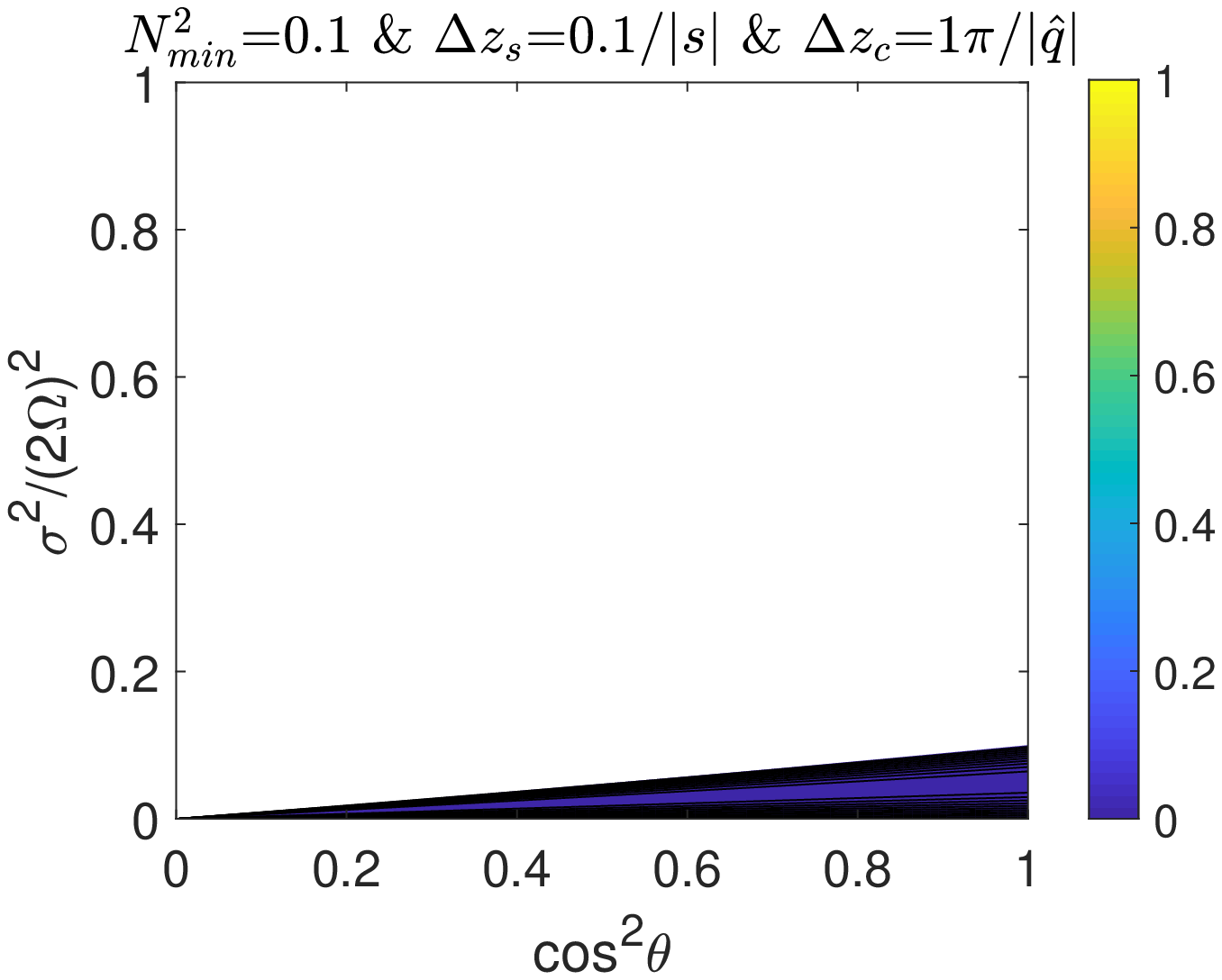}
\caption{}
\end{subfigure}

\caption{Contour plots of transmission ratios for the tunneling of inertial wave. The upper panel is for the three-layer structure; and the middle and the lower panels are for the multi-layer structures. (a-c)Transmission ratios for $N_{min}^2=10$ and $\Delta z'_{1}=(0.1,1,10)/|\hat{s}|$ in a three-layer structure. (d-f)Transmission ratios for $N_{min}^2=(10,1,0.1)$, $\Delta z_{c}=0.5\pi/|\hat{q}|$, and $\Delta z_{s}=0.1/|s|$ in a 101-layer structure. (g-i)Transmission ratios for $N_{min}^2=(10,1,0.1)$, $\Delta z_{c}=\pi/|\hat{q}|$, and $\Delta z_{s}=0.1/|s|$ in a 101-layer structure. Tunneling of inertial waves can only occur in colored regions. Regions are left white if tunneling of gravity waves is prohibited.\label{fig:f9}}
\end{figure}

Now we discuss the wave transmission of tunneling of inertial wave. Again, we first consider a three-layer structure. It is not difficult to obtain the transmission ratio
\begin{eqnarray}
\eta=16\left\{4(e^{-\hat{s}_{1}\Delta z'_{1}}+e^{\hat{s}_{1}\Delta z'_{1}})^{2}+\left[\left(\sqrt{\frac{\hat{s}_{1}^2}{q_{1}^2}}
+\sqrt{\frac{q_{1}^2}{\hat{s}_{1}^2}}\right)^{2}-4\right]\left[(e^{-\hat{s}_{1}\Delta z'_{1}}+e^{\hat{s}_{1}\Delta z'_{1}})^{2}-4\right]\right\}^{-1}~.
\end{eqnarray}
Again, we can show that efficient transmission requires that the thickness of the stable layer is much smaller than the e-folding decay distance $1/|\hat{s}_{1}|$. Figs.~\ref{fig:f9}(a-c) show the transmission ratios in the three-layer structure when the stable layer is strongly stratified. It clearly shows that the transmission is only efficient when $\hat{s}_{1}\Delta z'_{1}$ is small.

Now we discuss the tunneling of inertial wave in structure with more layers. Similarly, we can use the recursive relations to calculate the transmission ratio. Here we only give the result without showing details. Under the condition $|q|\Delta z_{c}=(\ell'+1/2)\pi$, we find the transmission ratio of tunneling of inertial waves in a multi-layer structure is
\begin{eqnarray}
\eta=\left\{\cos^2 M\beta+\frac{1}{4}\left[(\frac{\hat{s}}{q}-\frac{q}{\hat{s}})(e^{-\hat{s}\Delta z_{s}}+e^{\hat{s}\Delta z_{s}})\right]^2 \left[\left(\frac{\hat{s}}{q}+\frac{q}{\hat{s}}\right)^2 - (e^{-\hat{s}\Delta z_{s}}+e^{\hat{s}\Delta z_{s}})^2\right]^{-1}\sin^2 M\beta\right\}^{-1}~,
\end{eqnarray}
where
\begin{eqnarray}
\beta=\arg\left((e^{-\hat{s}\Delta z_{s}}+e^{\hat{s}\Delta z_{s}})+i\left[\left(\frac{\hat{s}}{q}+\frac{q}{\hat{s}}\right)^2 - (e^{-\hat{s}\Delta z_{s}}+e^{\hat{s}\Delta z_{s}})^2\right]^{1/2}\right),
\end{eqnarray}

The middle and lower panels of fig.~\ref{fig:f9} show the transmission ratios in a 101-layer structure. Again, we see that the efficiency of transmission mainly depends on the thicknesses of the convective and stable layers. For the transmission to be efficient, it requires that the thickness of stable layer ($\Delta z_{s}$) is much smaller than the e-folding decay distance $1/|\hat{s}|$, and the thickness of each convective layer is close to a multiple-and-a-half of the half wavelength. This result is similar to that obtained in the tunneling of gravity waves.

\section{Non-traditional effects}
\citet{sutherland2016internal} has discussed wave transmission in a multi-layer structure in traditional approximation. It is necessary to investigate the non-traditional effects on wave transmission in the multi-layer structure. Under the traditional approximation ($f_{s}^2=0$), the critical frequencies can be written as
\begin{eqnarray}
&\sigma_{1}^2=0, &\quad \sigma_{4}^2=f^2~, \\
&\sigma_{2}^2=\min(f^2,N_{min}^2), & \quad \sigma_{5}^2=\max(f^2,N_{min}^2)~,\\
&\sigma_{3}^2=\min(f^2,N_{max}^2), & \quad \sigma_{6}^2=\max(f^2,N_{max}^2)~.
\end{eqnarray}

\begin{figure}
\centering
\includegraphics[width=\linewidth]{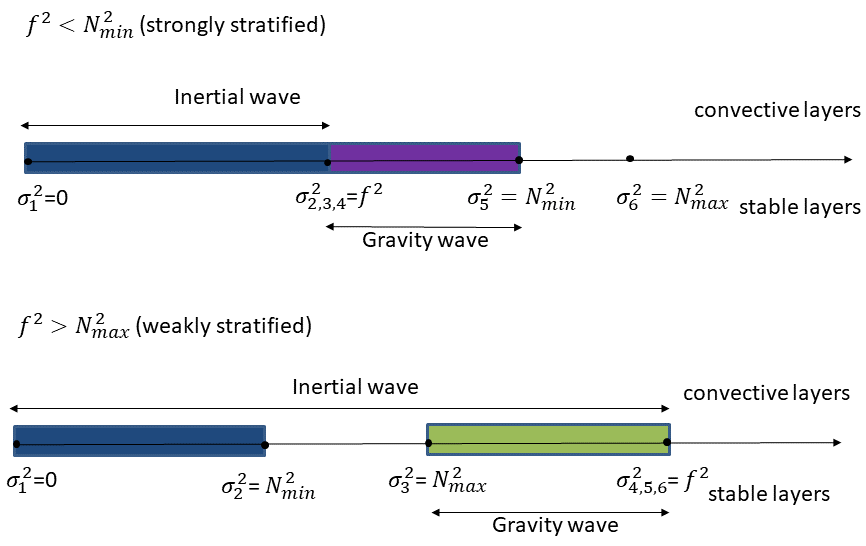}
\caption{Sketch plots of wave propagations in multiple convectively stable and unstable layers. Wave frequency ranges are shown above and below the middle arrow for convective and stable layers, respectively. Wave propagation occurs in the green region, tunneling of inertial wave occurs in the blue region, and tunneling of gravity wave occurs in the purple region. \label{fig:f10}}
\end{figure}

If stable layers are strongly stratified with $f^2<N_{min}^2$, then $\sigma_{2}^2=\sigma_{3}^2=\sigma_{4}^2=f^2$, $\sigma_{5}^2=N_{min}^2$, and $\sigma_{6}^2=N_{max}^2$. In such case, wave cannot propagate in both convective and stable layers, while tunneling of gravity waves occurs at $f^2<\sigma^2<N_{min}^2$ and tunneling of inertial waves occurs at $0<\sigma^2<f^2$ (see the upper panel of fig.~\ref{fig:f10}).

If stable layers are weakly stratified with $f^2>N_{max}^2$, then $\sigma_{4}^2=\sigma_{5}^2=\sigma_{6}^2=f^2$, $\sigma_{2}^2=N_{min}^2$, and $\sigma_{3}^2=N_{max}^2$. In such case, tunneling of gravity waves cannot occur, while wave can propagate in both convective and stable layers at $N_{max}^2<\sigma^2<f^2$, and tunneling of inertial waves occurs at $0<\sigma^2<N_{min}^2$ (see the lower panel of fig.~\ref{fig:f10}).

\begin{figure}
\centering
\begin{subfigure}{0.45\textwidth}
\includegraphics[width=\linewidth]{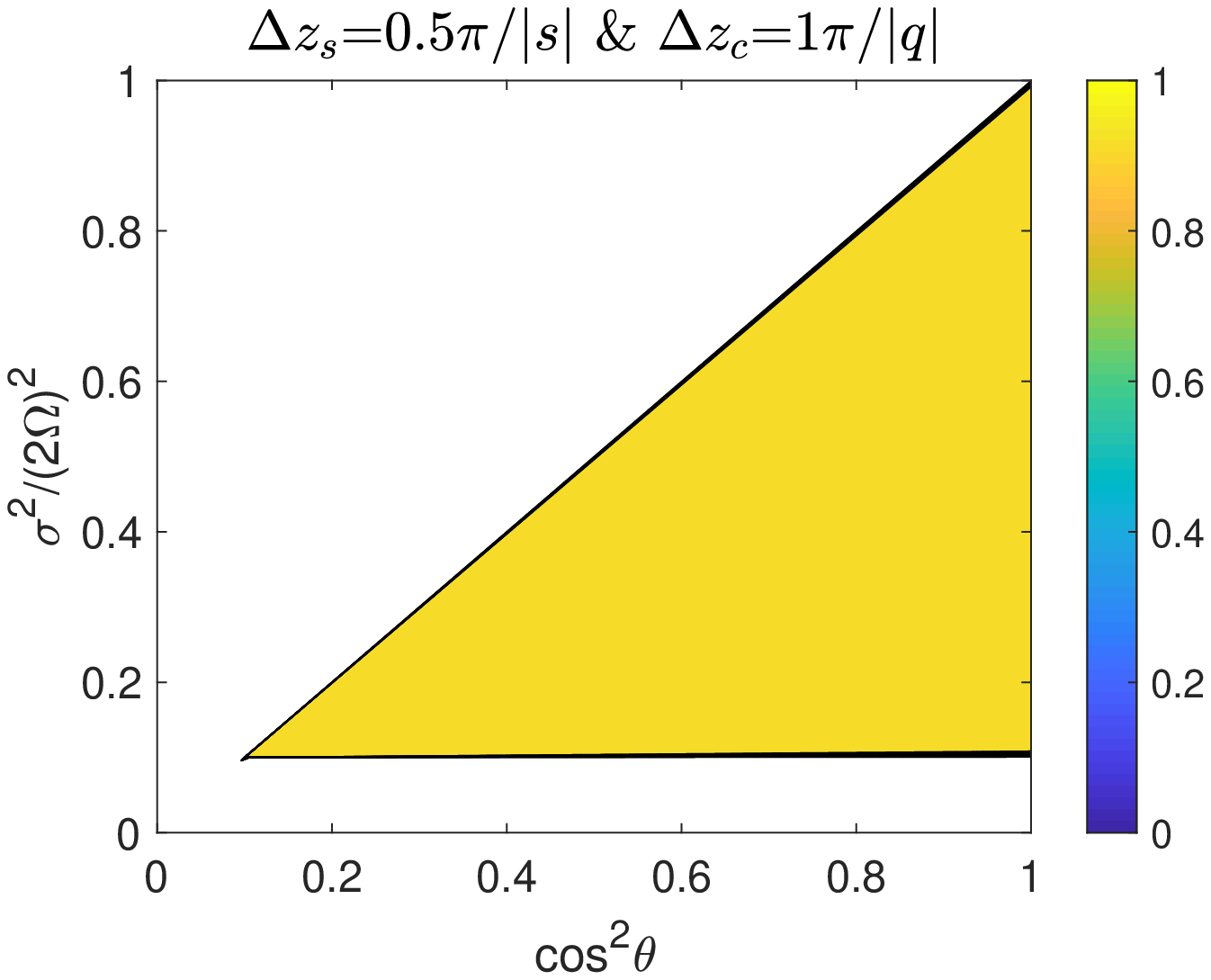}
\caption{}
\end{subfigure}
\begin{subfigure}{0.45\textwidth}
\includegraphics[width=\linewidth]{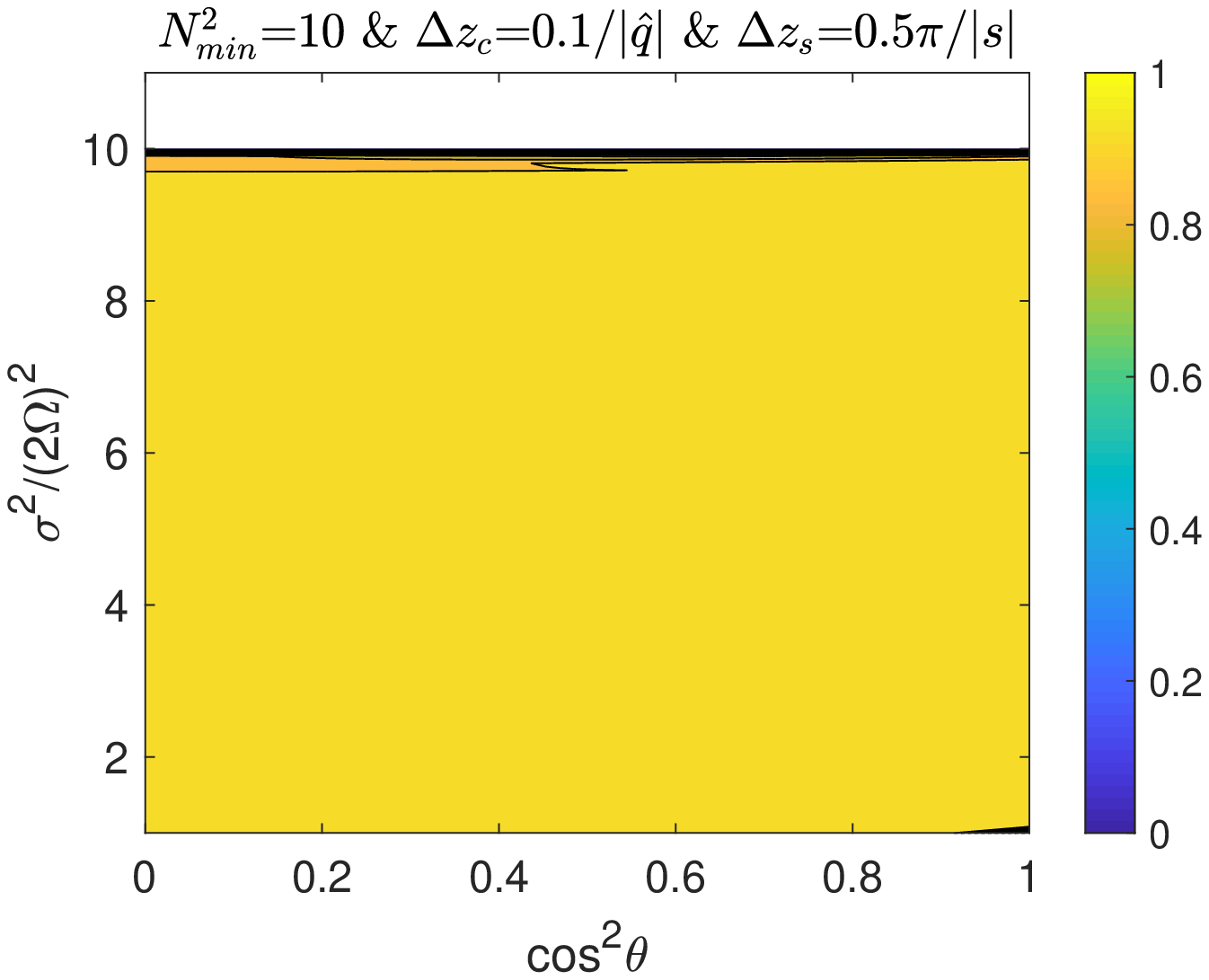}
\caption{}
\end{subfigure}

\medskip

\begin{subfigure}{0.45\textwidth}
\includegraphics[width=\linewidth]{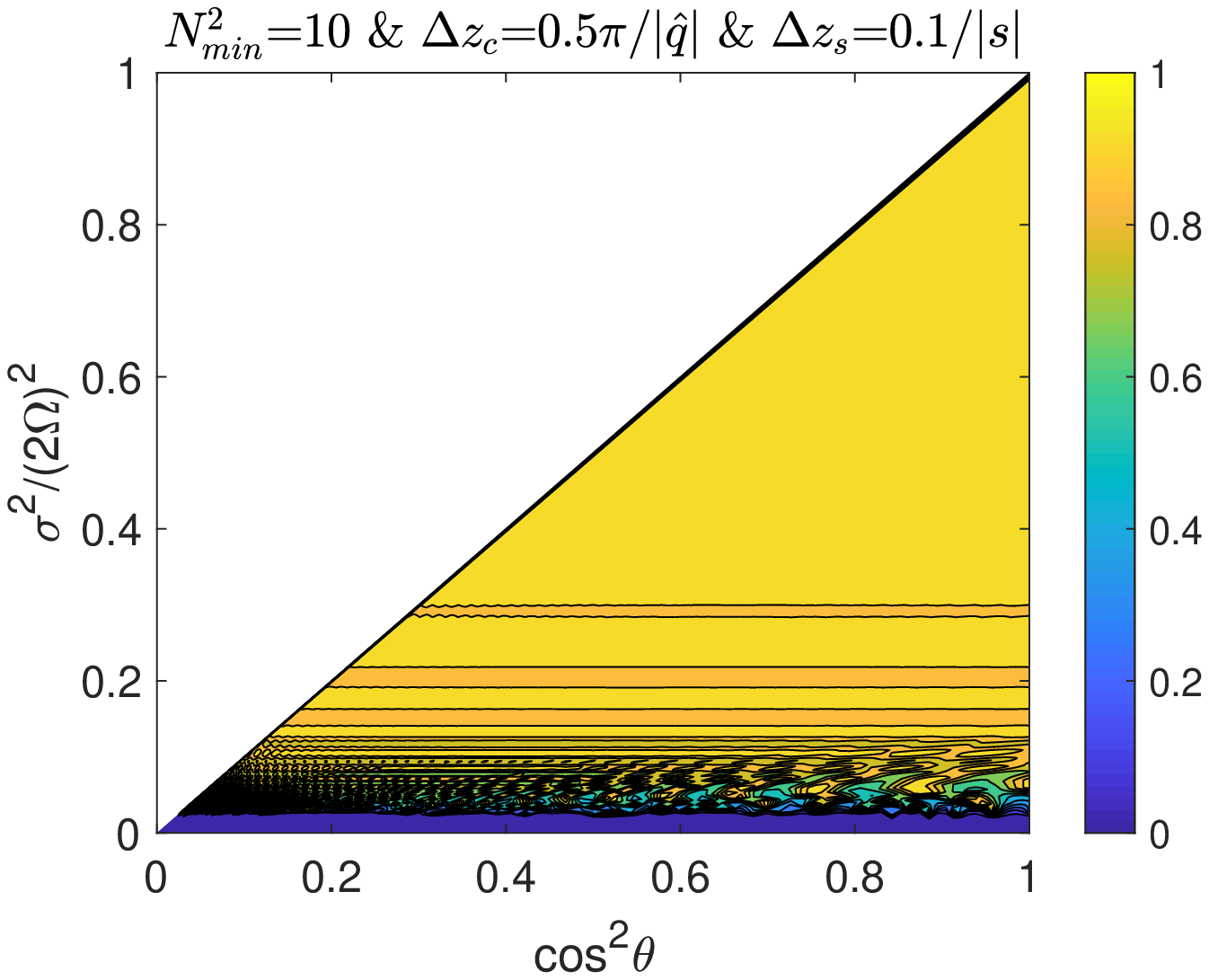}
\caption{}
\end{subfigure}
\begin{subfigure}{0.45\textwidth}
\includegraphics[width=\linewidth]{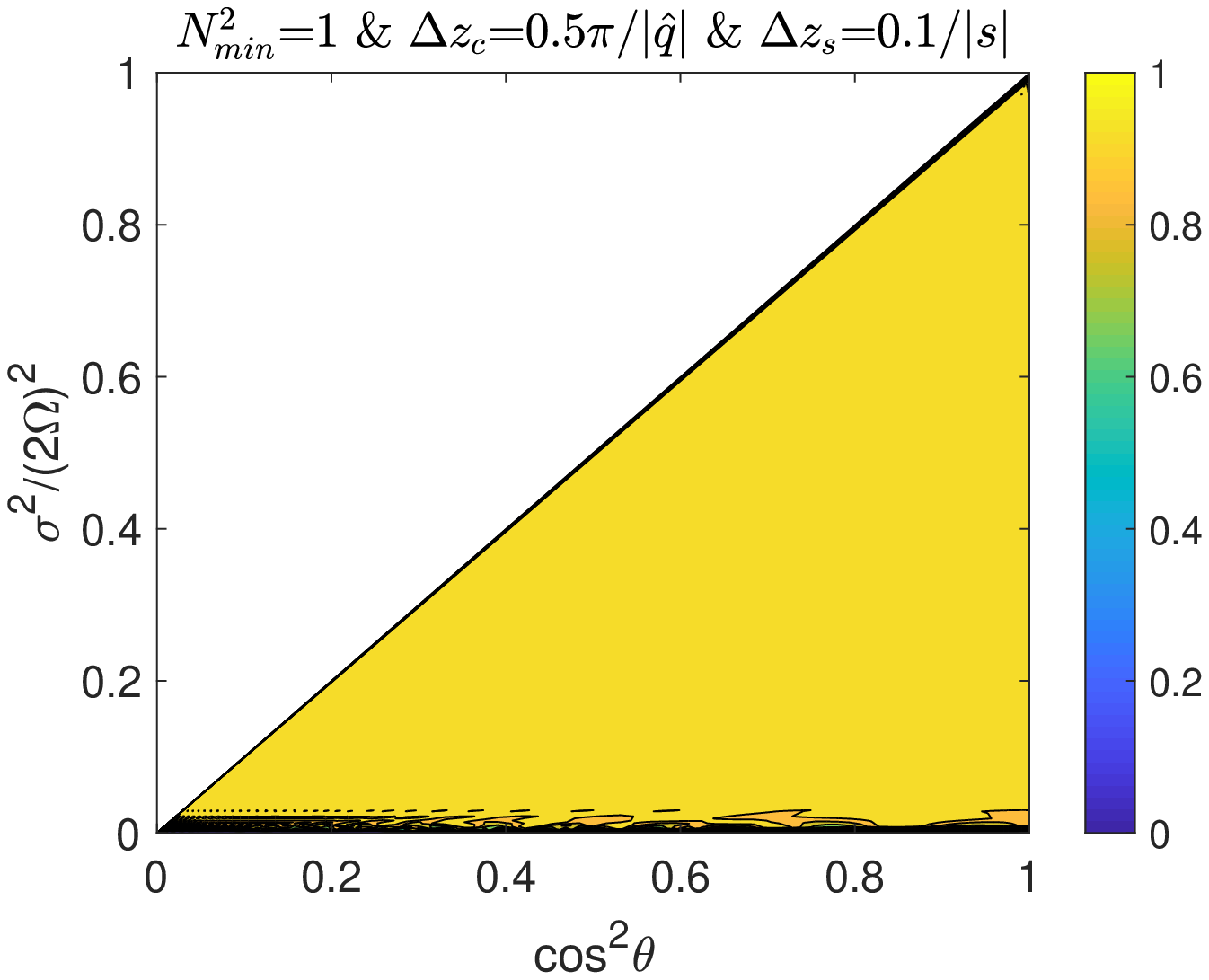}
\caption{}
\end{subfigure}

\medskip

\begin{subfigure}{0.45\textwidth}
\includegraphics[width=\linewidth]{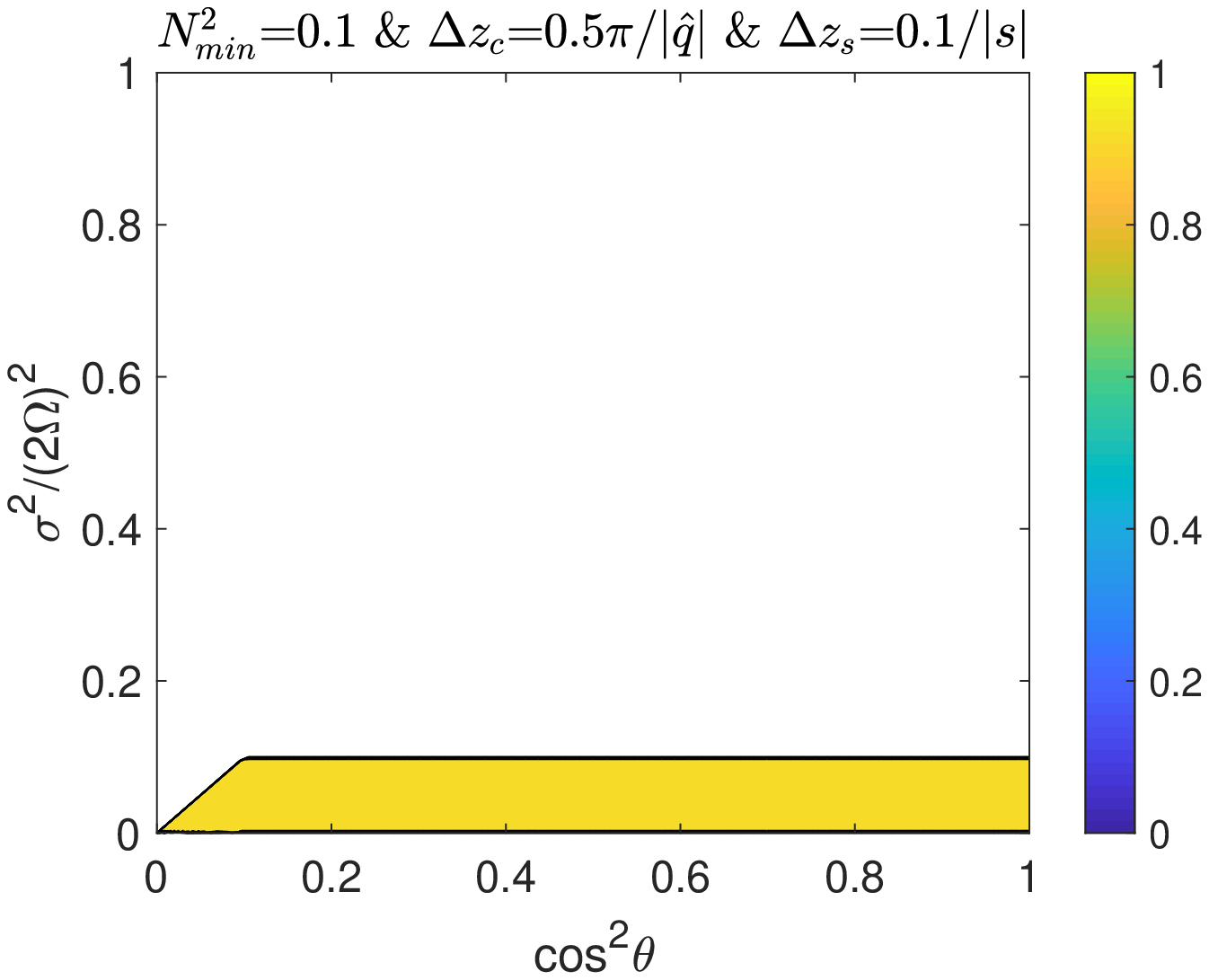}
\caption{}
\end{subfigure}

\caption{Wave transmissions with traditional approximation. (a)Transmission for wave propagation. The setting is identical to that of fig.~\ref{fig:f3}(g) except that $\tilde{f}_{s}^2=0$ in this figure. (b)Tunneling of gravity waves. The setting is the same as that in fig.~\ref{fig:f8}(a) except that $\tilde{f}_{s}^2=0$ in this figure. (c-e) Tunneling of inertial waves. The settings are the same as those in fig.~\ref{fig:f9}(d-f) except that $\tilde{f}_{s}^2=0$ in these figures. Colored and white regions have similar meaning as those mentioned in previous contour plots. \label{fig:f11}}
\end{figure}

If traditional approximation is made, wave propagation only occurs in weakly stratified flow, tunneling of gravity waves only occurs in strongly stratified flow, and tunneling of inertial waves can occur in both strongly and weakly stratified flows. When non-traditional effects are included, however, no similar restriction is obtained in wave propagation or tunneling. The non-traditional effects on wave propagation can already be seen by comparing the left panel of fig.~\ref{fig:f3} with other panels. In traditional approximation, $\tilde{f}_{s}$ is set to be zero and this can be achieved by setting $\sin\alpha=0$. The left panel of fig.~\ref{fig:f3} shows transmission ratios with small $\tilde{f}_{s}$, which are similar to situations with traditional approximation. If $\tilde{f}_{s}=0$, the colored regions in figs.~\ref{fig:f3}(a) and (d) vanish because wave propagation is prohibited when $N_{max}^2\geq f^2$; and the colored region in fig.~\ref{fig:f3}(c) will shrink into a triangle region below the diagonal line $\sigma^2/(2\Omega)^2=\cos^2\theta$ (see fig.~\ref{fig:f11}(a)). For weakly stratified flow ($N_{max}^2<f^2$), only sub-inertial waves ($\sigma^2<f^2$) can propagate with traditional approximation. However, super-inertial waves ($\sigma^2>f^2$) can propagate if non-traditional effects are taken into account.

With traditional approximation, tunneling of gravity waves can only occur when stable layers are strongly stratified($N_{min}^2>f^2$), and the wave frequency is smaller than buoyancy frequency ($\sigma^2<N_{min}^2$). When non-traditional effects present, we see from fig.~\ref{fig:f8} that tunneling of gravity waves can occur when stable layers are weakly stratified. Also, tunneling of gravity waves is possible for super-buoyancy-frequency waves.

For tunneling of inertial waves, comparing figs.~\ref{fig:f11}(c-e) with figs.~\ref{fig:f9}(d-f), we see that frequency ranges are overestimated in traditional approximation. It is especially true when stable layers are weakly stratified. From fig.~\ref{fig:f11}(c) and fig.~\ref{fig:f9}(d), we also see that traditional approximation has moderate effect on transmission ratio in the small frequency range.

\section{Summary}
In this paper, we have investigated wave transmissions in rotating stars or planets with multiple radiative and convective zones. Two situations have been considered: wave propagation and wave tunneling. For wave propagation, waves could propagate in both convective and stable layers. Previous studies on wave propagation in a two-layer structure with step function of stratification \citep{wei20,cai20} have shown that wave transmission is generally not efficient when the stable layer is strongly stratified (It is a typical behaviour despite that transmission can be efficient under certain conditions, such as at the critical latitude). In this work, however, we find that the wave transmission can be enhanced in a multiple-layer structure even though the stable layers are strongly stratified. We call this phenomenon `enhanced wave transmission'. Enhanced wave transmission can only occur when the top and bottom layers are both convective layers or stable layers. We have the following major findings on wave propagation:
\begin{itemize}
\setlength{\itemsep}{0pt}
\setlength{\parsep}{0pt}
\setlength{\parskip}{0pt}
\item[(1)] In a three-layer structure, transmission can be enhanced when the top and bottom layers (clamping layers) have similar buoyancy frequency, and the thickness of the middle layer is close to a multiple of the half wavelength of the propagating wave inside this layer.
\item[(2)] Enhancement of transmission can also take place in a multi-layer structure under similar conditions when clamping layers have similar properties, the thickness of each clamping layer is close to a multiple of the half wavelength of the propagating wave, and the total thickness of each embedded layer is close to a multiple of the half wavelength of the propagating wave. Efficient transmission can take place even when stable layers are strongly stratified. We call this phenomenon `resonant propagation'.
\end{itemize}

For wave tunneling, there are two cases: the tunneling of gravity wave, and the tunneling of inertial wave. In the first case, waves can propagate in stable layers but are evanescent in convective layers. In the second case, on the other hand, waves can propagate in convective layers but are evanescent in stable layers. We have the following major findings on wave tunneling:
\begin{itemize}
\setlength{\itemsep}{0pt}
\setlength{\parsep}{0pt}
\setlength{\parskip}{0pt}
\item[(4)] The tunneling of gravity wave can be efficient when stable layers have similar buoyancy frequencies, and the thickness of each embedded convective layer is much smaller than the corresponding e-folding decay distance, and the thickness of each stable layer is close to a multiple-and-a-half of half wavelength. The latter condition is unnecessary if the structure is three-layer. We call this `resonant tunneling of gravity waves'.
\item[(5)] The tunneling of inertial wave can be efficient when stable layers have similar buoyancy frequencies, and the thickness of each stable layer is much smaller than the corresponding e-folding decay distance, and each convective layer thickness is close to a multiple-and-a-half of half wavelength. The latter condition is unnecessary if the structure is three-layer. We call this `resonant tunneling of inertial waves'.
\item[(6)] The efficiency of the tunneling mainly depends on the layer thicknesses, the wavelengths, and the e-folding decay distances.
\end{itemize}

It would be interesting to investigate the tunneling of gravity waves in a non-rotating fluid. It is a special case with $f=0$ and $\tilde{f_{s}}=0$. In such case, we can easily obtain that $\hat{q}_{m}^2=k^2$ and $s_{m}^2=(N_{m}^2/\sigma^2-1)k^2$, where $\sigma^2<N_{m}^2$. Then the conclusion (4) can be directly applied to this special case.

Table~\ref{table:tab1} summarizes conditions on efficient transmissions of wave propagation and tunneling when all stable layers have similar buoyancy frequencies. For tunneling waves, clamping layers should have similar properties. Table~\ref{table:tab1} only lists the conditions in such structures. The first column of table~\ref{table:tab1} lists four types of wave transmissions: propagation with convective layers embedded, propagation with stable layers embedded, tunneling of gravity waves, and tunneling of inertial waves. The second column gives the frequency ranges. The third and fourth columns show the conditions for efficient transmissions. From the table, we see that the conditions on efficient wave transmissions are significantly different among tunneling and propagative waves.

\begin{table}
  \begin{center}
\def~{\hphantom{0}}
  \begin{tabular}{lccc}
      & $\sigma^2$ & $\Delta z_{c}$ & $\Delta z_{s}$ \\[3pt]
      Propagation with CLs embedded in SLs & $(\sigma_{3}^2,\sigma_{4}^2)$ & $M\Delta z_{c} \sim \ell\lambda_{c}/2$  & $\Delta z_{s} \sim \ell'\lambda_{s}/2$ \\
      Propagation with SLs embedded in CLs & $(\sigma_{3}^2,\sigma_{4}^2)$ & $\Delta z_{c} \sim \ell'\lambda_{c}/2$  &  $M\Delta z_{s} \sim \ell\lambda_{s}/2$\\
      Tunneling of GWs & $(\sigma_{4}^2,\sigma_{5}^2)$ & $\Delta z_{c}\ll \lambda_{c}$  &  $\Delta z_{s} \sim (\ell+1/2)\lambda_{s}/2$\\
      Tunneling of IWs & $(\sigma_{1}^2,\sigma_{2}^2)$  & $\Delta z_{c} \sim (\ell+1/2)\lambda_{c}/2$ & $\Delta z_{s}\ll \lambda_{s}$ \\
  \end{tabular}
  \caption{Summary on efficient wave transmissions in a multi-layer structure.}
  \label{table:tab1}
  \end{center}
  \begin{tablenotes}
\item Note: $\lambda_{c}$ is the wavelength or decay distance in convective layer. $\lambda_{s}$ is the wavelength or decay distance in the stable layer. $\Delta z_{c}$ and $\Delta z_{s}$ are the thicknesses of the convective and stable layers, respectively. $M$ is the number of embedded layers. $\ell$ and $\ell'$ is a non-negative integer. IW and GW denote inertial and gravity waves, respectively. CLs and SLs denote convective and stable layers, respectively. Here we only consider the situation that all stable layers have similar buoyancy frequencies, and clamping layers have similar properties.
\end{tablenotes}
\end{table}

Our findings have interesting implications in gaseous planets. In a multi-layer structure, \citet{belyaev15} found that the g-mode with vertical wavelengths smaller than the layer thickness are evanescent in gaseous planets. It is true for tunneling of g-mode waves. However, if considering wave propagation, g-mode waves can transmit efficiently even when the wavelength is smaller than the layer thickness. \citet{andre17} have made promising progress in the study of wave transmission in multi-layer structures, which reveals that wave transmission can be enhanced when the incident wave is resonant with waves in adjacent layers with half wavelengths equal to the layer depth. Their result is consistent with our derivations. By deriving a group of exact solutions of wave transmission coefficients in multi-layer structures, we provide a mathematical explanation for why the transmission can be enhanced in a multi-layer structure.
In addition, our analysis shows that wave transmission can also be enhanced in tunneling of gravity waves or inertial waves. Conditions on `resonant propagation' and `resonant tunneling' have been provided. \citet{pontin20} have conducted interesting research on wave propagation in a multi-layer structure in a non-rotating sphere, and found that wave transmissions are efficient for very large wavelength waves. It already shares some similarities with our analysis on tunneling of gravity waves in the {\it f}-plane. It has to be emphasized that our model is derived under the assumptions in local {\it f}-plane. Application of the results to global sphere should be very careful because of the following reasons. First, only short wavelengths are considered in the local model. Wave transmissions of global scale waves have not been discussed. Second, the geometrical effect has not been taken into account. This is important for waves propagating in a global sphere. For example, super-inertial waves propagating from equator to poles may change to sub-inertial waves across critical colatitudes, where waves can also be transmitted or reflected in the meridional direction \citep{gerkema2005near,shrira2010inertia,rieutord1997inertial,rieutord2001inertial}. Extending our work in full spherical geometry will be an improving direction in the future.

Our findings may also have implications in Earth oceans, and stars. It has been observed that there exists multi-layer structure in Arctic Ocean, with thin stratified layers separated by mixed layers created by double diffusion \citep{rainville2008mixing}. By performing a joint theoretical and experimental study, \citet{ghaemsaidi2016impact} revealed that the Arctic Ocean has a rich transmission behavior. With their data, we can have a simple estimation on wave transmission in Arctic Ocean by using our model. For the structure of the Arctic Ocean presented in \citet{ghaemsaidi2016impact}, $N$ in stratified layers are about $1.6\times 10^{-3}-3.2\times 10^{-3}s^{-1}$; the inertial frequency $f$ is about $1.4 \times 10^{-4}s^{-1}$; the thicknesses of embedded convective layers are of the order $O(1)$; The total depth of the multi-layer structure is about $30m$, which are approximately separated into 14 stable layers. Since $N\gg f$, gravity waves are expected to transport across the multi-layer structure by tunneling. From table~\ref{table:tab1}, we see that waves with wavelength $\lambda_{s}\sim 8.6m$ are possible to transmit efficiently by resonant tunneling. Near-inertial waves with wavelengths $10m-50m$ are also possible to transmit efficiently. This has been verified in \citet{ghaemsaidi2016impact}. Double diffusion also occurs in stars. Our model may also provide some insights for wave transmission in stars. The interior structures are different for different types of stars. For example, as studied in \citet{cai2014numerical}, late-type stars have a convectively stable-unstable-stable three-layer structure; A-F type stars generally have complicated internal structures with two separated convectively unstable layers (for example, some of them have a unstable-stable-unstable three-layer structure, and some of them have
a stable-unstable-stable-unstable-stable five-layer structure), and massive stars have a unstable-stable two-layer structure. For waves excited at the innermost layer, resonant wave propagation from the innermost to the outermost layers probably can be taken place in late-type and A-F stars since the top and bottom layers are both convectively stable or unstable. However, if waves are excited in the second innermost layer, enhanced wave transmission is unlikely to occur from this layer to the outermost layer, because the bottom and top layers of the interested region are different.
For massive stars, enhanced wave transmission are unlikely to take place because the properties of the top and bottom layers are different. We have to mention that the stratified structure specified in our model is ideal. In our Boussinesq model, density variation and viscous effect has been ignored. In real stars, however, density variation and viscous effect may be important. In addition, our model assumes that buoyancy frequency changes abruptly across interfaces between convective and stable layers. For real stars, the change is likely to be smoother. Previous investigations of wave transmission in two-layer structures with smoothly varying buoyancy frequencies have shown significant differences.
Models with more realistic settings are desirable in the future.

\section*{Acknowledgements}
T.C. has been supported by NSFC (No.11503097), the Guangdong Basic and Applied Basic Research Foundation (No.2019A1515011625), the Science and Technology Program of Guangzhou (No.201707010006), the Science and Technology Development Fund, Macau SAR (Nos.0045/2018/AFJ, 0156/2019/A3), and the China Space Agency Project (No.D020303). C.Y. has been supported by the National Natural Science Foundation of China (grants 11373064, 11521303, 11733010, 11873103), Yun-nan National Science Foundation (grant 2014HB048), and Yunnan Province (2017HC018). X.W. has been supported by National Natural Science Foundation of China (grant no.11872246) and Beijing Natural Science Foundation (grant no. 1202015). This work is partially supported by Open Projects Funding of the State Key Laboratory of Lunar and Planetary Sciences.\\

\noindent{\bf Declaration of Interests\bf{.}} The authors report no conflict of interest.\\


\appendix
\section{}\label{appendixa}
Equations (\ref{eq11}-\ref{eq14}) can be written in a matrix form
\begin{eqnarray}
\setlength{\arraycolsep}{2pt}
\renewcommand{\arraystretch}{1.3}
&&\bm{S}_{m}\left[
\begin{array}{c}
a_{m}\\
b_{m}
\end{array}
\right]=\bm{\Lambda}_{m} \bm{Q}_{m}\left[
\begin{array}{c}
c_{m}\\
d_{m}
\end{array}
\right]~,\\
&&\widetilde{\bm{\Lambda}}_{m} \widetilde{\bm{Q}}_{m}\left[
\begin{array}{c}
c_{m}\\
d_{m}
\end{array}
\right]= \widetilde{\bm{S}}_{m}\left[
\begin{array}{c}
a_{m+1}\\
b_{m+1}
\end{array}
\right]~.
\end{eqnarray}
where
\begin{equation}
\setlength{\arraycolsep}{2pt}
\renewcommand{\arraystretch}{1.4}
\bm{S}_{m} = \left[
\begin{array}{cc}
  e^{-is_{m}z_{m}} &  e^{is_{m}z_{m}} \\
  -e^{-is_{m}z_{m}} & e^{is_{m}z_{m}}
\end{array}
\right],~\quad
\bm{Q}_{m} = \left[
\begin{array}{cc}
  e^{-iq_{m}z_{m}} &  e^{iq_{m}z_{m}} \\
  -e^{-iq_{m}z_{m}} & e^{iq_{m}z_{m}}
\end{array}
\right],~\quad
\bm{\Lambda}_{m} = \left[
\begin{array}{cc}
  1 & 0 \\
  0 & q_{m}/s_{m}
\end{array}
\right]~,\\
\end{equation}

\begin{equation}
\widetilde{\bm{S}}_{m} = \left[
\begin{array}{cc}
  e^{-is_{m+1}z'_{m}} &  e^{is_{m+1}z'_{m}} \\
  -e^{-is_{m+1}z'_{m}} & e^{is_{m+1}z'_{m}}
\end{array}
\right],~\quad
\widetilde{\bm{Q}}_{m} = \left[
\begin{array}{cc}
  e^{-iq_{m}z'_{m}} &  e^{iq_{m}z'_{m}} \\
  -e^{-iq_{m}z'_{m}} & e^{iq_{m}z'_{m}}
\end{array}
\right],~\quad
\widetilde{\bm{\Lambda}}_{m} = \left[
\begin{array}{cc}
  1 & 0 \\
  0 & q_{m}/s_{m+1}
\end{array}
\right]~.
\end{equation}
Synthesizing these equations, we obtain the recursive relation
\begin{eqnarray}
\left[
\begin{array}{c}
a_{m}\\
b_{m}
\end{array}
\right]=\bm{T}_{m,m+1}
\left[
\begin{array}{c}
a_{m+1}\\
b_{m+1}
\end{array}
\right]~,
\end{eqnarray}
where the transfer matrix
\begin{eqnarray}
&&\bm{T}_{m,m+1}=\bm{S}_{m}^{-1}\bm{\Lambda}_{m} \bm{Q}_{m}\widetilde{\bm{Q}}_{m}^{-1}\widetilde{\bm{\Lambda}}_{m}^{-1}\widetilde{\bm{S}}_{m}=\left[
\begin{array}{cc}
  \widehat{T}_{11} & \widehat{T}_{12} \\
  \widehat{T}_{21} & \widehat{T}_{22}
\end{array}
\right]~,
\end{eqnarray}
and
\begin{eqnarray}
&&\widehat{T}_{11}=\frac{1}{4}e^{i(s_{m}z_{m}-s_{m+1}z'_{m})}[(1+\frac{q_{m}}{s_{m}}+\frac{s_{m+1}}{s_{m}}+\frac{s_{m+1}}{q_{m}})e^{iq_{m}\Delta z'_{m}}\\\nonumber
&& \quad +(1-\frac{q_{m}}{s_{m}}+\frac{s_{m+1}}{s_{m}}-\frac{s_{m+1}}{q_{m}})e^{-iq_{m}\Delta z'_{m}}]~,\\
&&\widehat{T}_{12}=\frac{1}{4}e^{i(s_{m}z_{m}+s_{m+1}z'_{m})}[(1+\frac{q_{m}}{s_{m}}-\frac{s_{m+1}}{s_{m}}-\frac{s_{m+1}}{q_{m}})e^{iq_{m}\Delta z'_{m}}\\\nonumber
&& \quad +(1-\frac{q_{m}}{s_{m}}-\frac{s_{m+1}}{s_{m}}+\frac{s_{m+1}}{q_{m}})e^{-iq_{m}\Delta z'_{m}}]~,
\end{eqnarray}
with $\widehat{T}_{21}=\widehat{T}_{12}^{*}$ and $\widehat{T}_{22}=\widehat{T}_{11}^{*}$.

\section{}\label{appendixb}
For the tunneling of gravity wave, the width of the frequency domain is
\begin{eqnarray}
\sigma_{5}^2-\sigma_{4}^2=\frac{1}{2}[(N_{min}^2 -f^2-\tilde{f}_{s}^2)+\sqrt{(N_{min}^2 -f^2-\tilde{f}_{s}^2)^2+4N_{min}^2 \tilde{f}_{s}^2}]~.
\end{eqnarray}
The monotonicity of the frequency width is equivalent to that of the function
\begin{eqnarray}
G(\theta,\mu_{1},\mu_{2})=G_{1}(\theta,\mu_{1},\mu_{2})+\sqrt{G_{1}^2 (\theta,\mu_{1},\mu_{2})+4G_{2}(\theta,\mu_{1},\mu_{2})}~,
\end{eqnarray}
where $G_{1}(\theta,\mu_{1},\mu_{2})=\mu_{2} -\cos^2\theta-\mu_{1}\sin^2\theta$, $G_{2}(\theta,\mu_{1},\mu_{2})=\mu_{1}\mu_{2}\sin^2\theta$, $\mu_{1}=\sin^2\alpha$, and $\mu_{2}=N_{min}^2/(2\Omega)^2$. To analyze the monotonicity of $G(\theta,\mu_{1},\mu_{2})$ on $\theta$, we compute
\begin{eqnarray}
G_{\theta}=G_{1\theta}+\frac{G_{1}G_{1\theta}+2G_{2\theta}}{\sqrt{G_{1}^2 +4G_{2}}}~.
\end{eqnarray}
Because $G_{1\theta}=\sin2\theta(1-\mu_{1})>0$ and $G_{2\theta}=\mu_{1}\mu_{2}\sin2\theta>0$, we have
\begin{eqnarray}
G_{\theta}=\frac{(\sqrt{G_{1}^2 +4G_{2}}+G_{1})G_{1\theta}+2G_{2\theta}}{\sqrt{G_{1}^2 +4G_{2}}}\geq 0~.
\end{eqnarray}
Therefore, the frequency width always increases with $\theta$.

To analyze the monotonicity of $G$ on $\mu_{1}$, we compute
\begin{eqnarray}
G_{\mu_{1}}&&=G_{1\mu_{1}}+\frac{G_{1}G_{1\mu_{1}}+2G_{2\mu_{1}}}{\sqrt{G_{1}^2 +4G_{2}}}\\
&&=-\sin^2\theta+\frac{\sqrt{G_{1}^2+4G_{2}+4\mu_{2}\cos^2\theta}}{\sqrt{G_{1}^2 +4G_{2}}}\sin^2\theta \geq 0~.
\end{eqnarray}
Therefore, the frequency width always increases with $\mu_{1}$.

To analyze the monotonicity of $G$ on $\mu_{2}$, we compute
\begin{eqnarray}
G_{\mu_{2}}&&=G_{1\mu_{2}}+\frac{G_{1}G_{1\mu_{2}}+2G_{2\mu_{2}}}{\sqrt{G_{1}^2 +4G_{2}}}\\
&&=\frac{\sqrt{G_{1}^2 +4G_{2}}+G_{1}+2\mu_{1}\sin^2\theta}{\sqrt{G_{1}^2 +4G_{2}}}>0~.
\end{eqnarray}
Therefore, the frequency width always increases with $\mu_{2}$.

\section{}\label{appendixc}
For the tunneling of inertial wave, the width of the frequency domain is
\begin{eqnarray}
\sigma_{2}^2-\sigma_{1}^2=\frac{1}{2}[(f^2+\tilde{f}_{s}^2+N_{min}^2)- \sqrt{(f^2+\tilde{f}_{s}^2+N_{min}^2)^2-4N_{min}^2 f^2}]~.
\end{eqnarray}
The monotonicity of the frequency width is equivalent to that of the function
\begin{eqnarray}
H(\theta,\mu_{1},\mu_{2})=H_{1}(\theta,\mu_{1},\mu_{2})-\sqrt{H_{1}^2(\theta,\mu_{1},\mu_{2})-4H_{2}(\theta,\mu_{1},\mu_{2})}~.
\end{eqnarray}
where $H_{1}(\theta,\mu_{1},\mu_{2})=\mu_{2}+\cos^2\theta +\mu_{1}\sin^2\theta$, $H_{2}(\theta,\mu_{1},\mu_{2})=\mu_{2}\cos^2\theta$, $\mu_{1}=\sin^2\alpha$, and $\mu_{2}=N_{min}^2/(2\Omega)^2$. The derivative of $H(\theta,\mu_{1},\mu_{2})$ to $\theta$ is
\begin{eqnarray}
H_{\theta}=H_{1\theta}-\frac{H_{1}H_{1\theta}-2H_{2\theta}}{\sqrt{H_{1}^2-4H_{2}}}~.
\end{eqnarray}
Since $H_{1\theta}=(\mu_{1}-1)\sin2\theta$ and $H_{2\theta}=-\mu_{2}\sin2\theta$, we have
\begin{eqnarray}
H_{\theta}=\frac{(\sqrt{H_{1}^2-4H_{2}}-H_{1})H_{1\theta}+2H_{2\theta}}{\sqrt{H_{1}^2-4H_{2}}} \leq 0~.
\end{eqnarray}
Therefore the width of the frequency domain decreases with $\theta$.
The derivative of $H(\theta,\mu_{1},\mu_{2})$ to $\mu_{1}$ is
\begin{eqnarray}
H_{\mu_{1}}=H_{1\mu_{1}}-\frac{H_{1}H_{1\mu_{1}}-2H_{2\mu_{1}}}{\sqrt{H_{1}^2-4H_{2}}}~.
\end{eqnarray}
Since $H_{1\mu_{1}}=\sin^2\theta$ and $H_{2\mu_{1}}=0$, we have
\begin{eqnarray}
H_{\mu_{1}}=\frac{(\sqrt{H_{1}^2-4H_{2}}-H_{1})\sin^2\theta}{\sqrt{H_{1}^2-4H_{2}}}\leq 0~.
\end{eqnarray}
Therefore the width of the frequency domain decreases with $\mu_{1}$.
The derivative of $H(\theta,\mu_{1},\mu_{2})$ to $\mu_{2}$ is
\begin{eqnarray}
H_{\mu_{2}}=H_{1\mu_{2}}-\frac{H_{1}H_{1\mu_{2}}-2H_{2\mu_{2}}}{\sqrt{H_{1}^2-4H_{2}}}~.
\end{eqnarray}
Since $H_{1\mu_{2}}=1$ and $H_{2\mu_{2}}=\cos^2\theta$, we have
\begin{eqnarray}
H_{\mu_{2}}&&=\frac{\sqrt{H_{1}^2-4H_{2}}-H_{1}+2\cos^2\theta}{\sqrt{H_{1}^2-4H_{2}}}~.
\end{eqnarray}
It is found that
\begin{eqnarray}
H_{1}^2-4H_{2}-(H_{1}-2\cos^2\theta)^2=\mu_{1}\sin^2 2\theta \geq 0~,
\end{eqnarray}
then we have
\begin{eqnarray}
H_{\mu_{2}}\geq 0~.
\end{eqnarray}
Thus the width of the frequency domain increases with $\mu_{2}$.

\bibliographystyle{jfm}
\bibliography{main}

\end{document}